\documentclass[12pt]{article}
\pdfoutput=1
\usepackage{graphicx}
\usepackage{amsfonts}
\usepackage{amsmath}
\usepackage[mathscr]{eucal}
\usepackage{booktabs}

\def\href#1#2{#2}   
\newif\ifdraft
\draftfalse	  
\reversemarginpar   
\makeatletter
\let\mlabel=\label
\let\adkendequation=\endequation%
\def\endequation{\adkendequation\adklabel\global\@ignoretrue}
\let\adkendeqnarray=\endeqnarray%
\def\endeqnarray{\adkendeqnarray\adklabel\global\@ignoretrue}
\newbox\marglabbox
\def\adklabel{\ifvoid\marglabbox\else\marginpar{\unhbox\marglabbox}\fi}
\def\label#1{\ifdraft\ifmmode%
  \global\setbox\marglabbox=\hbox{\hfill\fbox{\tiny\verb*~#1~}}%
  \else\ifinner\else\marginpar{\hfill\fbox{\tiny\verb*~#1~}}%
  \fi\fi\fi \mlabel{#1}}
\makeatother
\ifdraft%
\fi
\def\eusm{\mathscr}
\font\twelvefrak=eufm10 scaled 1200
\font\tenfrak=eufm10
  \newfam\frakfam
  
  \textfont\frakfam=\twelvefrak
  \scriptfont\frakfam=\tenfrak
  \scriptscriptfont\frakfam=\scriptfont\frakfam
\def\sqr#1#2{{\vcenter{\hrule height.#2pt
   \hbox{\vrule width.#2pt height#1pt \kern#1pt
      \vrule width.#2pt}
   \hrule height.#2pt}}}

\def\bsqr#1#2{{\vrule width #1pt height#2pt}}
\def\bsquare{{\mathchoice\bsqr66\bsqr66\bsqr33\bsqr33}}
\def\badbreak{\penalty1000}

\def\sgn{\mathop{\rm sgn}}                  
\newcommand{\cP}{{\cal P}}                  
\newcommand{\euT}{{\eusm T}}                
\newcommand{\psibar}{{\bar\psi}}            
\def\rpc{x}                                 
\def\Xg{X}                                  
\def\Fg{{\eusm X}}                          
\def\xd{P}                                  
\def\cop{{C}}                               
\def\chps{{\Lambda_{ch}}}                   
\def\df{{\cP}_f}                            
\def\db{{\cP}_b}                            
\def\dop{{\xd}_r}                           
\def\Fgr{{\Fg}_r}                           
\begin{document}

\begin{center}
{\Large{\bf Chiral Symmetry Breaking and Chiral Polarization:}} \\
\vspace*{.14in}
{\Large{\bf Tests for Finite Temperature and Many Flavors}} \\
\vspace*{.24in}
{\large{Andrei Alexandru$^1$ and Ivan Horv\'ath$^2$}}\\
\vspace*{.24in}
$^1$The George Washington University, Washington, DC, USA\\
$^2$University of Kentucky, Lexington, KY, USA

\vspace*{0.15in}
{\large{Nov 18 2014}}

\end{center}

\vspace*{0.05in}

\begin{abstract}

\noindent
It was recently conjectured that, in SU(3) gauge theories with fundamental 
quarks, {\em valence} spontaneous chiral symmetry breaking is equivalent 
to condensation of local dynamical chirality and appearance of chiral 
polarization scale $\chps$. Here we consider more general association 
involving the low--energy layer of chirally polarized modes which, 
in addition to its width ($\Lambda_{ch}$), is also characterized by volume 
density of participating modes ($\Omega$) and the volume density of total 
chirality ($\Omega_{ch}$). Few possible forms of the correspondence 
are discussed, paying particular attention to singular cases where $\Omega$ 
emerges as the most versatile characteristic. The notion of 
{\em finite--volume ``order parameter''}, capturing the nature of these 
connections, is proposed. We study 
the effects of temperature (in N$_f$=0 QCD) and light quarks (in N$_f$=12), 
both in the regime of possible symmetry restoration, and find agreement 
with these ideas. In N$_f$=0 QCD, results from several volumes indicate 
that, at the lattice cutoff studied, the deconfinement temperature $T_c$ 
is strictly smaller than the overlap--valence chiral transition temperature 
$T_{ch}$ in real Polyakov line vacuum. Somewhat similar intermediate phase 
(in quark mass) is also seen in N$_f$=12. It is suggested that deconfinement 
in N$_f$=0 is related to indefinite convexity of absolute \Xg--distributions.

\end{abstract}

\vspace*{-0.10in}

\section{Introduction and Summary}
\label{sec:intro}

Eigensystems of Dirac operator in equilibrium gauge backgrounds carry the 
information on fermionic aspects of quark--gluon dynamics. As an important 
example, inspection of Dirac spectral representation for scalar fermionic 
density immediately reveals that spontaneous chiral symmetry breaking (SChSB) 
is equivalent to mode condensation of massless Dirac operator. Ever since 
this association has been pointed out~\cite{Ban80A}, it became popular 
to think about SChSB in terms of quark near--zeromodes.

Contrary to SChSB, the mode condensation property is a well--defined notion 
in generic quark--gluon system, i.e. even when all quarks are massive. 
While there is no chiral symmetry of physical degrees of freedom to break
in this case, condensing dynamics can still be described via chiral symmetry 
considerations. Indeed, one can introduce e.g. a pair of fictitious fermionic 
fields (``valence quarks'') of degenerate mass $m_v$, and cancel their contribution 
to the action by also adding the associated bosonic partners~\cite{Mor87A}. 
This keeps the dynamics of physical quarks and gluons unchanged, but makes it 
meaningful to consider chiral rotations of valence fields in such extended system, 
and to inquire about ``valence SChSB'' in the $m_v \to 0$ limit. In this
language, 
\begin{equation}
  \mbox{\bf vSChSB} \quad \Longleftrightarrow \quad \mbox{\bf QMC}
  \label{eq:005} 
\end{equation} 
i.e. quark mode condensation (QMC) in arbitrary quark--gluon system is equivalent 
to valence spontaneous chiral symmetry breaking (vSChSB): dynamics supports 
condensing modes if and only if it supports valence chiral condensate and valence 
Goldstone pions.

It is useful to think about SChSB in the above more general sense, especially 
when inquiring about the mechanism underlying the phenomenon~\cite{Ale12D}. 
Indeed, the response of massless valence quarks to gauge backgrounds of various
quark--gluon systems provides a relevant point of dynamical distinction for 
associated theories: they either support ``broken'' or ``symmetric'' dynamics 
of the external massless probe. Moreover, valence SChSB is readily observed
in lattice simulations with physically relevant flavor arrangements, and the 
associated dynamical characteristics change smoothly in the light quark 
regime~\cite{Ale12D}. Valuable lessons on SChSB can thus be learned by 
studying its valence version with massive dynamical quarks: vSChSB becomes 
SChSB as dynamical chiral limit is approached. 

Unfortunately, the equivalence of quark mode condensation and valence SChSB 
does not provide 
window into specifics of broken quark dynamics. Indeed, the mode condensation 
property is merely a restatement of symmetry breakdown condition in Dirac 
spectral representation. However, it was recently proposed that another 
relation may hold, possibly with similar scope of validity, but with 
non--trivial dynamical connection to inner workings of the breaking 
phenomenon~\cite{Ale12D}. In particular, it was suggested that 
\begin{equation}
  \mbox{\bf vSChSB} \quad \Longleftrightarrow \quad \mbox{\bf DChC}
  \label{eq:015} 
\end{equation} 
i.e. that valence SChSB is equivalent to {\em dynamical chirality condensation} 
(DChC). This offers an intuitively appealing notion that the vacuum effect of 
chiral symmetry breaking is in fact the phenomenon of chirality condensation. 
In light of Eq.~(1), the above relation carries the same information
as QMC--DChC equivalence, which may be preferable for explicit checks. 

While entities involved in the above relations will all be defined in 
Sec.~\ref{sec:background}, it should be pointed out now that DChC 
relates to {\em dynamical} notion of local chirality in modes~\cite{Ale10A}: 
it expresses the tendency for asymmetry in magnitudes of left--right components 
(local chiral polarization), measured with respect to the baseline of statistical 
independence. The associated quantifier, the correlation coefficient of polarization 
$C_A \in [-1,1]$, is invariant with respect to the choice of parametrization for 
the asymmetry. It thus provides the information on the quark--gluon system that 
is inherently dynamical. DChC occurs when near--zeromodes are chirally 
polarized ($C_A>0$) and sufficiently abundant, namely when {\em chiral polarization
density} $\rho_{ch}(\lambda) \equiv \rho(\lambda) \,C_A(\lambda)$ is positive 
at $\lambda=0$ in infinite volume.

\begin{figure}[t]
\begin{center}
    \centerline{
    \hskip 0.00in
    \includegraphics[width=18.0truecm,angle=0]{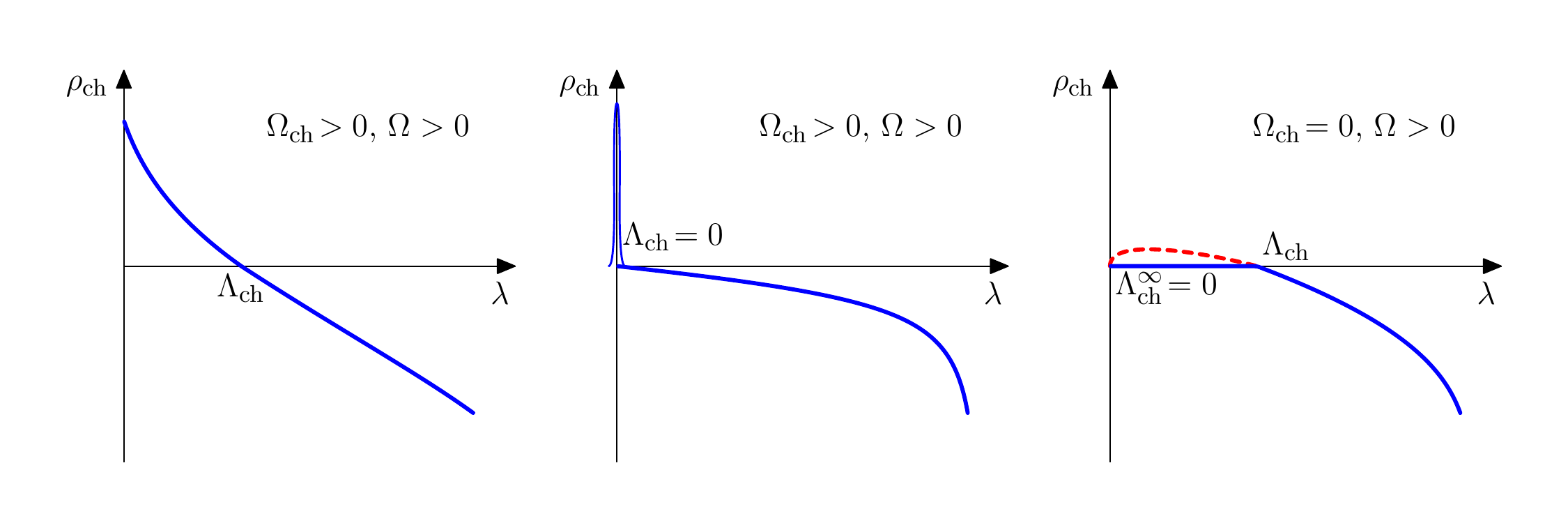}
     }
     \vskip -0.10in
     \caption{Layer of chirally polarized modes around $\lambda=0$ 
     is visible e.g. in the low--energy behavior of $\rho_{ch}(\lambda)$. Standard
     situation (left) and two singular ones (middle, right) are shown in infinite 
     volume. The dashed segment indicates that infinite volume limit
     was approached from chirally polarized side. See Sec.~\ref{ssec:chidefs} for
     definition of $\Lambda_{ch}^\infty$.}
     \label{fig:illus}
    \vskip -0.45in
\end{center}
\end{figure}

Important aspect of DChC is that it manifests itself in chiral spectral properties 
away from strictly infrared limit. Indeed, it was shown~\cite{Ale12D,Ale10A} that  
mode--condensing theories of physical interest exhibit {\em chiral polarization scale} 
$\chps$, marking the spectral point where functions 
$C_A(\lambda)$, $\rho_{ch}(\lambda)$ change sign and modes become anti--polarized 
(Fig.~\ref{fig:illus}(left)). The existence of $\chps$ 
in chirally broken asymptotically free theories can be intuitively understood from 
the fact that free modes are strictly anti--chiral and $C_A(\lambda)$ is thus 
expected to assume negative values at sufficiently high $\lambda$. 
The simultaneous occurrence of DChC and $\chps$ is thus not viewed as 
accidental but rather generic. This is, in fact, an integral part of the vSChSB--DChC 
conjecture as formulated in Ref.~\cite{Ale12D}, so that
\begin{equation}
  \mbox{\bf vSChSB} \quad \Longleftrightarrow \quad 
  \mathbf{\chps > 0}
  \label{eq:025}
\end{equation} 
with suitably general definition of $\chps$. Chiral polarization scale can
thus be viewed as a non--standard ``order parameter'' of the breaking phenomenon,
and is expected to be naturally tied to the mechanism of SChSB~\cite{Ale12D}.

Relations (\ref{eq:015},\ref{eq:025}) acquire their full meaning only when
the corresponding range of theories is specified. This is relevant since the larger 
the scope of theories conforming to vSChSB--DChC correspondence, the deeper the connection 
of local chirality to the symmetry breaking phenomenon. Indeed, if the association 
is generic, then it must be ascribed to the very nature of quark--gluon interaction. 
The original vSChSB--DChC conjecture was formulated in the context of SU(3) gauge theories 
with arbitrary number of fundamental quark flavors of arbitrary masses, and at arbitrary 
temperature. Thus both broken and symmetric theories are included in this landscape 
with corresponding transitions providing for most interesting tests of the conjecture. 
While the maximal range of validity may be significantly 
larger,\footnote{For example, extension to general SU(N) gauge groups may hold as well.} 
this is a physically relevant setup currently associated with the above statements. 

\medskip

\noindent The focus of this work involves two main aspects.

\medskip

{\em (I)} In Sec.~\ref{sec:background} we provide more complete description of chiral 
polarization phenomenon, and formulate the proposed connections to vSChSB within such 
wider context. Refined description reflects the premise that the vacuum feature we 
are associating with vSChSB is the layer of chirally polarized modes around 
$\lambda\!=\!0$ (``surface of the Dirac sea''). Apart from its width ($\Lambda_{ch}$), 
global characterization of this structure also includes volume density of modes 
involved ($\Omega$), and the volume density of total chirality generated by 
them ($\Omega_{ch}$). 
Incorporating these characteristics allows us to distinguish and discuss various 
special/singular cases that might arise. For example, the layer could 
approach zero width in the infinite volume limit, but acquire sufficiently singular 
abundance of modes so as to generate positive $\Omega_{ch}$ and $\Omega$ 
(see Fig.~\ref{fig:illus}(middle)). Another possibility is that polarization 
density asymptotically vanishes on the layer, due to modes being concentrated 
on a subset of space--time with measure zero, but $\Omega$ remains positive in 
the infinite volume limit. This is schematically shown in 
Fig.~\ref{fig:illus}(right). Even if behaviors of the above type won't appear 
in target continuum theories, they are likely to show up at low lattice cutoffs. 
Working within a framework capturing such cases is beneficial to make the proposed 
connections as universal as possible.

\medskip

\noindent Related to the above is a suggestion, described in Sec.~\ref{ssec:finvol}, 
whose conceptual content goes beyond that in Ref.~\cite{Ale12D}. To convey this,
note that our discussion implicitly proceeded in infinite volume where the notions 
of condensate and symmetry breaking are defined. For example,
the generic form of relation \eqref{eq:025} in the extended framework 
(see {\sl Conjecture 2''}) is
\begin{equation}
  \mbox{\bf vSChSB} \quad \Longleftrightarrow \quad 
  \mathbf{\Omega > 0}
  \tag{{\ref{eq:025}}'}
  \label{eq:035}
\end{equation} 
implying that appearance of chirally polarized layer around the surface of Dirac 
sea is considered to be physically significant if 
$\Omega \equiv \lim_{V\to\infty} \Omega(V) > 0$. Theories without polarized low 
modes in large finite volumes ($\Omega(V)$ identically zero) are then predicted 
to be unbroken, but theories with polarized low modes ($\Omega(V)>0$) could also be 
unbroken if $\lim_{V\to\infty} \Omega(V)\!=\!0$. Consistently with available data, we 
propose in {\sl Conjecture 3} that the latter possibility does not occur: 
if large--volume dynamics generates chiral polarization in low end of the spectrum, 
then valence chiral condensate appears in the infinite volume limit and vice versa. 
This provides for the closest possible relationship between vSChSB and chiral 
polarization (ChP), and it is in this sense that we refer to $\Lambda_{ch}$, 
$\Omega_{ch}$ and $\Omega$ as {\em finite--volume ``order parameters''}.\footnote{As 
discussed in Sec.~\ref{ssec:finvol}, $\Lambda_{ch}$, $\Omega_{ch}$ and $\Omega$ are 
zero/non--zero simultaneously in finite volume: they effectively represent a single 
finite--volume order parameter.} 
In this context, the correspondence analogous to \eqref{eq:015} is expressed as
\begin{equation}
  \mbox{\bf vSChSB} \quad \Longleftrightarrow \quad \mbox{\bf ChP} \;\,
  \mbox{\rm in large finite volumes}
  \label{eq:045}
  \tag{{\ref{eq:015}}'} 
\end{equation} 
and the right-hand side of relation \eqref{eq:035} modifies to:
$\,\Omega(V)>0$ for $V_0<V<\infty$. 

\medskip

{\em (II)} We present new lattice data supporting the above ideas. It is known 
that, at zero temperature, N$_f$=0 theory~\cite{Ale12D,Ale10A} as well as 
N$_f$=2+1 theory at physical point~\cite{Ale12D} exhibit chiral polarization, 
in accordance with the presence of vSChSB, and thus with the proposed equivalence. 
There is also an initial evidence that subjecting 
N$_f$=0 QCD to thermal agitation, chiral polarization and vSChSB cease to exist 
at common temperature T$_{ch}$ \cite{Ale12D}. When contemplating the validity of 
the vSChSB--ChP relationship over the vast theory landscape considered, it is useful 
to think of N$_f$=0, T=0 theory as a reference point~\cite{Ale12D}. Indeed, this 
dynamics produces maximal breaking of valence chiral symmetry, with thermal effects 
and the effects of light dynamical quarks providing two possible routes to symmetry 
restoration. Thus, in pilot investigations, it is natural to examine these two 
deformations of quenched theory independently of one another, 
both to ascertain conjecture's validity in such instances, as well as to learn 
about specific features associated with the two qualitatively different 
effects. This work is the first step in that direction with finite--temperature 
aspect examined in Sec.~\ref{sec:fintemp}, and many--flavor dynamics 
in Sec.~\ref{sec:light}. All of our results are consistent with the vSChSB--ChP 
equivalence.

\medskip

Few noteworthy byproducts of our main inquiry are also discussed here. 
{\em (i)} Performing a volume analysis at the lattice 
cutoff studied, we show the existence of the (lattice) phase $T_c < T < T_{ch}$ 
in N$_f$=0, simultaneously exhibiting vSChSB and 
deconfinement~\cite{Edw99A,Ale12A,Ale12B}. This chiral polarization dynamics  
appears to be of the singular type 
$(\Lambda_{ch}\!=\!0,\, \Omega_{ch}\!>\!0,\, \Omega\!>\!0)$. 
{\em (ii)} One of the characteristic
features of the above ``mixed phase'' is the appearance of very inhomogeneous 
near--zeromodes, well distinguished from the bulk of the spectrum. We find 
an intermediate phase (in quark mass) with such properties also in N$_f$=12. 
The relevance of these phases for continuum physics remains an open issue in both 
cases. {\em (iii)} Elementary analysis of the detailed chiral polarization 
characteristic, namely absolute \Xg--distribution, is also performed. Among 
other things, our data in N$_f$=0 indicate that deconfinement is characterized by 
the appearance of distributions with indefinite convexity. 

\section{The Background and the Conjectures}
\label{sec:background}

We start by reviewing the relevant background, and formulating the connections we 
aim to test. Our discussion will be fairly detailed and self--contained, in part
to sufficiently extend pertinent parts of Letter~\cite{Ale12D} which were necessarily
rather brief. In addition, we build an extended formalism for description of chiral
polarization, which allows us to include special/singular behaviors and has some 
benefits for continuum--limit considerations.  

\subsection{Valence Chiral Symmetry and Mode Condensation}
\label{ssec:qmc}

Implicitly assumed in what follows is the setup involving SU(3) gluons interacting 
with N$_f$ fundamental quarks of masses $M=(m_1,m_2,\ldots,m_{N_f})$.
To formally include valence chiral symmetry considerations, the system is 
augmented by a pair of degenerate valence quarks of mass $m_v$, and a pair of
complex commuting fields (pseudofermions) compensating for the dynamical effect 
of these fictitious particles~\cite{Mor87A}. This schematically corresponds to the action,  
\begin{equation}
S \,=\, S_g \,+\, 
        \sum_{f=1}^{N_f} \psibar_f \Bigl( D_{(1)} + m_f \Bigr) \psi_f \,+\,
        \sum_{i=1}^2 \bar{\eta}_i  \Bigl( D_{(2)} + m_v \Bigr) \eta_i \,+\,
        \sum_{i=1}^2 \phi^{\dagger}_i \Bigl( D_{(2)} + m_v \Bigr) \phi_i 
\label{eq:2.015}
\end{equation}
where $S_g$ is the pure glue contribution.
When viewing the above as expression in the continuum, then $D_{(1)}=D_{(2)}=D$, 
namely the continuum Dirac operator. However, on the lattice it is possible, and 
sometimes desirable, to consider different discretizations for dynamical and 
valence quarks. In particular, the role of $D_{(2)}$ in our case is to describe 
the response of physical vacuum to external chiral probe. It is thus desirable 
that it provides for exact lattice chiral symmetry, even though some numerically 
cheaper $D_{(1)}$ could have been used to simulate physical quarks 
and to define the theory.

Flavored chiral rotations of valence quark fields in the above extended system 
become the symmetry of the action in the $m_v \to 0$ limit, and one can  
meaningfully ask whether this symmetry is broken by the vacuum. If so, we speak
of {\em valence chiral symmetry breaking} (vSChSB). It has the usual consequence 
of being associated with the triplet of massless valence pions. While these are 
not physical states, they express the ability of the physical vacuum 
to support a specific type of long range order: the same kind of order that is 
required for physical chiral symmetry to be broken in the dynamical massless 
limit. vSChSB is thus a relevant vacuum characteristic of QCD--like theories.

Following the steps involved in derivation of the standard Banks--Casher 
relation~\cite{Ban80A}, one can inspect that the valence chiral condensate 
in theory (\ref{eq:2.015}) is given by  
\begin{equation}
   \Sigma(M) \, \equiv \, 
   \lim_{m_v\to 0} \, \lim_{V \to \infty} 
   \langle\, \bar{\eta} \eta \, \rangle_{M,m_v,V} \;=\; 
   \pi \lim_{\epsilon \to 0^+} \, \frac{1}{\epsilon} \,
   \lim_{V \to \infty} \sigma_{(2)}(\epsilon,M,V)
   \label{eq:2.025}
\end{equation}
where $\eta$ is one (arbitrary) of the valence flavors, 
$\sigma_{(2)}=\sigma_{(2)}(\lambda,M,V)$ the cumulative eigenmode density of 
$D_{(2)}$ and $V$ the 4--volume. 
In the continuum, $\sigma_{(2)}\to \sigma$ is the cumulative eigenmode density of 
continuum Dirac operator. On the lattice, we implicitly assume that $D_{(2)}$ 
is the overlap Dirac operator~\cite{Neu98BA} which is our discretization of choice 
in this study, and for which the relation such as (\ref{eq:2.025}) can be 
straightforwardly derived~\cite{Cha98A}. 

The notion of {\em cumulative eigenmode density}, used above, is defined as 
\begin{equation}
  \sigma(\lambda,M,V) \equiv \frac{1}{V} \, 
  \langle \, \sum_{0 \le \lambda_k < \lambda} \,1\;\rangle_{M,V}   
  \label{eq:2.030}
\end{equation} 
where $\lambda$ (real number) represents the imaginary part of Dirac eigenvalue. 
In case of overlap Dirac operator one can also take it to be the magnitude of 
the eigenvalue multiplied by the sign of the imaginary part. $\lambda_k$ are 
the values associated with given gauge background and ordered appropriately.
Note that $\sigma(\lambda,M,V)\equiv 0$ for $\lambda \le 0$. The differential 
version of $\sigma$, referred to as {\em eigenmode density}, is generally 
available and frequently useful, namely
\begin{equation}
  \rho(\lambda,M,V) \equiv \lim_{\epsilon \to 0^+} \, 
  \frac{\sigma(\lambda+\epsilon,M,V) - \sigma(\lambda,M,V)}{\epsilon}
  \,=\, \frac{\partial}{\partial_+ \lambda} \, \sigma(\lambda,M,V)
  \label{eq:2.035}
\end{equation} 
Note that we chose to distinguish the above eigenmode density from the usual
\begin{equation}
  \bar{\rho}(\lambda,M,V) \equiv \frac{1}{V} \, 
  \sum\limits_k \langle \, \delta(\lambda - \lambda_k) \,\rangle_{M,V}
  \label{eq:2.040}
\end{equation} 
which, in some singular cases, has to be represented by a generalized function 
with at most countably many ``atoms'' ($\delta$--functions), while $\rho$ is 
always an ordinary function which simply takes the value ``$\infty$'' 
at the position of the atoms. 

The infinite volume limit of $\rho(\lambda,M,V)$ will be defined as
\begin{equation}
  \rho(\lambda,M) \equiv \frac{\partial}{\partial_+ \lambda}
  \lim_{V \to \infty}  \sigma(\lambda,M,V)
  \label{eq:2.045}
\end{equation} 
rather than as point--wise limit of $\rho(\lambda,M,V)$, but the two can only 
differ in certain singular points of the spectrum. The theory is said to exhibit 
{\em quark mode condensation} (QMC) if
\begin{equation}
   \lim_{\epsilon \to 0^+}  \, \frac{1}{\epsilon} \, 
   \lim_{V \to \infty} \,\sigma(\epsilon,M,V)  
   \;=\; \rho(0,M) > 0
   \label{eq:2.050}
\end{equation}
i.e. when the abundance of ``infinitely infrared'' modes scales as the total 
number of modes, namely with space--time volume. Relation \eqref{eq:2.025} then 
implies 
\begin{equation}
  \Sigma(M) > 0 \quad \Longleftrightarrow \quad \rho(0,M) > 0
  \tag{{\ref{eq:005}}'}
  \label{eq:2.055} 
\end{equation} 
which is an explicit representation of equivalence (\ref{eq:005}) between vSChSB 
and QMC.

\medskip

\noindent Few remarks regarding the above should now be made.

\medskip

\noindent {\em (i)} The discussion has been carried out at a rather general 
level mainly to accommodate the possibility of most singular behavior at the origin 
in the infinite volume limit. If $\lambda=0$ is the only ``atom'' in that situation, 
then $\bar{\rho}(\lambda,M) = C(M) \delta(\lambda) + \hat{\rho}(\lambda,M)$, with 
$\hat{\rho}$ being an ordinary function, and 
$C(M) = \lim_{\epsilon\to 0} \int_0^\epsilon \bar{\rho}(\lambda,M)\, d\lambda$. 
We then have
\begin{equation}
   \rho(0,M) \; = \; 
   \lim_{\epsilon\to 0^+} \frac{C(M)}{\epsilon}  
   \;+\; \lim_{\lambda \to 0^+} \rho(\lambda,M)
  \label{eq:2.060} 
\end{equation} 
since $\lim_{\lambda \to 0^+} \hat{\rho}(\lambda,M) = 
\lim_{\lambda \to 0^+} \rho(\lambda,M)$. Therefore, in addition to the usual 
second term, mode condensate acquires an infinite value if there is an atom 
of spectral mode density at the origin.\footnote{Note that $\rho(0,M)$ can in 
principle be infinite even when $C(M)=0$ since $\hat{\rho}(\lambda)$ could still 
have an ordinary integrable divergence at $\lambda=0$.} This is in accordance 
with diverging chiral condensate in such instance, namely
\begin{equation}
   \Sigma(M) \; = \; 
   \lim_{m_v \to 0} \, \frac{2 C(M)}{m_v}  
   \;+\; \pi \lim_{\lambda \to 0^+} \rho(\lambda,M)
   \label{eq:2.065} 
\end{equation} 
We emphasize that the singular accumulation of modes discussed above is not due 
to exact zero--modes of finite volumes: their contribution is well--known to 
vanish in the infinite volume limit. Rather, $C(M)>0$ would have to be generated 
by modes that are ``squeezed'' into near--zeromodes at arbitrary finite volume 
but become ``infinitely infrared'' as volume is taken to infinity. It is not known
whether there are continuum theories generating such dynamics, but we will discuss 
the behavior that might be of this type at finite cutoff. 

\medskip

\noindent {\em (ii)} It should be noted that, at finite cutoff, it is in principle 
possible to obtain contradictory answers on vSChSB (QMC) with different choices of 
$D_{(2)}$. However, such discrepancies, if any, are expected to disappear 
sufficiently close to the continuum limit. 

\medskip

\noindent {\em (iii)} In the above considerations we assumed $T=0$ for simplicity. 
Incorporating theories on 3--volume $V_3$ at finite temperature $T$ is 
straightforward and simply involves replacing $V \rightarrow (T,V_3)$ in labels, 
and $V \rightarrow V_3/T$ in volume factors.

\subsection{Spectral Measures of Dynamical Chirality}
\label{ssec:chidefs}

The term {\em ``dynamical local chirality''} refers to characterization of local 
asymmetry in left--right values of Dirac eigenmodes using absolute polarization 
method of Ref.~\cite{Ale10A}. Dynamical nature of this approach mainly stems 
from the fact that it quantifies polarization relative to the population of 
statistically independent left--right components. The basic characteristic
is the correlation coefficient of polarization $\cop_A \in [-1,1]$, 
assigned to a given eigenmode $\psi_\lambda$. It is linearly related to 
the probability that the local value of $\psi_\lambda$ is more polarized 
than value chosen from associated distribution of statistically independent 
left--right components. In case of correlation ($\cop_A>0$), dynamics enhances 
polarization and the mode is referred to as chirally polarized, while 
anti--correlation ($\cop_A<0$) indicates that dynamics suppresses polarization 
and the mode is chirally anti--polarized. Such comparison of polarization can 
also be performed in a detailed differential manner, resulting in absolute 
$\Xg$--distribution $\xd_A(\Xg)$, with $\Xg \in [-1,1]$. A concise 
introduction to these concepts together with precise definitions can be found 
in Appendix~\ref{app:chirality}.

Spectral characteristics of a given theory based on dynamical chirality measures 
$\cop_A$ and $\xd_A(\Xg)$ can be defined as follows~\cite{Ale12D}. The cumulative 
dynamical chirality per unit volume ({\em cumulative chiral polarization density}), 
is given by
\begin{equation}
  \sigma_{ch}(\lambda,M,V) \equiv \frac{1}{V} \, 
  \langle \, \sum_{0 \le \lambda_k < \lambda} \,\cop_{A,k}\;\rangle_{M,V}   
  \label{eq:4.005}
\end{equation} 
where $\cop_{A,k}$ is chiral polarization (correlation) in the k--th mode. The associated 
differential contribution due to modes at scale $\lambda$, namely
\begin{equation}
  \rho_{ch}(\lambda,M,V) \,\equiv\, \lim_{\epsilon \to 0^+} \, 
  \frac{\sigma_{ch}(\lambda+\epsilon,M,V) - \sigma_{ch}(\lambda,M,V)}{\epsilon}
  \;=\; \frac{\partial}{\partial_+ \lambda} \, \sigma_{ch}(\lambda,M,V)
  \label{eq:4.015}
\end{equation} 
is referred to as {\em chiral polarization density} as is its formal companion
\begin{equation}
  \bar{\rho}_{ch}(\lambda,M,V) \equiv \frac{1}{V} \, 
  \sum\limits_k \langle \, \delta(\lambda - \lambda_k) \, \cop_{A,k} \,\rangle_{M,V}
  \label{eq:4.025}
\end{equation} 
which has to be represented by a generalized function in certain singular cases.

The average chiral polarization at scale $\lambda$ is given by
\begin{equation}
   \cop_A(\lambda,M,V) \,\equiv\, \lim_{\epsilon \to 0^+}
   \frac{\sigma_{ch}(\lambda+\epsilon,M,V) - \sigma_{ch}(\lambda,M,V)}
        {\sigma(\lambda+\epsilon,M,V) - \sigma(\lambda,M,V)}
   \;=\;
   \frac{\rho_{ch}(\lambda,M,V)}{\rho(\lambda,M,V)}
    \label{eq:4.035}
\end{equation}
where the second equation only applies when $0 < \rho(\lambda,M,V) < \infty$. We 
similarly define the average absolute $\Xg$--distribution at scale $\lambda$ which,
written without shorthands, reads
\begin{equation}
   \xd_A(\Xg,\lambda,M,V) \,\equiv\, \lim_{\epsilon \to 0^+} 
   \frac{\langle \; \sum\limits_{\lambda \le \lambda_k < \lambda+\epsilon} 
         \xd_{A,k}(\Xg) \; \rangle_{M,V}}
        {\langle \; \sum\limits_{\lambda \le \lambda_k < \lambda+\epsilon} 
         1 \; \rangle_{M,V}}
    \label{eq:4.045}
\end{equation}
where $\xd_{A,k}(\Xg)$ is the absolute $\Xg$--distribution of the eigenmode  
associated with $\lambda_k$. 

The infinite volume limit of $\rho_{ch}(\lambda,M,V)$ is defined to be
\begin{equation}
  \rho_{ch}(\lambda,M) \equiv \frac{\partial}{\partial_+ \lambda} \,
  \lim_{V \to \infty}  \sigma_{ch}(\lambda,M,V)
  \label{eq:4.055}
\end{equation} 
We say that the theory exhibits {\em dynamical chirality condensation}~\cite{Ale12D} if 
\begin{equation}
   \lim_{\epsilon \to 0^+}  \, \frac{1}{\epsilon} \, 
   \lim_{V \to \infty} \,\sigma_{ch}(\epsilon,M,V)  
   \;=\; \rho_{ch}(0,M) > 0
   \label{eq:4.065}
\end{equation}
i.e. if its ``infinitely infrared'' Dirac modes are chirally polarized and their 
contribution to dynamical chirality scales with space--time volume. Similarly, 
dynamical anti--chirality condensation occurs when $\rho_{ch}(0,M)<0$. Note that 
(anti-)chirality condensation implies mode condensation but not vice--versa. 

Since $\rho_{ch}(\lambda)$ is a real--valued function of indefinite sign, we can assign 
{\em chiral polarization scale} $\chps \ge 0$ to it as the largest $\Lambda$ such 
that $\rho_{ch}(\lambda)>0$ on $[0,\Lambda)$ except for isolated zeros.
The provision for ``isolated zeros'' has two rationales. First, even with such zeros
present, $\chps$ retains its intended meaning as a spectral range of dynamical 
chirality around the surface of Dirac sea. Second, it ensures that defining chiral 
polarization scale via $\cop_A(\lambda)$ leads to the same scale in finite volume. 
Indeed, if $\Lambda'_{ch} \ge 0$ is the largest $\Lambda$ such that 
$\cop_A(\lambda)>0$ on $[0,\Lambda)$ except for isolated zeros, then 
$\chps = \Lambda'_{ch}$ in finite volume.\footnote{What is relevant here 
is that a zero of $\cop_A$ is also a zero of $\rho_{ch}$ but not necessarily 
vice-versa.} Note that when positive $\chps$ doesn't exist, the above definition is 
vacuously true for $\chps=0$, which is then the assigned chiral polarization 
scale. Also, $\chps = \infty$ is associated with $\rho_{ch}(\lambda)$ that is positive 
on $[0,\infty)$ except for possible isolated zeros. Given $\chps=\chps(M,V)$,
the {\em total dynamical chirality} of Dirac spectrum at low energy 
$\Omega_{ch}=\Omega_{ch}(M,V)$ is
\begin{equation}
   \Omega_{ch} \,\equiv\, 
   \max \,\{\, \sigma_{ch} (\chps),\;  
   \lim_{\epsilon \to 0^+} \sigma_{ch}(\chps+\epsilon)  \,\}
   \label{eq:4.060}
\end{equation}
where the possibility of discontinuity at $\chps$ has been taken into account. 
The associated {\em total number of chirally polarized modes} $\Omega=\Omega(M,V)$ is
\begin{equation}
   \Omega \,\equiv\, \begin{cases}
   \sigma(\chps) & 
   \text{if $\Omega_{ch}=\sigma_{ch} (\chps)$}\\  
   \lim\limits_{\epsilon \to 0^+} \sigma(\chps+\epsilon)  & 
   \text{if $\Omega_{ch}=\lim\limits_{\epsilon \to 0^+} \sigma_{ch}(\chps+\epsilon)$}
   \end{cases}
   \label{eq:4.062}
\end{equation}
Note that the characteristics $\Omega \ge 0$, $\Omega_{ch} \ge 0$ are volume densities 
while the condensates $\rho(0)$, $\rho_{ch}(0)$ are both volume and spectral densities.

\begin{figure}[tbp]
\begin{center}
    \centerline{
    \hskip -0.20in
    \includegraphics[width=14.5truecm,angle=0]{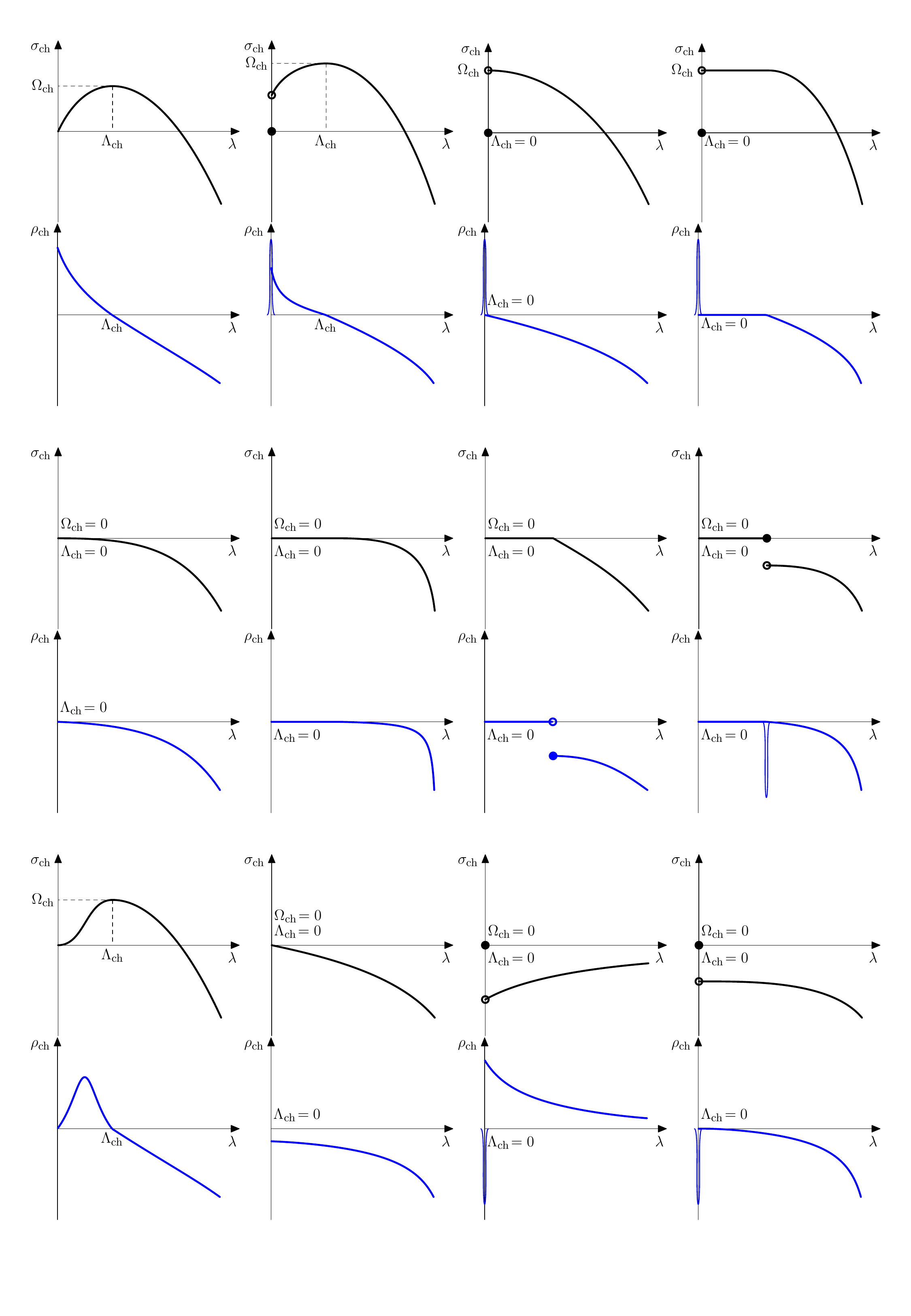}
     }
     \vskip -0.60in
     \caption{Examples of assigning $\Lambda_{ch}$ and $\Omega_{ch}$ to 
     $\sigma_{ch}(\lambda)$. The associated $\rho_{ch}(\lambda)$ is shown in 
     the lower figure for each case. For theory in {\em infinite volume}, 
     {\sl Conjecture 2'} identifies the first/second pair of rows as options
     that can only occur in broken/symmetric vacuum, while the behavior 
     in the third pair of rows is predicted to be impossible for theories 
     in ${\eusm T}$. The possibilities in each category are not meant 
     to be exhaustive or guaranteed to occur.}
     \label{fig:sigch_vs_lam_illus}
\end{center}
\end{figure}

\smallskip
Infinite volume limits of the above objects require some attention since the order
of operations might be relevant in certain cases. In discussion of the condensation 
phenomena, we emphasized the primary role of cumulative densities, with 
their infinite volume limits 
\begin{equation}
   \sigma(\lambda,M)\equiv \lim_{V\to \infty} \sigma(\lambda,M,V) \qquad\quad
   \sigma_{ch}(\lambda,M)\equiv \lim_{V\to \infty} \sigma_{ch}(\lambda,M,V)
   \label{eq:4.064}
\end{equation}
being the basis for computation of the condensates. However, one virtue of chiral 
polarization framework is that such insistence on the order of operations
ceases to be crucial. In fact, our default view of infinite volume limits 
for $\chps$, $\Omega$ and $\Omega_{ch}$ is just the direct limit of their 
finite--volume versions, rather than corresponding functionals of 
$\sigma_{ch}(\lambda,M)$ and $\sigma(\lambda,M)$. Thus, while we explicitly 
distinguish the two options, for example
\begin{equation}
   \Omega_{ch}(M) \,\equiv\, \lim_{V\to \infty} \Omega_{ch}(M,V)
   \qquad\quad
   \Omega_{ch}^\infty(M) \,\equiv\, \Omega_{ch}[\sigma_{ch}(\lambda,M)]
   \label{eq:4.070}
\end{equation}
it is the first definition that is implicitly understood if not stated otherwise.
We use analogous notational convention also in case of $\chps$ and $\Omega$.

\vfill\eject

\noindent Few remarks regarding these concepts should be made.

\medskip

\noindent {\em (i)} The above definitions of $\chps$ and $\Omega_{ch}$ assume 
certain analytic properties of $\sigma_{ch}(\lambda)$, such as existence 
of $\rho_{ch}(\lambda)$ or the limit in Eq.~\eqref{eq:4.060}. In case of
cumulative mode density $\sigma(\lambda)$, such properties are, for most
part, inherent in its definition, i.e. it is continuous except possibly at 
countably many finite jumps, and differentiable almost everywhere. While these
properties are expected to hold also for $\sigma_{ch}(\lambda)$ in any theory, 
it is comforting that $\chps$ and $\Omega_{ch}$ can be assigned to 
$\sigma_{ch}(\lambda)$ that is completely generic. The corresponding definition 
and related considerations are discussed in Appendix~\ref{app:spectral}. 

\medskip

\noindent {\em (ii)} In terms of $\sigma_{ch}(\lambda)$, chiral polarization
scale corresponds to the largest $\Lambda$ such that $\sigma_{ch}(\lambda)$ is
strictly increasing on $[0,\Lambda]$, and $\Omega_{ch}$ its associated maximal 
value. Fig.~\ref{fig:sigch_vs_lam_illus} shows various behaviors of 
$\sigma_{ch}(\lambda)$, illustrating how $\chps$ and $\Omega_{ch}$ are assigned 
via above definitions. For the current purpose, one should only view 
$\sigma_{ch}(\lambda)$, $\rho_{ch}(\lambda)$ shown as admissible pairs of 
functions: many of these situations are not expected to occur in theories
of interest.

\medskip

\noindent {\em (iii)} In the above considerations we assumed $T=0$ for 
notational simplicity. Extension of all definitions to finite temperatures 
is straightforward (see remark {\em (iii)} of Sec.~\ref{ssec:qmc}).

\subsection{Conjecture Formulations}
\label{ssec:conjectures}

We will consider and extend {\em Conjecture 2} of Ref.~\cite{Ale12D}, which ties 
the phenomenon of valence spontaneous chiral symmetry breaking to that of 
dynamical local chirality in low--lying modes. Such vSChSB--DChC correspondence 
was proposed to hold over the set ${\eusm T}$ containing SU(3) gauge theories 
in 3+1 space--time dimensions with any number ($N_f$) of fundamental quarks of 
arbitrary masses $M\equiv (m_1,m_2,\ldots,m_{N_f})$, and at arbitrary 
temperature $T$. Different continuum theories are thus labeled by $(M,T)$. 
Statements of this section are formulated in infinite volume (infinite extent of 
all spatial dimensions) so that spontaneous symmetry breaking and the condensation 
concepts have definite meaning. To begin with, we formulate {\em Conjecture 2}
using more precise language and concepts developed here. 
By ``chiral polarization characteristics'' we mean the parameters $\Lambda_{ch}$, 
$\Lambda_{ch}^\infty$, $\Omega_{ch}$, $\Omega_{ch}^\infty$, $\Omega$, 
$\Omega^\infty$. 

\medskip

\noindent {\bf Conjecture 2}

\smallskip
\noindent {\sl The following holds in every lattice--regularized $(M,T) \in {\eusm T}$ 
at sufficiently large ultraviolet cutoffs $\Lambda_{lat}$ and infinite volume.} 
\begin{description}
\item[{\sl (I)}] {\sl Chiral polarization characteristics exist and are zero or 
non--zero simultaneously. Moreover, $\chps = \Lambda_{ch}^\infty$ and
$0 \le \chps \ll \Lambda_{lat}$.
If $\Lambda_{ch}>0$ then $\rho_{ch}(\lambda)$ is positive on $[0,\Lambda_{ch})$.}
\item[{\sl(II)}] {\sl The property of valence spontaneous chiral symmetry breaking 
is characterized by}
\begin{equation}
  \Sigma > 0    \quad  \Longleftrightarrow   \quad   
  \rho(\lambda=0) > 0    \quad  \Longleftrightarrow   \quad \Lambda_{ch} > 0  
  \label{eq:5.020}
\end{equation}
\item[{\sl(III)}] {\sl $\rho_{ch}(\lambda)$ is non--positive for $\lambda > \Lambda_{ch}$,
except possibly in the vicinity of $\Lambda_{lat}$.}
\end{description}

\vfill\eject

\noindent The statement is thus split into three parts which we now discuss. 

\bigskip

\noindent {\sl (I)} $\Lambda_{ch}\!=\!\Lambda_{ch}^\infty$ implies that 
$\Omega_{ch}=\Omega_{ch}^\infty$ and $\Omega=\Omega^\infty$. Neither this pairwise 
equality nor the simultaneous positivity of all six parameters follows from definitions 
alone. Rather, they represent anticipated constraints on actual dynamical behavior in 
theories under consideration close to the continuum. Note that the statement rules
out the possibility of chiral polarization over the whole Dirac spectrum. Also, while 
$\Lambda_{ch}>0$ implies positivity of $\rho_{ch}(\lambda)$ on $[0,\Lambda_{ch})$ only 
up to isolated points, such points are not expected to be present in infinite volume, 
as explicitly stated. 

\medskip

\noindent {\sl (II)} Relation (\ref{eq:5.020}) asserts that chiral polarization
scale $\Lambda_{ch}$ is an unconventional order parameter of vSChSB on ${\eusm T}$:
chiral symmetry is broken if and only if this dynamical scale is generated.
Note that from {\sl (I)} and {\sl (II)} it follows that 
$\rho(0) > 0 \, \Leftrightarrow \, \rho_{ch}(0) > 0$, which is the mathematical 
representation of relation (\ref{eq:015}): quark mode condensation (hence vSChSB) 
is equivalent to dynamical chirality condensation. Anti--chirality doesn't 
condense~\cite{Ale12D}.

\medskip

\noindent {\sl (III)} This part reflects the expectation that dynamical chirality can 
only occur in the low end of the Dirac spectrum, characterized by $\chps$. In lattice
theory, chiral polarization could exist in the vicinity of the cutoff (due to lattice
artifacts) but will scale out in the continuum limit. While generic anti--chirality 
in the ultraviolet is supported by asymptotic freedom, the assertion 
that chiral polarization cannot dominate at intermediate scales, i.e. in a spectral band  
separated from origin, is not easy to verify directly. Indeed, reaching such 
intermediate parts of Dirac spectra via numerical lattice QCD sufficiently close to 
the continuum limit can be computationally demanding. Nevertheless, at least at zero 
temperature, there is little doubt that the above scenario holds since running of the gauge 
coupling, well understood, is monotonic across scales. The situation at finite temperature 
is more involved though. The effects of asymmetry between magnetic and electric 
couplings~\cite{Com97A} and the influence of infrared fixed point in dimensionally 
reduced (3-d) theory~\cite{Bra05A} could improve prospects for more complicated behavior 
at sufficiently high temperatures. However, the available data is not hinting 
the existence of intermediate--scale chirality, as reflected in the statement.

\medskip

The meaning of {\sl Conjecture 2} is directly tied to the lattice definition 
of the theory: following some line of constant physics in the parameter space of 
a regularized setup, it is claimed that the statement becomes valid in associated lattice 
theories with sufficiently large cutoff.\footnote{In the continuum language, masses 
$M \equiv (m_1,m_2,\ldots,m_{N_f})$ label different continuum theories (different lines
of constant physics) and should thus be viewed as renormalized quark masses in some 
fixed scheme.} It is thus useful to attempt a formulation of the
vSChSB $\leftrightarrow$ DChC correspondence valid for the widest range of cutoffs 
possible. Similarly to vSChSB $\leftrightarrow$ QMC connection, which holds at 
arbitrary cutoff, the range of validity may include even symmetry breaking instances 
due to lattice artifacts. To formulate the alternative version of the conjecture, 
we were mostly guided by results of the numerical study at finite temperature, described 
in Sec.~\ref{sec:fintemp}. 
These results suggest the viability of the scenario in which 
chirally broken dynamics generates chiral polarization only via discrete contribution from 
strictly infrared modes. To incorporate these cases, it is necessary to abandon the notion 
that chiral polarization characteristics are non--zero simultaneously. {\sl Conjecture 2'}
stated below provides for the minimal extension of this type. 

\bigskip

\noindent {\bf Conjecture 2'}

\smallskip
\noindent {\sl The following holds in every lattice--regularized $(M,T) \in {\eusm T}$
at sufficiently large ultraviolet cutoffs $\Lambda_{lat}$ and infinite volume.}  
\begin{description}
\item[{\sl (I)}] {\sl Chiral polarization characteristics exist,
$\Lambda_{ch} = \Lambda_{ch}^\infty$ and $0 \le \chps \ll \Lambda_{lat}$.
If $\Lambda_{ch}>0$ then $\rho_{ch}(\lambda)$ is positive on $[0,\Lambda_{ch})$.}
\item[{\sl(II)}] {\sl The property of valence spontaneous chiral symmetry breaking 
is characterized by}
\begin{equation}
  \Sigma > 0    \quad  \Longleftrightarrow   \quad   
  \rho(\lambda=0) > 0    \quad  \Longleftrightarrow   \quad \Omega_{ch} > 0  
  \tag{{\ref{eq:5.020}}'}
  \label{eq:5.020a}
\end{equation}
\item[{\sl(III)}] {\sl $\rho_{ch}(\lambda)$ is non--positive for $\lambda > \Lambda_{ch}$,
except possibly in the vicinity of $\Lambda_{lat}$.}
\end{description}

\medskip

\noindent Discussion following {\em Conjecture 2} mostly applies here as well but 
it is $\Omega_{ch}$ that serves as the order parameter of vSChSB on ${\eusm T}$: 
chiral symmetry is broken if and only if total low energy chirality 
per unit volume is nonzero. Singular cases motivating this extension arise when 
positive core in $\rho_{ch}(\lambda,V)$ becomes proportional to $\delta(\lambda)$ in 
the infinite volume limit, leading to $\Lambda_{ch}=0$, $\Omega_{ch}>0$ as shown in
Fig.~\ref{fig:illus}(middle).\footnote{The possibility of transition 
from $\Lambda_{ch}$ to $\Omega_{ch}$ has been discussed in Ref.~\cite{Ale12D} already. 
However, $\Omega_{ch}$ was denoted as $\Omega$ in that work.} Note that 
vSChSB $\leftrightarrow$ DChC correspondence still follows from {\sl (I)} and {\sl (II)}.
Part {\sl (III)} implies that $\Omega_{ch}$ is not just the total low--energy chirality 
but the total chirality of the entire Dirac spectrum. The essential content of 
{\sl Conjecture 2'} can then be summarized by saying that quark--gluon dynamics breaks 
chiral symmetry if and only if it generates volume density of dynamical chirality.

It is interesting to look back at Fig.~2 in light of the above statement. Indeed, 
assuming that various $\sigma_{ch}(\lambda)$, $\rho_{ch}(\lambda)$ shown represent 
infinite volume limits, the first pair of rows corresponds to options for chirally 
broken vacuum, the second pair is associated with symmetric vacuum, and the cases in 
the third pair do not occur on ${\eusm T}$ if {\sl Conjecture 2'} is valid.

With the same rationale that motivated {\sl Conjecture 2'}, we now put forward yet 
more generic association of chiral polarization and vSChSB. In particular, it is 
possible to drop the notion that DChC is a necessary companion of vSChSB, while 
still maintaining the connection to chiral polarization. This can arise when, in 
addition to simultaneous positivity, the pairwise equality of chiral polarization 
parameters is abandoned as well. The prototypical situation we have in mind is when 
$\Lambda_{ch}(V)>0$ converges to $\Lambda_{ch}>0$ in the infinite volume limit, but
$\rho_{ch}(\lambda, V)$ scales to zero on interval $[0,\Lambda_{ch})$. This results 
in $\Lambda_{ch}^\infty\!=\!0$ and $\Omega_{ch}\!=\!0$ 
(see Fig.~\ref{fig:illus}(right)). Note that 
$\rho_{ch} = \rho\, \cop_A$ can approach zero due to {\sl (i)} $\cop_A \to 0$ or 
{\sl (ii)} $\rho \to 0$ or {\sl (iii)} both. 
Option {\sl (i)} is quite interesting since it can occur when 
low--lying Dirac modes are dimensionally reduced.\footnote{Loosely speaking, 
eigenmodes are {\em dimensionally reduced} when their effective support comprises
a vanishing fraction of the associated domain in the infinite--volume limit.}
Indeed, the ``active'' part of the eigenmode can be chirally polarized and 
induce vSChSB, but its contribution to total polarization gets overwhelmed by 
the uncorrelated bulk in the infinite volume limit. While $\Omega_{ch}=0$ in this 
situation, $\Omega>0$ still signals the association of chiral symmetry breaking 
and chiral polarization. Thus, in the form below, the conjecture states that 
vSChSB proceeds if and only if there is a volume density of chirally polarized 
modes around the surface of the Dirac sea.

\medskip

\noindent {\bf Conjecture 2''}

\smallskip
\noindent {\sl The following holds in every lattice--regularized $(M,T) \in {\eusm T}$
at sufficiently large ultraviolet cutoffs $\Lambda_{lat}$ and infinite volume.}  
\begin{description}
\item[{\sl (I)}] {\sl Chiral polarization characteristics exist and 
$0 \le \Lambda_{ch} + \Lambda_{ch}^\infty \ll \Lambda_{lat}$.}
\item[{\sl(II)}] {\sl The property of valence spontaneous chiral symmetry breaking 
is characterized by}
\begin{equation}
  \Sigma > 0    \quad  \Longleftrightarrow   \quad   
  \rho(\lambda=0) > 0    \quad  \Longleftrightarrow  \quad \Omega > 0  
  \tag{{\ref{eq:5.020}}''}
  \label{eq:5.020b}
\end{equation}
\item[{\sl(III)}] {\sl $\rho_{ch}(\lambda)$ is non--positive for 
$\lambda > \Lambda_{ch}^\infty$, except possibly in the vicinity of $\Lambda_{lat}$.}
\end{description}

\medskip

It should be emphasized that all three versions of the conjecture may be valid 
simultaneously. Indeed, they do not necessarily exclude one another, but rather 
express varied degrees of detail in which chiral polarization could manifest itself 
in vSChSB. In fact, it is entirely feasible that the differences are only relevant 
at sufficiently low lattice cutoffs.

\subsection{Finite Volume}
\label{ssec:finvol}

Discussion in the previous section has been carried out in the infinite volume
which is a native setting for vSChSB and various condensates. However, it is
both relevant practically and interesting conceptually, to examine how chiral 
polarization concepts enter the finite volume considerations.

To begin with, it is useful to fix a convention regarding exact zero--modes. 
Indeed, the default lattice setup in this discussion involves standard 
(anti--)periodic boundary conditions and the overlap Dirac operator as 
a chiral probe. Consequently, exact zero--modes can appear in finite volume
whether infinite--volume theory breaks chiral symmetry or not. In either case 
though, their abundance doesn't scale with volume. This renders them inessential 
for valence condensate and there is a choice whether to include them in 
specific considerations. For our purposes it is more convenient to leave
zero--modes out which is what will be assumed from now on in this article. 
In fact, one useful advantage of using overlap operator is that it cleanly 
separates out topological modes, ensuring that their {\em a priori} local 
chirality doesn't contaminate that of near--zeromodes which are of actual 
interest.

The definition of chiral symmetry breaking in Dirac eigenmode representation
involves a strictly infrared condition: the existence of mode condensate (QMC).
This is of course not surprising: the strictly infrared nature of relevant mass 
($m_v=0$) gets translated into strictly infrared corner of the Dirac spectrum 
($\lambda=0$). However, applying QMC condition to detect vSChSB in finite volume 
is futile since it is never satisfied. Indeed, $\rho(\lambda \!\to\! 0,V)$ is 
zero identically. This is symptomatic of the fact that 
vSChSB $\leftrightarrow$ QMC correspondence is kinematic in nature: since 
the definition of vSChSB demands infinite volume, QMC condition, being its 
equivalent, reacts to avoid conflict with the presence of infrared cutoff.

However, if the mechanism of vSChSB was known, it would be reasonable to expect
that other properties of quark modes, those suggested by the mechanism, 
could be used to signal vSChSB even in given finite volume. After all, broken 
and symmetric dynamics should be distinguishable in any volume. In fact, if 
infrared cutoff is smaller than relevant {\em finite} scales in the theory, 
the distinction from single volume should be essentially unambiguous.

While the mechanism of vSChSB has not been satisfactorily clarified yet, one of 
the driving motivations for developing chiral polarization framework was to 
provide a possible indicator of the above type. It is important in that regard 
that, unlike $\rho(0)$, chiral polarization parameters $\chps$, $\Omega_{ch}$ 
and $\Omega$ can be readily non--zero in finite volume. If forming chirally 
polarized layer around the surface of Dirac sea provides sufficient and necessary 
condition for producing vSChSB, as conjectures of the previous section propose,
then chiral polarization characteristics represent viable candidates for such 
{\em finite--volume ``order parameters''} indeed.
 
It is interesting that the situation in finite volume is in fact simpler than 
what we dealt with in the previous section. There are just three 
characteristics which, taking into account the convention on exact zero--modes, 
can only be non--zero simultaneously, effectively yielding a single 
finite--volume order parameter. Theory $(M,T,V)$ is said to be in 
{\em chirally polarized phase} if chiral polarization characteristics are 
positive, say $\Omega(M,T,V)>0$. We emphasize again that this provides 
a well--defined dynamical distinction between theories in finite volume based 
on their chiral behavior. However, in addition to this and the fact that 
$\Omega(M,T)$ is expected to be a valid order parameter in traditional sense,
our notion of finite--volume order parameter for vSChSB involves another 
feature contained in the following statement consistent with available data.

\medskip

\noindent {\bf Conjecture 3}

\smallskip
\noindent {\sl The following holds in every lattice--regularized $(M,T) \in {\eusm T}$
at sufficiently large ultraviolet cutoffs $\Lambda_{lat}$.}  
\medskip

\noindent{\sl (I)} {\sl Chiral polarization characteristics exist and 
$0 \le \Lambda_{ch}(V) \ll \Lambda_{lat}$ for sufficiently large $V$.}
\medskip

\noindent{\sl(II)} {\sl The property of valence spontaneous chiral symmetry breaking 
is characterized by}
\begin{equation}
  \Sigma > 0    \quad  \Longleftrightarrow   \quad   \Omega(V) > 0 
  \quad  \mbox{\rm for} \quad V_0 < V < \infty
  \label{eq:5.030}
\end{equation}
\noindent {\sl i.e. it occurs if and only if the theory is in chirally polarized phase 
in large finite volumes.} 
\medskip

\noindent{\sl(III)} {\sl $\rho_{ch}(\lambda,V)$ is non--positive for 
$\lambda > \Lambda_{ch}(V)$, except possibly in the vicinity of $\Lambda_{lat}$.}

\bigskip

The conceptual novelty in the above statement is that the right--hand side 
of Eq.~\eqref{eq:5.030} does not involve explicit infinite volume limit.
This is different e.g. from {\em Conjecture 2''} wherein the infinite volume
limit of $\Omega(M,T,V)$, while significantly easier to deal with than QMC, 
still has to be investigated: the theory could stay in chirally polarized 
phase for arbitrary large volumes, but with $\Omega(M,T,V)$ scaling to zero. 
However, {\em Conjecture 3} proposes that it is impossible to approach chirally 
symmetric physics in infinite volume via finite--volume physics in chirally 
polarized phase.\footnote{Note that, for discussion in paragraph 
preceding and motivating {\em Conjecture 2''}, this implies that options 
{\sl (ii)} and {\sl (iii)} don't occur.}  
In other words, it suggests that, at least sufficiently close to the continuum 
limit, vSChSB and dynamical chirality are inextricable.

We emphasize that the above is not to say that it is impossible to have
a ``finite volume correction'' to prediction on vSChSB based on chiral
polarization in single volume. However, in the realm of {\sl Conjecture 3}, 
the false positive corresponds to missing out on the whole dynamically--defined
phase. In other words, it is expected that only in very small volumes, when 
other aspects of physics are severely mutilated as well, is such occurrence likely. 

Finally, we remark that while $\rho(\lambda \!=\!0,V)$ and 
$\rho_{ch}(\lambda \!=\!0,V)$ always vanish and cannot be utilized as 
indicators of vSChSB, this is not necessarily true for all strictly infrared
constructs in finite volume. For example, in chirally broken case, 
$\lambda \!=\! 0$ is typically an isolated zero of $\rho(\lambda,V)$, but
$\cop_A(\lambda\!=\!0,V) \equiv \lim_{\lambda \to 0} \rho_{ch}(\lambda,V)/\rho(\lambda,V)$ 
is expected to be well defined and positive. In fact, the strictly infrared 
version of {\sl Conjecture 3} with $\Omega(M,T,V)$ on the right--hand side of 
Eq.~\eqref{eq:5.030} replaced by 
$\cop_A(\lambda \!=\! 0,M,T,V)$ is an equally viable representation of 
the correspondence, albeit less appealing from practical standpoint.

\begin{figure}[t]
\begin{center}
    \centerline{
    \hskip 0.00in
    \includegraphics[width=18.0truecm,angle=0]{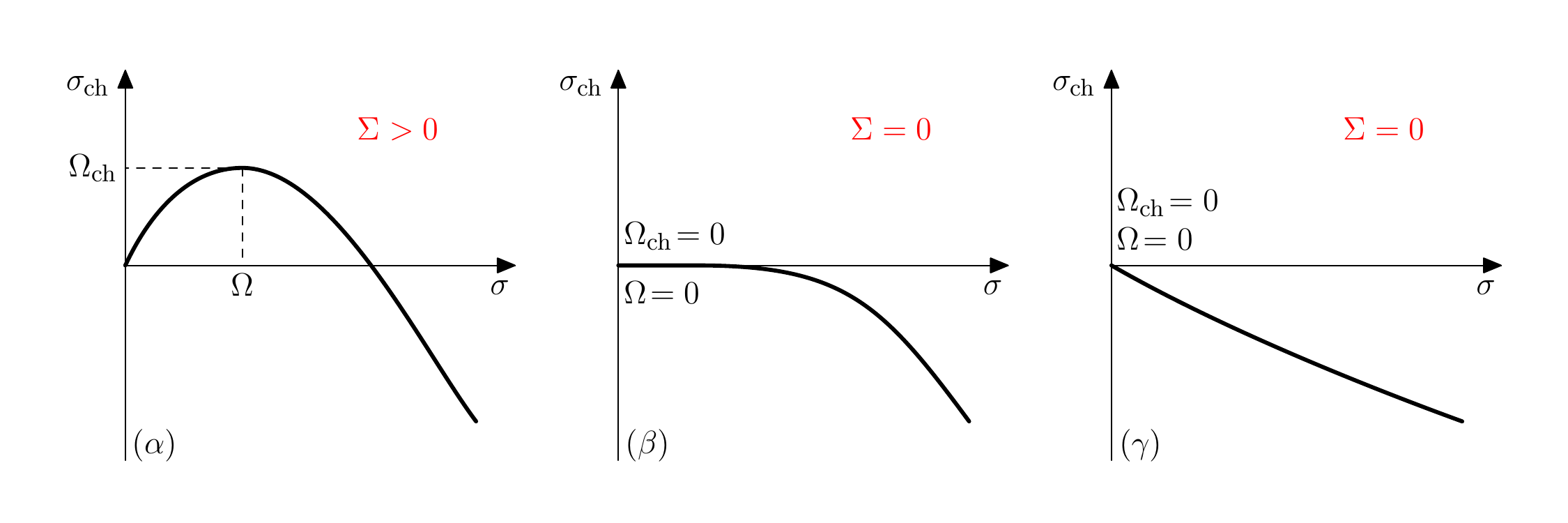}
     }
     \vskip -0.10in
     \caption{Possible behaviors of $\sigma_{ch}(\sigma)$ in the vicinity of
     $\sigma=0$. In finite volume, case $(\alpha)$ predicts vSChSB of infinite 
     volume theory, while $(\beta)$ and $(\gamma)$ entail valence 
     chiral symmetry.}
     \label{fig:sigch_vs_sig_illus}
    \vskip -0.45in
\end{center}
\end{figure}

\subsection{$\sigma$--Parametrization and the Universal Scale of vSChSB}
\label{ssec:universal}

For certain purposes, some of which are discussed below, it is useful to  
parametrize chiral polarization properties by cumulative eigenmode density 
$\sigma$, rather than Dirac spectral parameter $\lambda$. Recall that 
the description discussed so far is based on two cumulative densities, 
namely $\sigma(\lambda)$ and $\sigma_{ch}(\lambda)$: the behavior of 
$\sigma_{ch}(\lambda)$ defines polarization parameters 
$\Lambda_{ch}$ and $\Omega_{ch}$, while $\Omega$ emerges ``a posteriori'' 
from $\sigma(\lambda)$ and $\Lambda_{ch}$.

We wish to eliminate $\lambda$ from the pair $\sigma(\lambda)$, 
$\sigma_{ch}(\lambda)$ and consider $\sigma_{ch}=\sigma_{ch}(\sigma)$. 
The latter is certainly a well-defined function when $\sigma(\lambda)$ is 
one-to-one, but this is also true in general case. Indeed, since 
$\sigma(\lambda)$ is non--decreasing and can have finite jumps, there are only 
two special circumstances to examine. 
({\em a}) If $\sigma(\lambda)$ is constant on interval $[\lambda_1,\lambda_2]$ 
then multiple values of $\sigma_{ch}$ could in principle be associated with 
$\sigma_1=\sigma(\lambda_1)$. However, it follows from its definition that
$\sigma_{ch}(\lambda)$ is then also constant on $[\lambda_1,\lambda_2]$, 
and $\sigma_{ch}(\sigma_1)$ is thus unique.   
({\em b}) If $\sigma(\lambda)$ has a jump at $\bar{\lambda}$ with $\sigma^l$, 
$\sigma^r$ being the left and right values respectively, then $\sigma_{ch}(\sigma)$ 
is a priori undefined on the interval $(\sigma^l,\sigma^r)$. However, since 
the population of modes associated with interval $(\sigma^l,\sigma^r)$ is assigned 
a common value of chiral correlation, namely $\cop_A(\bar{\lambda})$, there is 
a natural unique definition of $\sigma_{ch}(\sigma)$ on $(\sigma^l,\sigma^r)$:
the linear dependence whose graph connects the points $(\sigma^l, \sigma_{ch}^l)$ 
and $(\sigma^r, \sigma_{ch}^r)$. Here $\sigma_{ch}^l$, $\sigma_{ch}^r$ are the left 
and right values of $\sigma_{ch}(\lambda)$ at $\bar{\lambda}$. Note that the slope 
of $\sigma_{ch}(\sigma)$ is $\cop_A(\sigma)$ for every $\sigma$.

\medskip

\noindent There are few points we wish to emphasize regarding the utility of 
the above.

\medskip

\noindent {\em (i)} Since $\sigma_{ch}(\lambda)$ can only have jumps at arguments
where $\sigma(\lambda)$ does (see Appendix~\ref{app:spectral}), the function 
$\sigma_{ch}(\sigma)$ is not only well--defined but always continuous on its 
domain.\footnote{In lattice units, this domain is in fact the interval $[0,12]$ 
since $N_c=3$ for theories in ${\eusm T}$. This is irrespective of volume,
cutoff or lattice Dirac operator used.} 
Given that $\sigma_{ch}(\sigma\!=\!0)\!=\!0$, there are then three possible behaviors 
of $\sigma_{ch}(\sigma)$ in the vicinity of zero: it can $(\alpha)$ turn positive, 
$(\beta)$ remain identically zero or $(\gamma)$ turn negative. While it simply 
classifies theories as chirally polarized, unpolarized or anti--polarized at lower 
spectral end, this dynamical distinction acquires deeper meaning in light of our 
conjectures. Indeed, in finite volume, the first case indicates that the theory is 
in chirally polarized phase, and is thus predicted to be chirally broken, as opposed 
to the remaining two options (see Fig.~\ref{fig:sigch_vs_sig_illus}). 
In this way, the ``low $\sigma$'' behavior of $\sigma_{ch}(\sigma)$ provides for 
rather succinct and elegant representation of the ideas discussed here. 
\medskip

\noindent One noteworthy point in this regard is that when presented with 
$\sigma_{ch}(\sigma)$ in infinite volume only, the behavior $(\beta)$ in itself 
is indefinite with regard to vSChSB. Indeed, it could correspond to the situation 
$\Omega_{ch}=0$, $\Omega>0$ with $\Omega$ undetermined by $\sigma_{ch}(\sigma)$
alone. In such case the information on large--volume behavior is needed to resolve 
the ambiguity.  
\medskip

\noindent {\em (ii)} In our discussion we paid little attention to the issues 
related to precise behavior of various new constructs in the continuum limit.
In fact, this is not essential for our purposes. Indeed, our goal is to understand
the dynamics of vSChSB, which means gaining insight in the regime of lattice theory
where it becomes insensitive to the ultraviolet cutoff. Various ``order parameters'' 
we discussed are well--defined at arbitrary finite cutoff, and serve as indicators 
of vSChSB which itself is well--defined at arbitrary finite cutoff if chirally 
symmetric Dirac operator (overlap operator) is used.
\medskip

\noindent Nevertheless, it is appealing to characterize chiral polarization via 
parameters with well--defined continuum values. As mentioned in~\cite{Ale12D}
already, $\Lambda_{ch}$ is expected to require a normalization factor to define 
its unique continuum limit e.g. in N$_f$=2+1 zero temperature QCD at physical 
point.\footnote{Note that in this discussion we implicitly assume that 
$D_{(1)}=D_{(2)}=D_{overlap}$ in Eq.~\eqref{eq:2.015}, but we expect our points
to apply whenever $D_{(2)}=D_{overlap}$.}  
However, the renormalization properties of mode density described 
in~\cite{Del05A,Giu07A} imply that $\sigma_{ren}(\lambda_{ren})=\sigma(\lambda)$ 
for renormalized/bare cumulative density. Hence, $\Omega$ is expected to be free 
of renormalization factors and have a universal continuum limit. The same would hold 
also for $\Omega_{ch}$ if the renormalization of $\rho(\lambda)$ and
$\rho_{ch}(\lambda)$ proceeded via the same factor. This, however, would have 
to be established. Note that, in such case, the whole function $\sigma_{ch}(\sigma)$ 
would be universal. 
\medskip

\noindent {\em (iii)} Removing the reference to spectral parameter $\lambda$ 
underscores the dynamical nature of the vSChSB $\leftrightarrow$ ChP equivalence. 
Indeed, in this form any explicit connection to ``strictly infrared'' scales, which 
has its roots in kinematic considerations, is eliminated. Rather, the correspondence 
relies exclusively on local dynamical behavior of lowest modes, irrespective of 
precise Dirac eigenvalues they are labeled with. Given that $\sigma_{ch}(\sigma)$ 
can be conveniently computed directly, without invoking the spectral representation, 
it is worthwhile to formulate the proposed connection directly in this language. 
In fact, the formulation becomes somewhat more concise: $\Omega$ is defined 
as the maximal $\bar{\sigma}$ such that $\sigma_{ch}(\sigma)$ is strictly 
increasing on $[0,\bar{\sigma}]$, while $\Omega_{ch}$ is always simply 
$\sigma_{ch}(\sigma=\Omega)$ due to continuity. The analog of {\sl Conjecture 3}
is then as follows, with $\Omega(V)$ in Eq.~\eqref{eq:5.040} replaceable by 
$\Omega_{ch}(V)$ if so desired.

\medskip

\noindent {\bf Conjecture 3'}

\smallskip
\noindent {\sl The following holds in every lattice--regularized $(M,T) \in {\eusm T}$
at sufficiently large ultraviolet cutoffs $\Lambda_{lat}$.}  
\medskip

\noindent{\sl (I)} {\sl Function $\sigma_{ch}(\sigma,V)$ is continuous on its 
domain $\sigma \in [0,12 \Lambda_{lat}^4]$, and $\Omega(V) \ll 12\Lambda_{lat}^4$ for 
sufficiently large $V$.}

\medskip

\noindent{\sl(II)} {\sl The property of valence spontaneous chiral symmetry breaking 
is characterized by}
\begin{equation}
  \Sigma > 0    \quad  \Longleftrightarrow   \quad   \Omega(V) > 0 
  \quad  \mbox{\rm for} \quad V_0 < V < \infty
  \label{eq:5.040}
\end{equation}
\medskip
\noindent{\sl(III)} {\sl $\sigma_{ch}(\sigma,V)$ is non--increasing for 
$\sigma > \Omega(V)$, except possibly in the vicinity of $\sigma=12 \Lambda_{lat}^4$.}

\medskip

\noindent {\em (iv)} Relative to discussion in Ref.~\cite{Ale12D}, our description
of chiral polarization around the surface of Dirac sea became more detailed.
In addition to the width of the polarized layer ($\Lambda_{ch}$), the phenomenon
is also characterized by the volume density of total number of modes involved 
($\Omega$) and the associated total chirality ($\Omega_{ch}$). In the spirit 
of Refs.~\cite{Ale12D,Ale10A} and the conjectures discussed here, we consider 
the corresponding scales, namely $\Lambda_{ch}$, $\Omega^{1/4}$ and 
$\Omega_{ch}^{1/4}$, to be the dynamical scales associated with the phenomenon of 
vSChSB. In the massless light--quark limit of ``real--world QCD'' (i.e. N$_f$=2+1), 
they become the scales of SChSB. In Ref.~\cite{Ale12D} we estimated the 
(unrenormalized) value of chiral polarization scale in N$_f$=2+1 QCD at physical 
point to be $\Lambda_{ch} \approx 80$ MeV. The estimated values of the other 
two parameters in this case are $\Omega^{1/4} \approx 150$ MeV (expected to be 
universal) and $\Omega_{ch}^{1/4} \approx 60$ MeV.

\section{Temperature Effects}
\label{sec:fintemp}

In this section, we present results of the finite--temperature study in 
N$_f$=0 QCD. This theory has a well--established deconfinement transition
temperature $T_c$ defined via expectation value of the Polyakov loop.
According to the standard scenario, vSChSB disappears in close vicinity
of $T_c$, but in general at some other temperature $T_{ch}$. We find that
mode condensation and chiral polarization of valence overlap quarks exactly 
follow each other, in accordance with vSChSB--ChP correspondence. 
The data also shows that, at the fixed lattice cutoff used 
($\Lambda_{lat} \simeq 2.3\,$GeV), chiral transition temperature $T_{ch}$ is 
strictly larger than $T_c$.

\subsection{Lattice Setup and Polyakov Loop Sectors}

We simulate pure--glue SU(3) theory with Wilson action at $\beta=6.054$. 
The non--perturbative parametrization of Ref.~\cite{Gua98A} is used to set 
the lattice scale, resulting in $a/r_0=0.170$ at the aforementioned gauge 
coupling. Using the standard value $r_0\!=\!0.5\,$fm for reference scale, this 
translates into $a=0.085\,$fm. To perform a basic temperature scan, we vary 
the ``time'' extent of the lattice between $N_t\!=\!4$ and $N_t\!=\!20$ which 
corresponds to temperatures $T=1/(N_ta)$ in the range 116--579 MeV. The spatial
extent of the lattice is kept fixed at $N\!=\!20$, corresponding to volume 
$V_3=(N a)^3=(1.7\,$fm $\!\!)^3$. The information about these ensembles is 
summarized in Table~\ref{tab:fint_ensemb} with some relevant explanations 
provided below.

\begin{table}[t]
   \centering
   \begin{tabular}{@{} ccccccccc @{}} 
      \toprule
      Ensemble  &  $N_t$  &  $T/T_c$  &  $T$[MeV]  &  $N_{cfg}$  & 
      $|\lambda|_{min}^{av}$  &  $|\lambda|_{min}$ &  
      $|\lambda|_{max}^{av}$  &  $|\lambda|_{max}$\\
      \midrule
     $E_{1}$  &  20  & 0.42  &  116  &  100  &  0.0204 & 0.0027 & 0.6128 & 0.6160 \\
     $E_{2}$  &  12  & 0.70  &  193  &  200  &  0.0320 & 0.0065 & 0.7241 & 0.7270 \\
     $E_{3}$  &  10  & 0.84  &  232  &  200  &  0.0379 & 0.0018 & 0.7658 & 0.7701 \\
     $E_{4}$  &   9  & 0.93  &  258  &  200  &  0.0402 & 0.0039 & 0.7912 & 0.7944 \\
     $E_{5}$  &   8  & 1.05  &  290  &  400  &  0.0859 & 0.0011 & 0.8208 & 0.8246 \\ 
     $E_{6}$  &   7  & 1.20  &  331  &  400  &  0.2473 & 0.0006 & 0.8631 & 0.8675 \\
     $E_{7}$  &   6  & 1.39  &  386  &  100  &  0.4038 & 0.0498 & 0.9233 & 0.9283 \\
     $E_{8}$  &   4  & 2.09  &  579  &  100  &  0.7868 & 0.7129 & 1.1608 & 1.1673 \\
      \bottomrule
   \end{tabular}
   \caption{20$^3 \times N_t$ ensembles of N$_f$=0 theory with Wilson gauge  
   action ($\beta=6.054$), used in the overlap eigenmode calculations. 
   $|\lambda|_{min}^{av}$ is the average magnitude of smallest non--zero 
   eigenvalue in a configuration, while $|\lambda|_{max}^{av}$ that of  
   the largest one. The magnitudes of all computed non--zero eigenvalues in 
   an ensemble satisfy $|\lambda|_{min} \le |\lambda| \le |\lambda|_{max}$.}
   \label{tab:fint_ensemb}
   \vskip -0.06in
\end{table}

Distinctive aspect of studying chiral issues in N$_f$=0 theory relates
to the fact that, while the deconfinement transition is associated
with spontaneous breakdown of Z$_3$ symmetry signaled by the expectation
value of Polyakov loop~\cite{Pol78A,Sve82A}, Dirac spectral properties in 
the deconfined phase depend on which vacuum broken theory happens 
to visit. In particular, it was pointed out that valence chiral symmetry
in the ``real sector'' might be restored at lower temperature than in 
the ``complex sectors''~\cite{Cha95A}, with the former transition 
being the one occurring close to $T_c$. While this conclusion remained somewhat 
controversial~\cite{Gat02A}, it became common to study the real sector in 
connection with chiral symmetry restoration in N$_f$=0 QCD. This is also tied 
to the fact that dynamical fermions tend to bias gauge fields correspondingly 
(see e.g.~\cite{Kov08A}), bringing up the expectation that the dynamics of 
the real--sector vacuum better resembles the behavior in real--world QCD. 

\begin{table}[b]
   \centering
   \begin{tabular}{@{} cccccccc @{}} 
      \toprule
      Ensemble  &  $N$  &  $L$[fm]  &  $N_{cfg}$  & 
      $|\lambda|_{min}^{av}$  &  $|\lambda|_{min}$ &  
      $|\lambda|_{max}^{av}$  &  $|\lambda|_{max}$\\
      \midrule
     $G_{1}$  &  16  & 1.36  &  400  &  0.301513 & 0.005507 & 1.029540 & 1.034761 \\
     $G_{2}$  &  24  & 2.04  &  400  &  0.200300 & 0.000089 & 0.746919 & 0.753360 \\
     $G_{3}$  &  32  & 2.72  &  200  &  0.067461 & 0.000046 & 0.598446 & 0.607929 \\
      \bottomrule
   \end{tabular}
   \caption{$N^3\times$7 ensembles of N$_f$=0 theory with Wilson gauge plaquette action
   ($\beta=6.054$), used to study finite volume effects in the $N_t=7$ system ($E_6$).}
   \label{tab:fint_ensemb2}
\end{table}

Here we adopt the above point of view and
present results for the real Polyakov loop sector. While in infinite volume the Z$_3$ 
distinction is only relevant above $T_c$, it is reasonable to keep the separation 
on both sides of the transition for finite system due to tunneling. To that 
effect, we rotated each generated configuration into complementary Z$_3$ sectors, but 
$N_{cfg}$ of Table~\ref{tab:fint_ensemb} refers to independent unrotated 
configurations. Except for ensembles $E_1$ and $E_2$, Dirac eigensystems were 
calculated in all Z$_3$ sectors for each configuration, and the statistics for real--phase 
results is thus $N_{cfg}$. For $E_1$ and $E_2$, only configurations originally 
generated in real sector, 31 and 61 of them respectively, were included. 
Overlap Dirac operator with Wilson kernel ($r\!=\!1$) and $\rho\!=\!26/19$ was used 
in these valence quark calculations, and the quoted spectral 
bounds refer to the real Polyakov--loop sector. For all ensembles used in this work,
200 lowest eigenmodes with non--negative imaginary part were computed.

\begin{figure}[t]
\begin{center}
    \centerline{
    \hskip 0.00in
    \includegraphics[width=9.5truecm,angle=0]{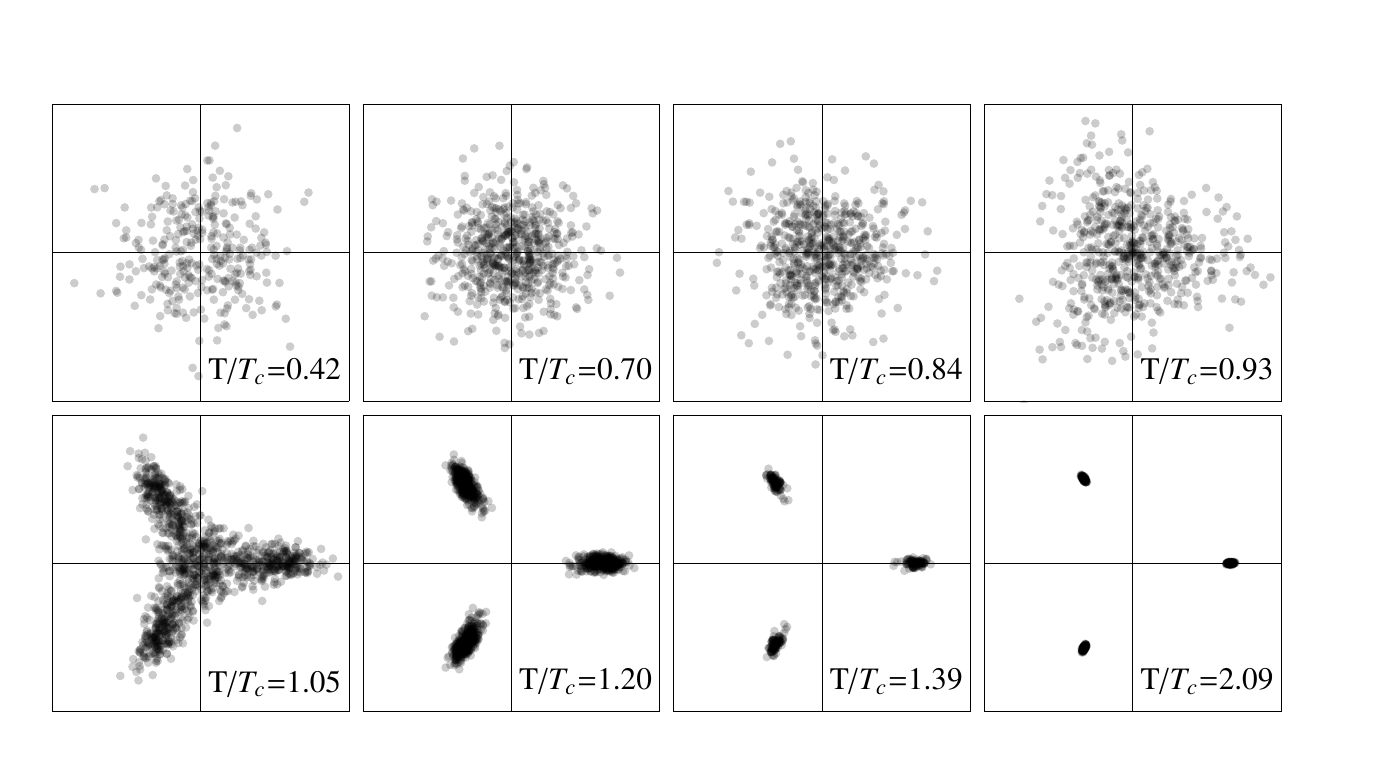}
    \hskip 0.0in
    \includegraphics[width=7.5truecm,angle=0]{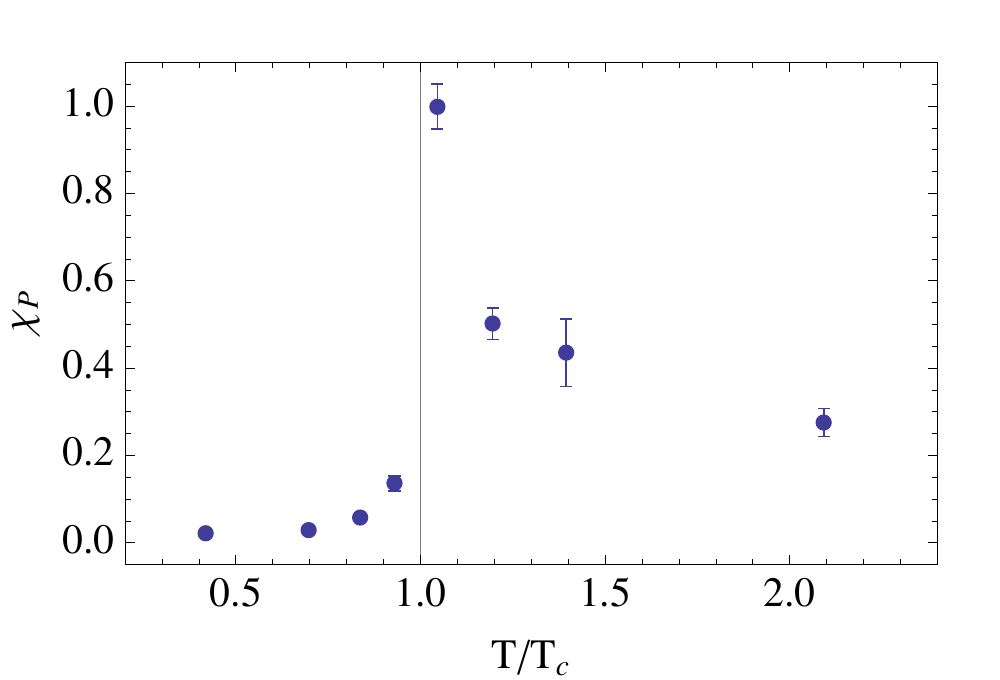}
     }
     \vskip 0.00in
     \caption{Left: scatter plots of Polyakov loop for ensembles $E_1$--$E_8$. Each point 
     corresponds to a configuration with Z$_3$ symmetrization included. Sliding scale has 
     been adjusted in each case so as to contain all points. Right: Polyakov--loop 
     susceptibilities.}
    \label{fig:z3}
    \vskip -0.45in
\end{center}
\end{figure}

We use $T_c/\sqrt{\sigma}=0.631$, quoted in Ref.~\cite{Kar97A}, as a reference value 
for infinite--volume continuum--limit transition temperature to label our ensembles. 
With string tension value $\sigma=(440\, \mbox{\rm MeV})^2$ this translates into 
$T_c=277\,$MeV. 
Since the volume and lattice cutoff are finite, and there is also a small uncertainty 
in the determination of the lattice scale, it is prudent to check whether Z$_3$ 
transition in our system occurs in the expected range of temperatures. To do so, we 
show the scatter plot of the Polyakov loop in Fig.\ref{fig:z3} (left). As can be seen 
quite clearly, the expected symmetric distribution below $T_c$ is contrasted with 
Z$_3$--concentrated population above $T_c$. This is also confirmed by the behavior 
of Polyakov--loop susceptibility shown in Fig.\ref{fig:z3} (right). There is thus little 
doubt that ensembles $E_1$--$E_4$ and $E_5$--$E_8$ represent the system in confined and 
deconfined phases respectively.

In the course of our analysis it will turn out that, while clearly in the deconfined 
phase, the $N_t\!=\!7$ system still exhibits chiral polarization and vital signs of 
vSChSB. To ascertain this, we study the finite--volume behavior on additional 
lattices with details given in Table~\ref{tab:fint_ensemb2}.  
The ensemble $G_3$ with $V_3=(2.72\,$fm $\!\!)^3$ corresponds to the largest system 
studied. All the above notes concerning $E$--ensembles also apply to $G$--ensembles.

\subsection{Raw Data}

\begin{figure}[t]
\begin{center}
    \centerline{
    \hskip 0.00in
    \includegraphics[width=17.0truecm,angle=0]{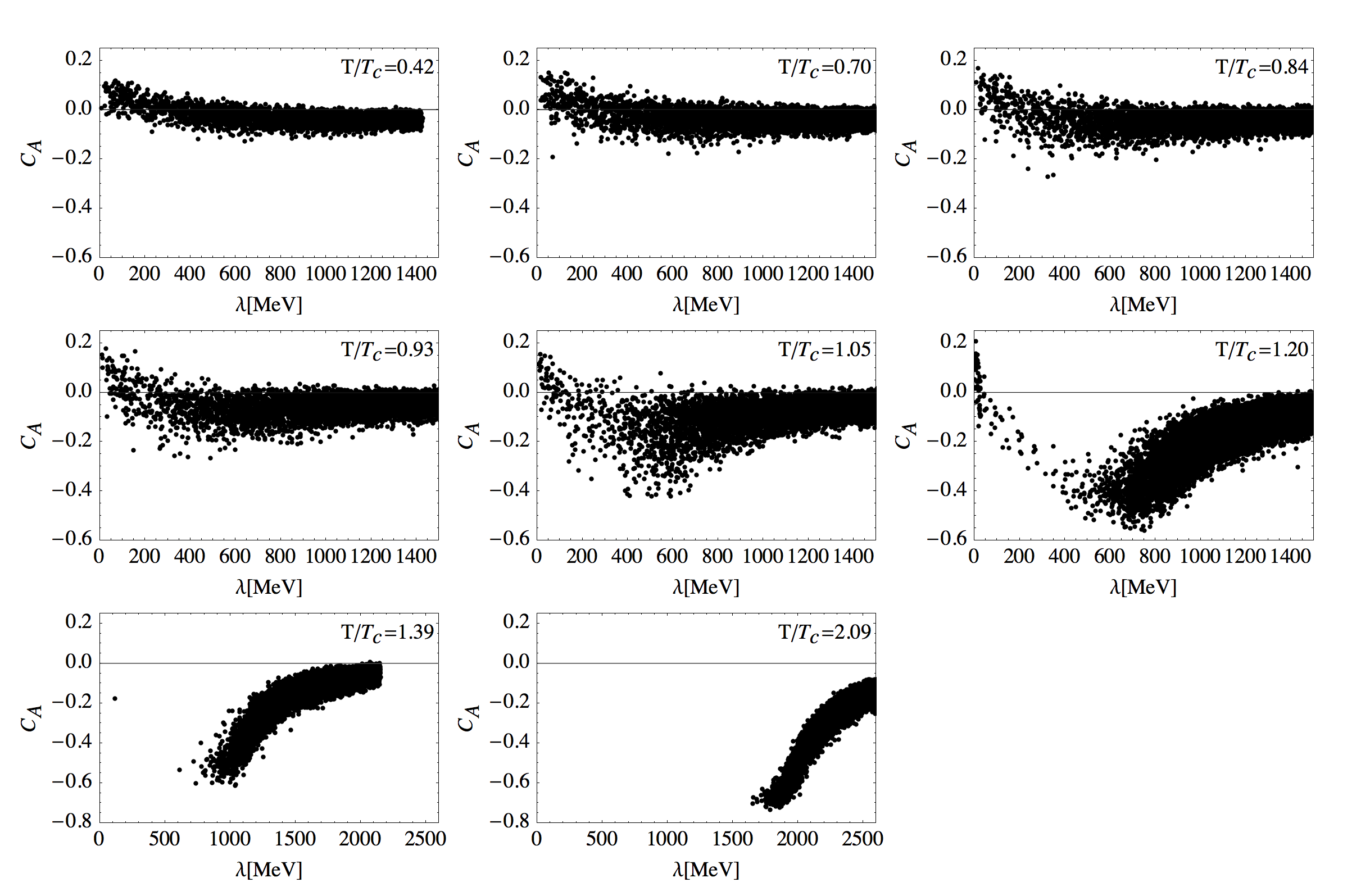}
     }
     \vskip -0.10in
     \caption{Scatter plots of chiral polarization correlation ($\cop_A$) versus 
     eigenvalue magnitude ($\lambda$) for $E$--ensembles of 
     Table~\ref{tab:fint_ensemb}. Note the change of scale for $E_7$, $E_8$ relative
     to $E_1$--$E_6$.}
     \label{fig:T_raw_CA}
    \vskip -0.45in
\end{center}
\end{figure}

One useful way to obtain a quick overview of the situation at hand is to examine 
the scatter plots of $\cop_A$ versus $\lambda$ since they provide a simultaneous
qualitative picture of spectral abundance and chiral polarization. Thus, each
eigenmode associated with given ensemble contributes a point to the plot, specified 
by the magnitude of the eigenvalue ($\lambda$) and its correlation coefficient of 
polarization ($\cop_A$). In Fig.~\ref{fig:T_raw_CA} we show these plots for
$E$--ensembles. Note that the temperature increases in lexicographic order. 

Regarding the vSChSB$\leftrightarrow$ChP correspondence, the main aspect to examine 
is whether presence of near--zero modes is always associated with tendency for chiral 
polarization at low energy, and that chiral polarization is always absent in 
accessible spectrum if near--zero modes are not produced. Such qualitative
correlation is clearly observed in Fig.~\ref{fig:T_raw_CA}.

Quick look at scatter plots also suggests three kinds of qualitative behavior as the
temperature is increased. First, the ``low--temperature dynamics'', exemplified by 
$E_1$ ($T/T_c=0.42$) and discussed extensively in \cite{Ale12D}, appears to apply
throughout the confined phase. Second, the ``transition dynamics'', exemplified 
by $E_6$ ($T/T_c=1.20$) extends approximately from $T_c$ to chiral transition point
$T_{ch}$. It is characterized by the spectral separation of near--zero modes from 
the bulk and the creation of mode--depleted region between them. Finally, 
the "high--temperature dynamics'' turns on above $T_{ch}$ as near--zeromodes can 
no longer be supported in sufficient numbers, and the anti--polarization of the bulk 
takes over.

\begin{figure}[t]
\begin{center}
    \centerline{
    \hskip 0.00in
    \includegraphics[width=17.5truecm,angle=0]{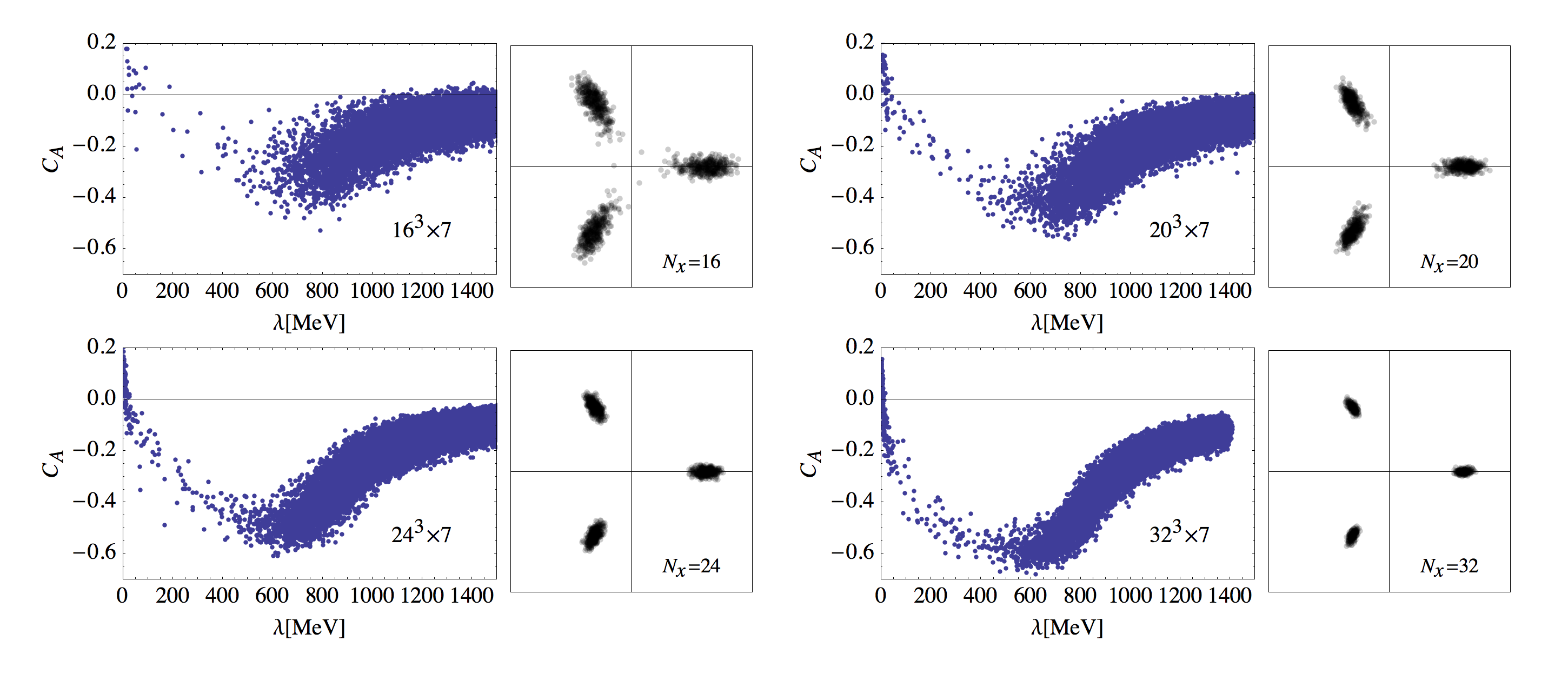}
     }
     \vskip -0.10in
     \caption{Scatter plots ($\cop_A$ vs $\lambda$) for $N_t\!=\!7$ system ($T/T_c=1.20$) 
     in varying spatial volume. The associated raw data for Polyakov loop is also shown.}
     \label{fig:T_raw_CA_nt7}
     \vskip -0.45in
\end{center}
\end{figure}

In the above temperature scan, the system associated with ensemble 
$E_6$ ($T/T_c=1.20$) is of prime interest: not only does it suggest itself as 
a lattice example of vSChSB with deconfined gauge fields~\cite{Edw99A} but, more 
importantly, it is the most borderline case where the vSChSB$\leftrightarrow$ChP 
association should be ascertained. To do this at the level of raw data, we show 
in Fig.~\ref{fig:T_raw_CA_nt7} the $\lambda$--$\cop_A$ scatter plots for this system 
in increasing 3--volumes. The abundance of near--zero modes clearly rises with volume,
with the layer of high concentration increasingly focused toward the origin. Thus, 
mode condensation appears all but inevitable. At the same time, chiral polarization 
persists as predicted. Decreasing width of the polarized layer suggests possible 
singular behavior in the vein of our discussion in Sec.~\ref{ssec:conjectures}. Note 
that we also show the associated scatter plots of Polyakov loop, clearing any suspicion 
that the system could be in confined phase.

\subsection{Chiral Polarization Transition}

We now start putting the observations of the previous section into more formal terms. 
The first task in this process is to determine the transition point (in temperature) for 
chiral polarization. It should be emphasized that whether the system is in the polarized 
phase or not is a well--defined question in any finite volume. The temperature scan 
will be performed for $E$--ensembles sharing the same 3--volume. Resulting transition 
point is thus associated with the corresponding ultraviolet cutoff and volume.

\begin{figure}[]
\begin{center}
    \centerline{
    \hskip 0.00in
    \includegraphics[width=16.5truecm,angle=0]{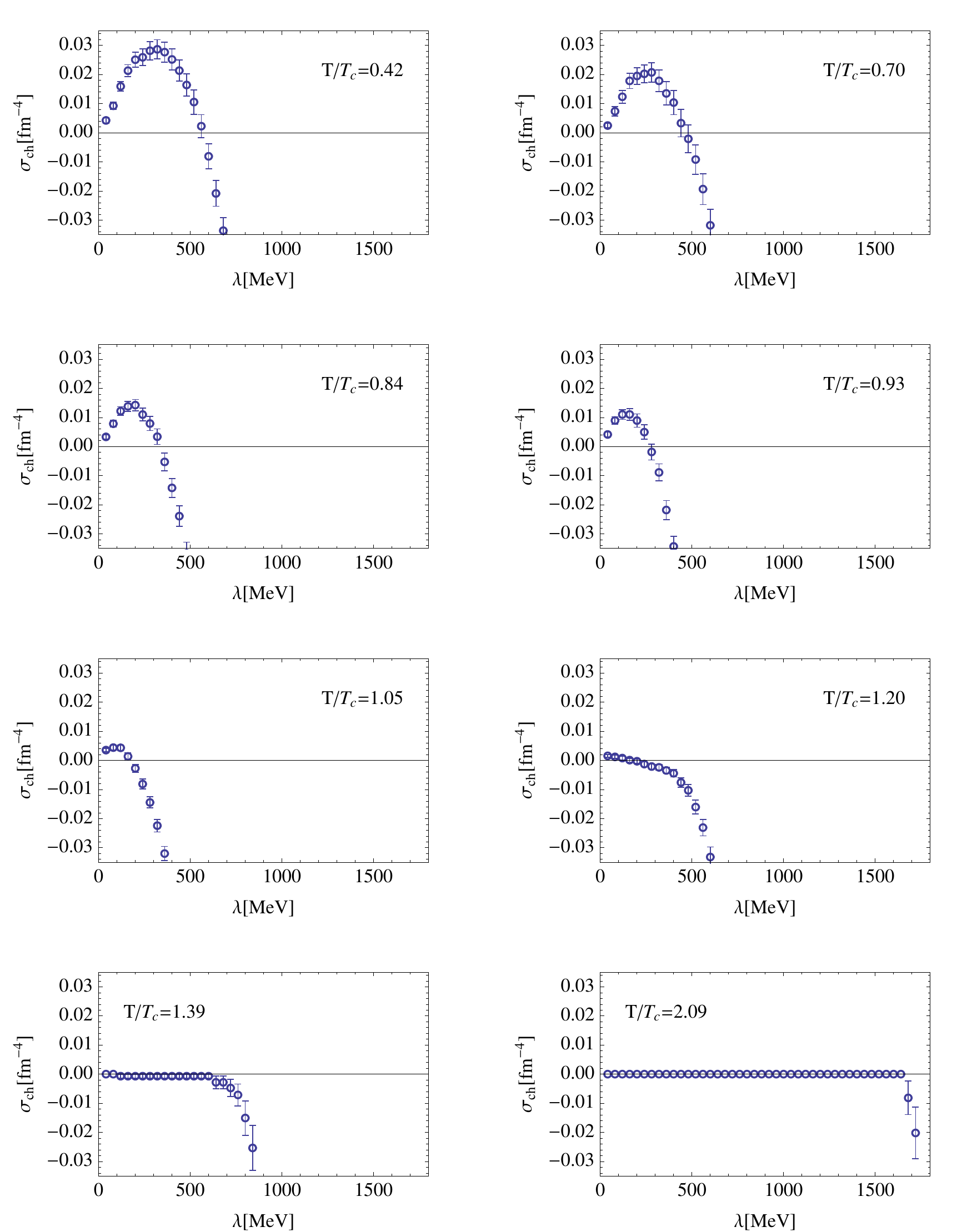}
     }
     \vskip -0.00in
     \caption{Cumulative chiral polarization density for all $E$--ensembles. Ranges
     are fixed.}
     \label{fig:T_sigmach_lam_all}
    \vskip -0.45in
\end{center}
\end{figure}

\begin{figure}[]
\begin{center}
    \centerline{
    \hskip 0.00in
    \includegraphics[width=16.5truecm,angle=0]{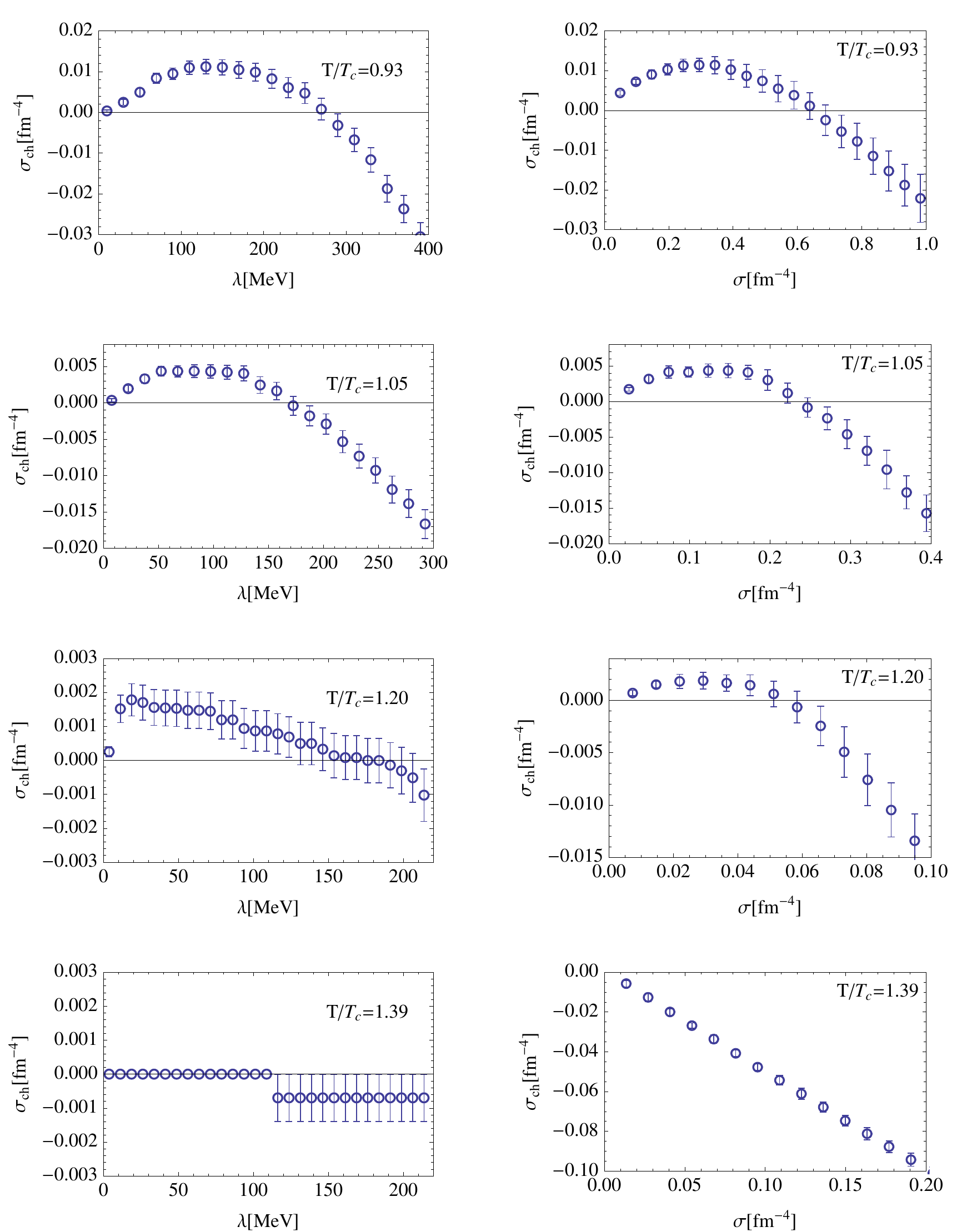}
     }
     \vskip -0.00in
     \caption{Low--energy closeup of $\sigma_{ch}(\lambda)$ (left column) and 
     $\sigma_{ch}(\sigma)$ (right column) for ensembles $E_4$--$E_7$. Theory is clearly 
     chirally polarized at $T/T_c=1.20$. Note the sharp transition in the latter 
     representation when changing to $T/T_c=1.39$.}
     \label{fig:T_sigmach_lam_sig}
    \vskip -0.45in
\end{center}
\end{figure}

To do this in a straightforward way, we monitor the behavior of cumulative chiral 
polarization density $\sigma_{ch}(\lambda)$. The result is shown in 
Fig.~\ref{fig:T_sigmach_lam_all} with temperature increasing in lexicographic order.
The characteristic positive bump (see discussion for Fig.~\ref{fig:sigch_vs_lam_illus})
appears at low temperatures signaling the creation of chirally polarized layer around 
the surface of the Dirac sea. Data is shown on identical scales for all ensembles
to see the changing position of the maximum ($\Lambda_{ch}$) as well as its value 
($\Omega_{ch}$). The polarization feature is clearly absent at $T/T_c=1.39$, while it
appears to be present at the borderline case ($T/T_c=1.20$) which, however would benefit
from better resolution.

In Fig.~\ref{fig:T_sigmach_lam_sig} we show closeups of the low part of the spectrum for 
ensembles $E_4$--$E_7$. Note that the scales are no longer fixed in order to properly 
resolve the polarization feature. 
Temperature grows from top to bottom with left column displaying 
$\sigma_{ch}(\lambda)$ and the right column the associated $\sigma_{ch}(\sigma)$.
As advertised, and suggested by the raw data, at $T/T_c=1.20$ the system is still
in chirally polarized phase. Both $\sigma_{ch}(\lambda)$ and $\sigma_{ch}(\sigma)$ 
tell the same story, but note how the latter effectively removes the depleted regions 
of the Dirac spectrum (ensembles $E_6$ and $E_7$) from consideration, focusing on 
polarization properties of existing modes.

\begin{figure}[h]
\begin{center}
    \centerline{
    \hskip 0.12in
    \includegraphics[height=3.98truecm,angle=0]{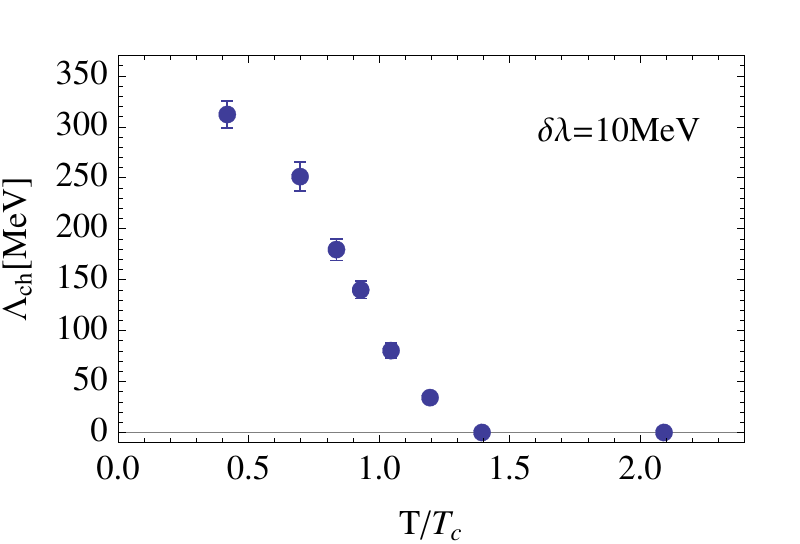}
    \hskip -0.16in
    \includegraphics[height=3.98truecm,angle=0]{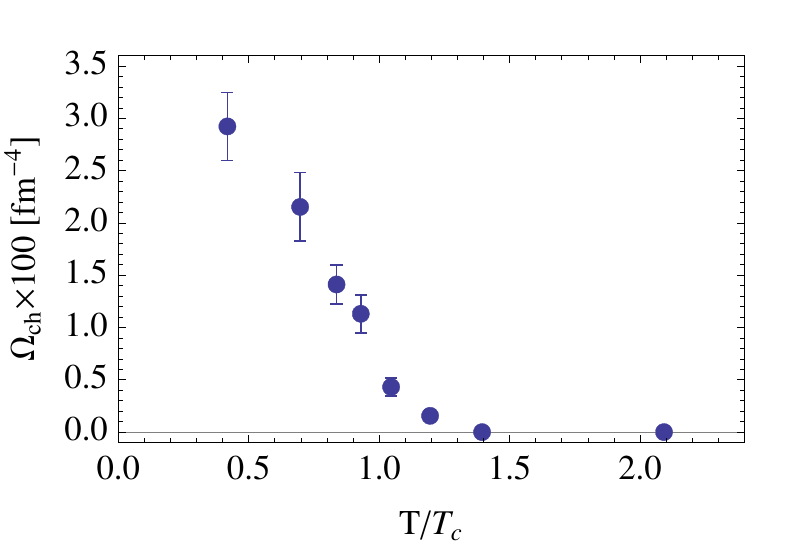}
    \hskip -0.16in
    \includegraphics[height=3.98truecm,angle=0]{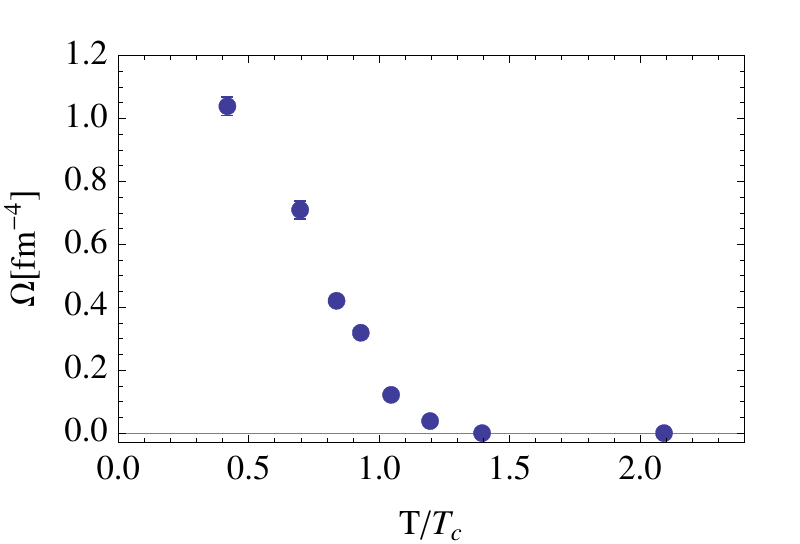}
     }
     \vskip -0.05in
     \caption{Global characteristics of chiral polarization in $E$--ensembles as functions 
     of temperature. They indicate the transition temperature $1.2\,T_c < T_{ch} < 1.39\,T_c$ 
     for the cutoff and volume in question. $\delta\lambda$ refers to coarse--graining 
     parameter used in determination of $\Lambda_{ch}$.}
     \label{fig:params_vs_T}
    \vskip -0.25in
\end{center}
\end{figure}

Finally, in Fig.~\ref{fig:params_vs_T} we show the temperature dependence for the three 
global characteristics of the chirally polarized layer, namely $\Lambda_{ch}$, 
$\Omega_{ch}$ and $\Omega$. As discussed in Sec.~\ref{ssec:finvol}, they are all 
equivalent indicators of chiral polarization in finite volume, and potentially 
{\em finite--volume order parameters} of vSChSB, as proposed by {\em Conjecture 3}. 

\begin{figure}[h]
\begin{center}
    \centerline{
    \hskip 0.04in
    \includegraphics[width=5.76truecm,angle=0]{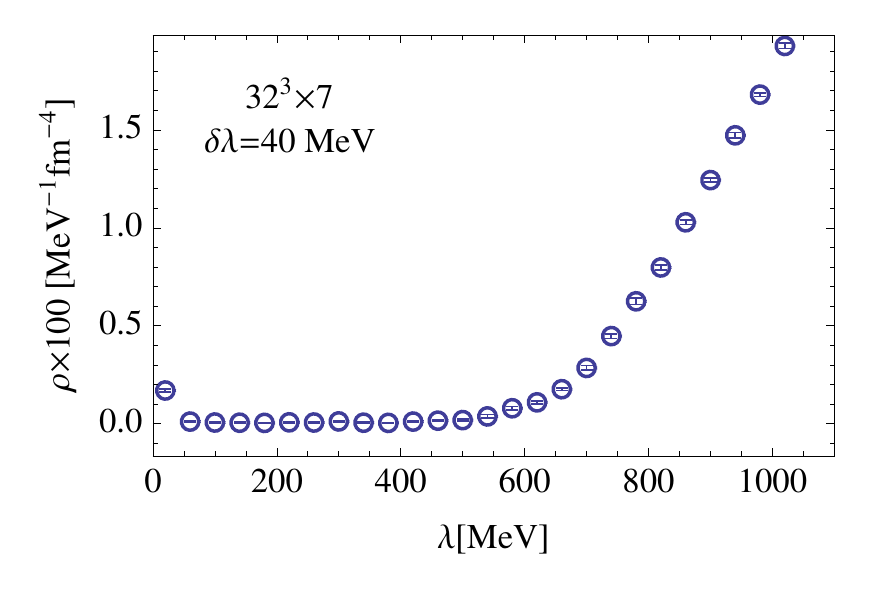}
    \hskip -0.14in
    \includegraphics[width=5.76truecm,angle=0]{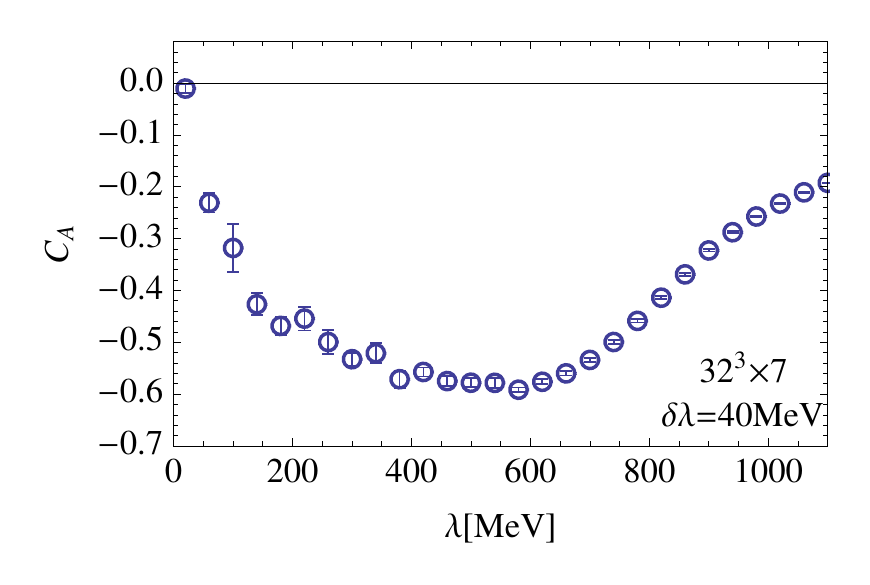}
    \hskip -0.14in
    \includegraphics[width=5.76truecm,angle=0]{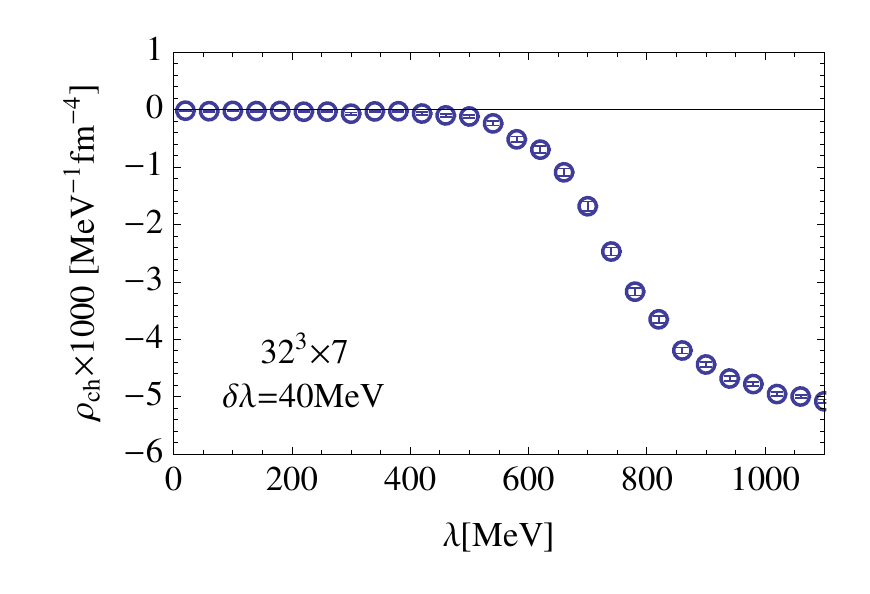}
     }
     \vskip -0.1in
    \centerline{
    \hskip 0.05in
    \includegraphics[width=5.8truecm,angle=0]{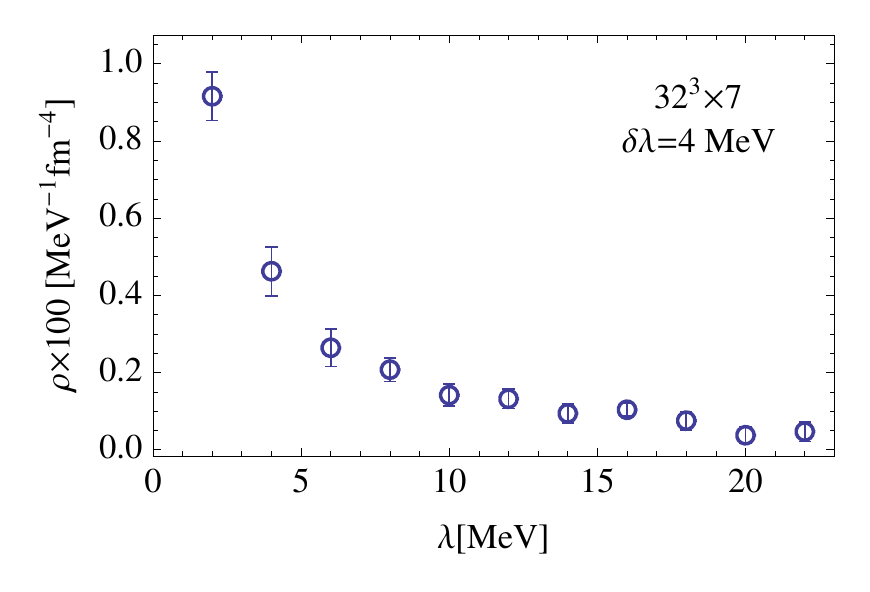}
    \hskip -0.14in
    \includegraphics[width=5.8truecm,angle=0]{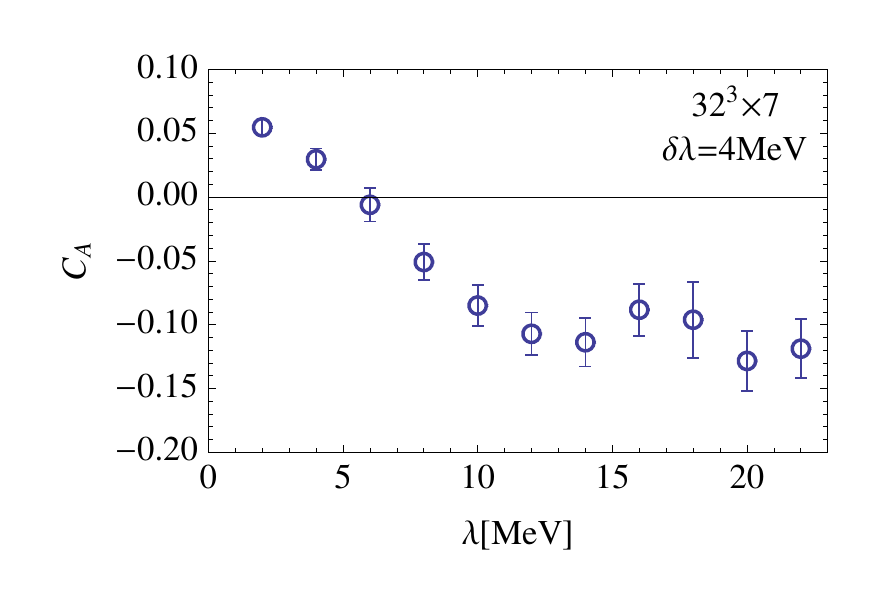}
    \hskip -0.14in
    \includegraphics[width=5.8truecm,angle=0]{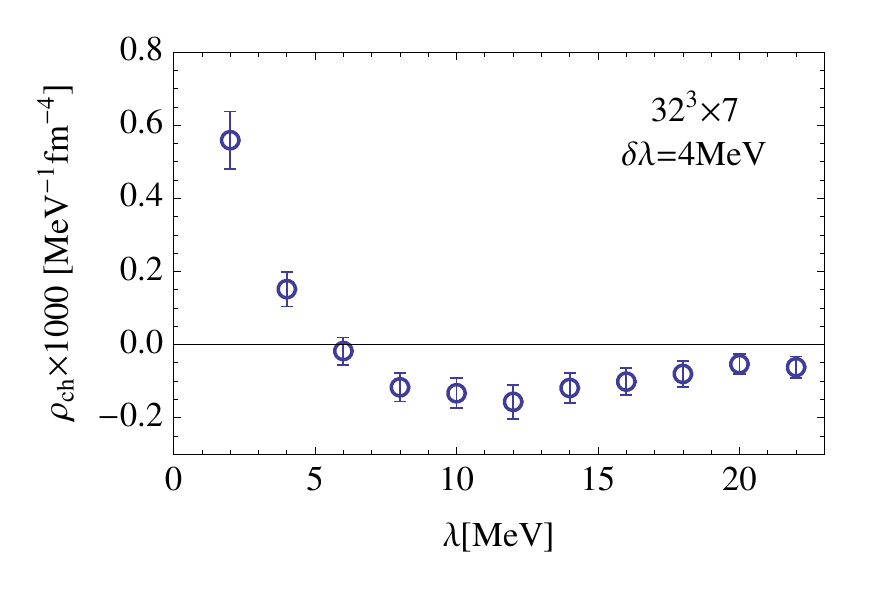}
     }
     \vskip -0.1in
     \caption{Large view (top) and closeup (bottom) of $\rho(\lambda)$, $\cop_A(\lambda)$, 
     $\rho_{ch}(\lambda)$ for ensemble $G_3$.}
     \label{fig:n32_nt7}
    \vskip -0.25in
\end{center}
\end{figure}

\subsection{Infinite Volume}

To test the proposed conjectures, it is necessary to deal with infinite volume
considerations since the definition of vSChSB explicitly relies on it.
In more concrete terms,
it is important to check which forms of {\em Conjecture 2} (if any) apply 
to the finite--cutoff situation at hand. Also, assessing possible merits in the notion
of finite--volume order parameter, associated with {\em Conjecture 3}, depends on 
the behavior of polarization observables in large volumes.

As is clear from our discussion in preceding sections, the relevant situation 
to study in this regard is $T/T_c=1.2$, represented by $E_6$ and the $G$--ensembles. 
Indeed, this is the borderline case, providing the most sensitive test for vSChSB--ChP 
correspondence. To overview the situation directly in the largest volume available 
($N\!=\!32$), we show in Fig.~\ref{fig:n32_nt7} (top) the triplet of characteristics 
$\rho(\lambda)$, $\cop_A(\lambda)$, $\rho_{ch}(\lambda)$ over a large spectral range, 
fully covering the depleted region observed in the raw data. While the increased 
density near the origin is clearly visible, chiral polarization feature is not 
recognizable at this resolution. However, closeup to the vicinity of the origin 
in Fig.~\ref{fig:n32_nt7} (bottom) reveals a clear polarized layer as anticipated. 
Note that the feature became quite thin in this largest volume.

\begin{figure}[t]
\begin{center}
    \centerline{
    \hskip -0.1in
    \includegraphics[width=15.5truecm,angle=0]{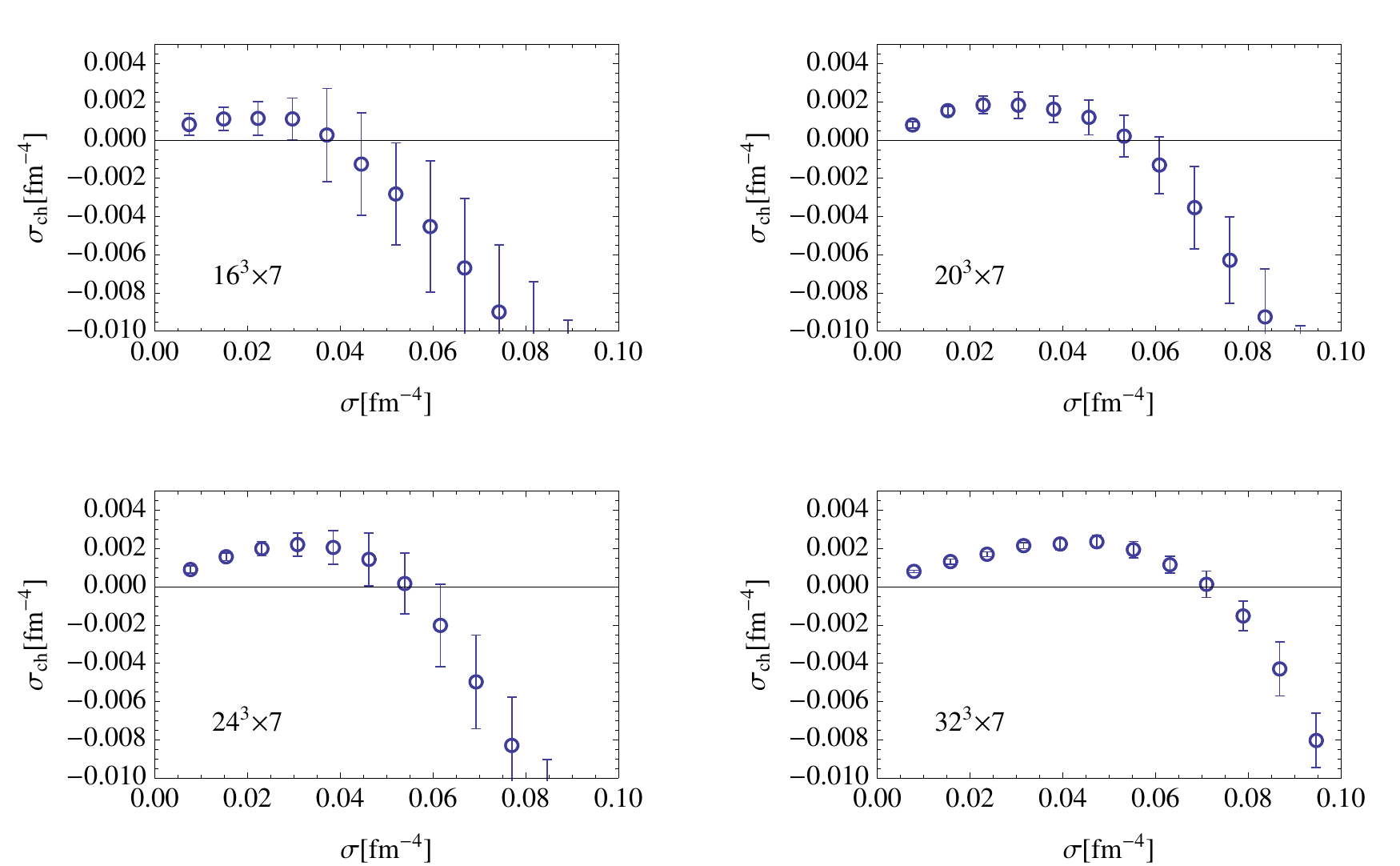}
     }
     \vskip -0.00in
     \caption{Polarization characteristic $\sigma_{ch}(\sigma)$ at $T/T_c=1.2$
     for increasing 3--volumes.}
     \label{fig:nt7_sigmach_sig}
    \vskip -0.45in
\end{center}
\end{figure}

To examine the situation in detail, we first focus on the ChP side of the correspondence.
In practical terms, this means determining the form of the chiral polarization layer 
for large and infinite volume. To that effect, we show in Fig.~\ref{fig:nt7_sigmach_sig} 
the behavior of $\sigma_{ch}(\sigma)$ for the four volumes available. As pointed out in 
the previous section, this is the most robust way of visualizing the layer in case of 
depleted spectra such as those we are dealing with in the transition region. Indeed, 
even though Fig.~\ref{fig:T_raw_CA_nt7} indicates severe depletion (even close to 
the spectral origin) for the $N\!=\!16$ system, the polarization layer is still quite
clearly visible in $\sigma_{ch}(\sigma)$. In larger volumes, the characteristic positive 
bump tends to grow both in $\sigma$ and $\sigma_{ch}$--directions, strongly suggesting 
that the ChP layer remains the dynamical feature of the system in the infinite 
volume limit.

\begin{figure}[t]
\begin{center}
    \centerline{
    \hskip 0.03in
    \includegraphics[height=4.0truecm,angle=0]{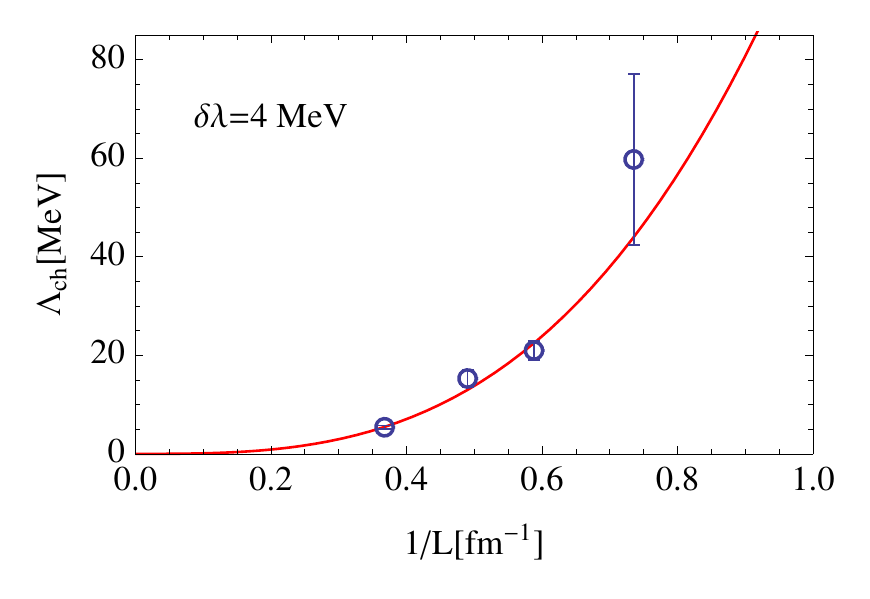}
    \hskip -0.15in
    \includegraphics[height=4.0truecm,angle=0]{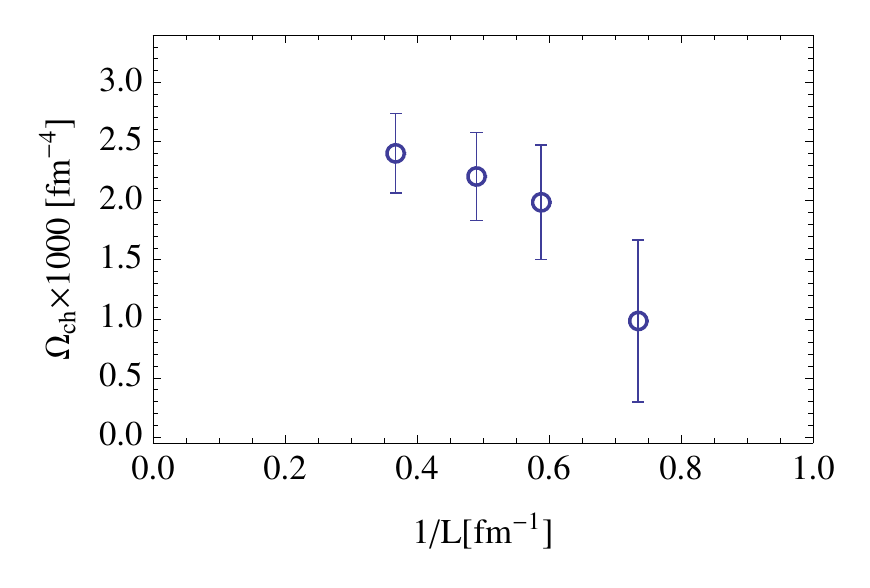}
    \hskip -0.15in
    \includegraphics[height=4.0truecm,angle=0]{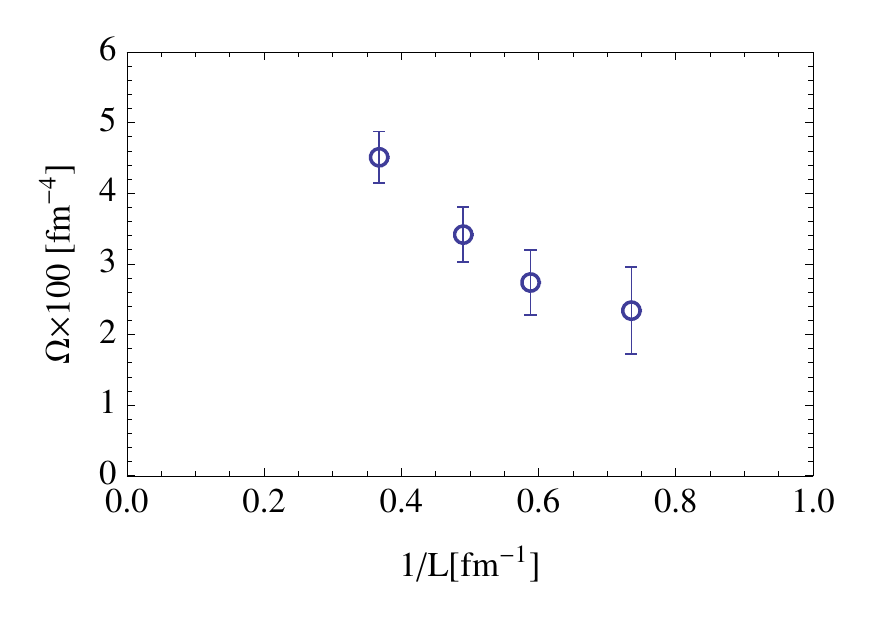}
     }
     \vskip -0.05in
     \caption{Global characteristics of chiral polarization at $T/T_c=1.2$ against
     infrared cutoff. Simple power--law fit was included in case of 
     $\Lambda_{ch}(1/L)$ to guide the eye.}
     \label{fig:params_vs_V}
    \vskip -0.20in
\end{center}
\end{figure}

There is, however, a difference in how infinite--volume ChP is realized at $T/T_c \!=\! 1.2$ 
and at zero or small temperatures. This is revealed in Fig.~\ref{fig:params_vs_V} 
where we show the dependence of chiral polarization parameters on the infrared cutoff. 
As foretold by Fig.~\ref{fig:nt7_sigmach_sig}, $\Omega_{ch}$ and $\Omega$ grow as 
the cutoff is removed, but $\Lambda_{ch}$ decreases, likely toward zero. 
Continuation of these trends in larger volumes is only possible when $\rho(\lambda,V)$ 
and $\rho_{ch}(\lambda,V)$ develop a $\delta(\lambda)$--core in $V \to \infty$ limit, 
to keep the observed volume density of polarized modes and volume density of dynamical 
chirality finite. We may thus be dealing with a singular case of the type 
($\Lambda_{ch}\!=\!0$, $\Omega_{ch} \!>\! 0$, $\Omega \!> \!0$) discussed repeatedly 
in Sec.~\ref{sec:background}. Including such ChP behaviors was the main motivation behind 
extending our formalism to current form.

Let's now examine the vSChSB side of the correspondence. This might seem unnecessary from
strictly logical standpoint since the result of our ChP analysis is such that it already 
carries with it the implication of vSChSB (with divergent condensate) at $T/T_c=1.2$.
However, it certainly helps the case if the same conclusion can also be reached on its own, 
independently of chiral polarization. We thus wish to check directly whether dynamics 
at $T/T_c=1.2$ is consistent with the definition of mode--condensing theory. To begin with, 
note that the behavior of $\rho(\lambda)$ for largest volume available, shown in the 
lower--left plot of Fig.~\ref{fig:n32_nt7}, is in itself strongly suggestive of mode 
condensation. Indeed, contrary to monotonically increasing function typical of zero and 
low temperatures, this $\rho(\lambda)$ is monotonically decreasing in the very infrared 
range shown. One thus naively expects non--zero mode density to survive at the spectral 
origin.

\begin{figure}[t]
\begin{center}
    \centerline{
    \hskip -0.0in
    \includegraphics[height=8.0truecm,angle=0]{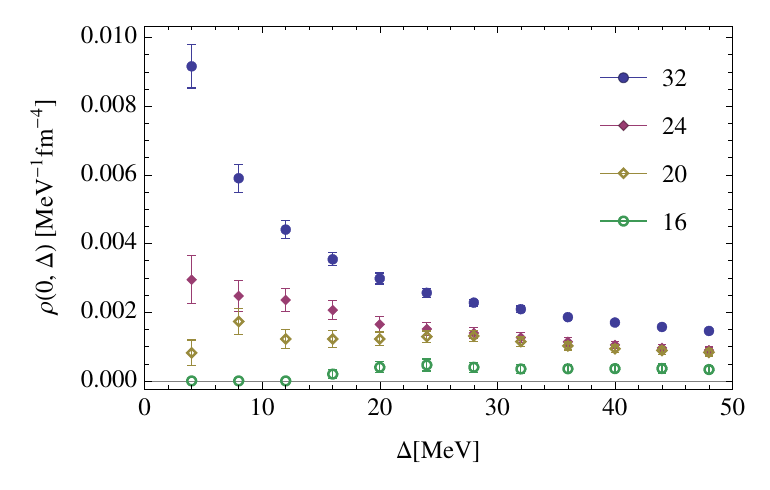}
     }
     \vskip -0.05in
     \caption{Coarse--grained mode condensate $\rho(0,\Delta)$ at $T/T_c=1.2$
     for increasing 3--volumes.}
     \label{fig:rho0_Delta_all}
    \vskip -0.15in
\end{center}
\end{figure}

However, the definition \eqref{eq:2.050} of mode condensation demands that volume 
trends be examined at fixed infrared spectral windows to make meaningful conclusions. 
In line with this definition, we consider the coarse--grained version of $\rho(0)$, namely
\begin{equation}
   \rho(\lambda\!=\!0, \Delta, V) \,\equiv\, \frac{1}{\Delta} \,
   \sigma(\lambda\!=\!\Delta,V)
\end{equation}
to see the associated volume tendencies for various values of $\Delta$.
The result of such calculation for a range of infrared windows down to 4 MeV is shown 
in Fig.~\ref{fig:rho0_Delta_all}. Note that for any fixed $V$, function 
$\rho(0,\Delta,V)$ will approach zero for $\Delta \to 0$. This is explicitly seen in 
$N\!=\!16$ and $20$ cases but not for the two larger 3--volumes where the downward bend 
occurs at yet smaller values of $\Delta$. Important feature of these results is 
that $\rho(0,\Delta,V)$ grows with volume for all $\Delta$ shown. In fact, the rate of
growth increases at small $\Delta$, and even before the occurrence of the bend, leaving 
little room for the possibility that 
$\lim_{\Delta\to 0} \lim_{V\to\infty} \rho(0,\Delta,V)$ vanishes. Rather, the data
is consistent with diverging mode condensate as expected from ChP analysis.

To summarize, we presented evidence that the layer of chirally polarized modes around 
the surface of the Dirac sea remains the feature of N$_f$=0 QCD at $T/T_c=1.2$ even in 
the infinite--volume limit. At the same time, the system was found to be mode--condensing,
in accordance with the general vSChSB--ChP correspondence. The specific form of ChP layer
conforms to the types described by {\em Conjectures 2',2''}, but most likely doesn't fall 
into the realm of {\em Conjecture 2}. We emphasize that this doesn't mean that 
{\em Conjecture 2} is invalid: this would only transpire if the concluded type of ChP 
behavior persisted at arbitrarily large ultraviolet cutoff. Note also that our
analysis is in agreement with {\em Conjectures 3,3'}, thus lending support to the concept 
of finite--volume order parameter. 

\subsection{Absolute \Xg--Distributions at Finite Temperature}
\label{ssec:Xd}

In the previous analysis of QCD phase transition, we focused on spectral properties 
based on the correlation coefficient of chiral polarization $\cop_A$. Indeed, this 
is sufficient for formulating and verifying the vSChSB--ChP correspondence which is 
our main focus here. However, for other purposes involving vacuum structure, more 
detailed information contained in absolute \Xg--distributions $\xd_A(\Xg)$ might be 
valuable. In Ref.~\cite{Ale10A} ({\em Proposition 1}) it was concluded that,
in N$_f$=0 QCD at zero temperature, $\xd_A(\Xg)$ has a simple behavior for
low--energy Dirac eigenmodes: it is either purely convex or purely concave, with 
the former being associated with polarization while the latter with anti--polarization. 
It is thus a natural question to ask whether something different happens in this 
regard due to thermal agitation.

To start such inquiry, we wish to establish how \Xg--distribution of lowest modes 
changes with temperature and, in particular, whether a qualitative shift occurs when
crossing the chiral transition point $T_{ch}$. To formalize this question, it is 
preferable to think in terms of $\xd_A(\Xg,\sigma)$ rather than the canonical
$\xd_A(\Xg,\lambda)$ of Eq.~\eqref{eq:4.045}. Indeed, this puts the low--temperature 
systems with abundance of small eigenvalues, and the high--temperature systems with 
low--energy depletion, on the same footing. We are then interested in 
$\xd_A(\Xg,\sigma \!\to\! 0)$ which in practice needs to be coarse--grained with 
respect to $\sigma$. The latter is accomplished by considering 
$\xd_A(\Xg,\sigma\!=\!0,\Delta)$, which represents $\xd_A(\Xg,\sigma)$ averaged
over $\sigma \in [0,\Delta]$.    

\begin{figure}[t]
\begin{center}
    \centerline{
    \hskip 0.0in
    \includegraphics[height=7.0truecm,angle=0]{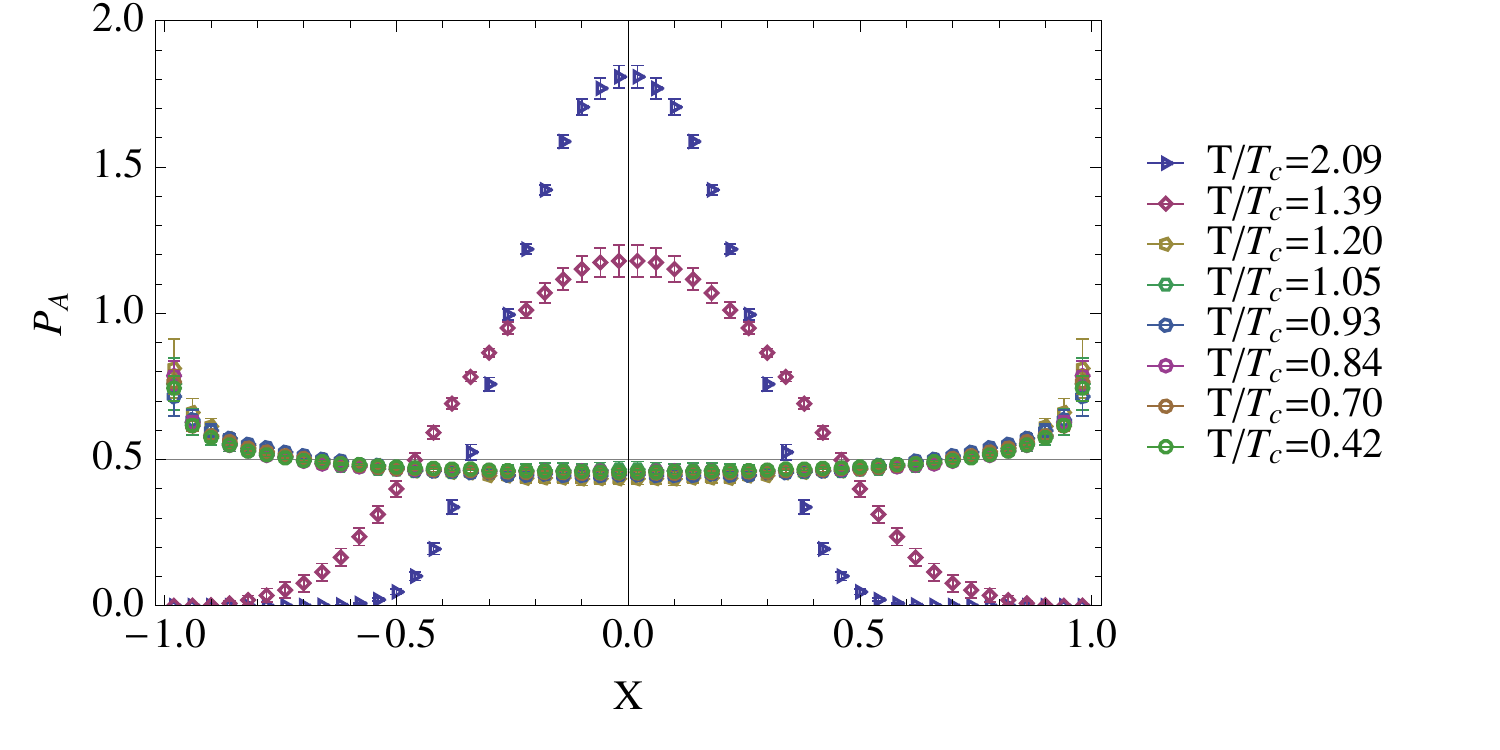}
     }
     \vskip -0.10in
     \caption{The absolute \Xg--distribution $\xd_A(\Xg,\sigma\!\to\!0)$, i.e. for 
     lowest modes in $E$--ensembles.}
     \label{fig:PA0_E_all}
    \vskip -0.35in
\end{center}
\end{figure} 

The result of such calculation for $E$--ensembles is shown in Fig.~\ref{fig:PA0_E_all}. 
We chose to fix the (total) number of included modes to 25 for each system, which
corresponds to small coarse--graining parameter $\Delta$ in the range 0.02--0.15 fm$^{-4}$.
This statistics is also sufficient to verify that further lowering this cut doesn't 
change the behavior of absolute $\Xg$--distribution. As one can see, chiral transition 
is dramatically reflected in $\xd_A(\Xg,\sigma\!=\!0)$: we observe a new type of functional 
behavior above $T_{ch}$ (ensembles $E_7$ and $E_8$), namely that of indefinite convexity. 
To specify the convexity properties more precisely, we will implicitly view $\xd_A(\Xg)$ 
as defined on a positive chiral branch, namely $\Xg \in [0,1]$, in what follows. The new
behaviors we found are then characterized by presence of a single inflection point 
$\Xg_0 \in (0,1)$ of the type concave-to-convex, i.e. $\xd_A(\Xg)$ is concave on 
$[0,\Xg_0]$ and convex on $[\Xg_0,1]$.

\medskip

\noindent {\bf Conjecture 4}

\smallskip

\noindent {\sl Consider lattice--regularized N$_f$=0 theory at arbitrary temperature
$T$. Let $T_{ch}\!=\!T_{ch}(\Lambda_{lat},V_3)$ be the temperature of chiral polarization 
transition at sufficiently large ultraviolet cutoff $\Lambda_{lat}$, and sufficiently 
large 3--volume $V_3$. Then absolute \Xg--distribution $\xd_A(\Xg,\sigma=0)$ is convex 
for $T<T_{ch}$, and has a single inflection point of type concave-to-convex for 
$T>T_{ch}$.}
\medskip

\noindent Note that in the fixed--scale approach, utilized in our numerical work, 
the set of accessible temperatures is discrete. Thus, rather than a unique $T_{ch}$ 
at fixed finite $\Lambda_{lat}$, there is a range associated with the two successive 
values of $N_t$ across which the transition occurs. This is implicitly understood in 
the above. A fine--grained question in this regard is whether there could be a brief 
phase above $T_{ch}$ where $\xd_A(\Xg,\sigma\!=\!0)$ is concave, and which cannot be 
resolved at the lattice cutoff we are using.   

The above finding naturally raises questions about the prevalence of modes
with indefinite convexity in the bulk of the finite--temperature spectrum. 
First of all, we have not found any modes of indefinite convexity for $T<T_c$. 
The absence of such modes at zero temperature was the the main part of 
{\em Propositions 1,3} in Ref.~\cite{Ale10A}, and it carries over to this wider 
regime. The proposed statement in the language used here is as follows.

\medskip

\noindent {\bf Conjecture 5a}

\smallskip

\noindent {\sl Consider lattice--regularized N$_f$=0 theory at temperature 
$T\!<\!T_c \!=\!T_c(\Lambda_{lat})$. If $\,\Omega\!=\!\Omega(\Lambda_{lat},V_3)$ is 
the density of chirally polarized modes, then the following holds at sufficiently 
large $\Lambda_{lat}$ and $V_3$. Absolute \Xg--distribution $\xd_A(\Xg,\sigma)$ is
(i) convex in $\Xg$ for $\sigma \in (0,\Omega)$, (ii) uniform for $\sigma=\Omega$, 
and (iii) concave at least for some band $\sigma > \Omega$.}

\medskip

\begin{figure}[t]
\begin{center}
    \centerline{
    \hskip 0.0in
    \includegraphics[width=12.0truecm,angle=0]{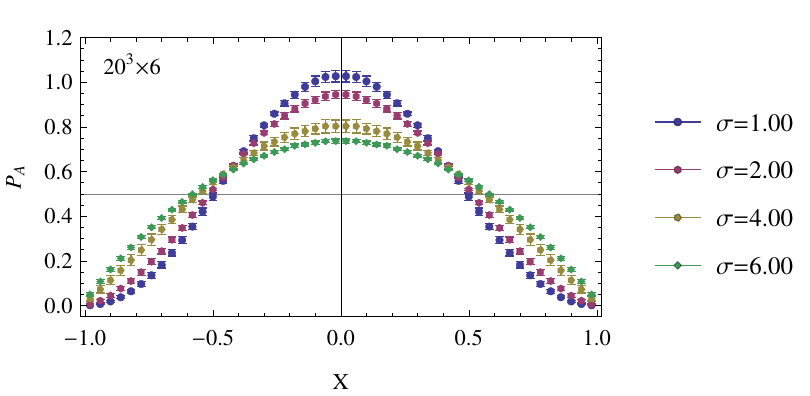}
     }
     \vskip -0.15in
     \caption{The absolute \Xg--distribution $\xd_A(\Xg,\sigma,\delta=0.2)$ for 
     ensemble $E_7$ ($T/T_c\!=\!1.39$) with increasing values of $\sigma$. 
     Value $\delta=0.2$ entails averaging over 46 eigenmodes.}
     \label{fig:PAsig_E_7}
    \vskip -0.40in
\end{center}
\end{figure}

\noindent It turns out that the situation is in fact analogous for $T>T_{ch}$ except
that the role of chirally polarized (convex) modes is assumed by modes with
indefinite convexity of $\xd_A(\Xg)$. To support this, we show in 
Fig.~\ref{fig:PAsig_E_7} the sequence of absolute \Xg--distributions with increasing 
$\sigma$ for ensemble $E_7$. Note that for non--zero values of $\sigma$ we use 
symmetric coarse--graining, averaging over the interval 
$(\sigma - \delta/2, \sigma + \delta/2)$. The data suggests the existence of 
a point $\sigma = \Omega_1$ where the behavior changes from convex--indefinite 
to concave. At this particular temperature the transition occurs in the vicinity of
$\,\Omega_1 \approx 3.0\,$fm$^{-4}$ and a precise determination can be performed if 
desired. We are thus led to formulate the following statement. 
 
\medskip

\noindent {\bf Conjecture 5b}

\smallskip

\noindent {\sl Consider lattice--regularized N$_f$=0 theory at temperature 
$T\!>\!T_{ch} \!=\!T_{ch}(\Lambda_{lat},V_3)$. At sufficiently large 
$\Lambda_{lat}$ and $V_3$ there exists $\,\Omega_1\!=\!\Omega_1(\Lambda_{lat},V_3,T)$ 
such that the following holds. Absolute \Xg--distribution $\xd_A(\Xg,\sigma)$ 
(i) has a single inflection point of type concave-to-convex for $0<\sigma<\Omega_1$ 
and (ii) is concave at least for some band $\sigma > \Omega_1$.}

\medskip

\begin{figure}[t]
\begin{center}
    \centerline{
    \hskip 0.00in
    \includegraphics[width=8.0truecm,angle=0]{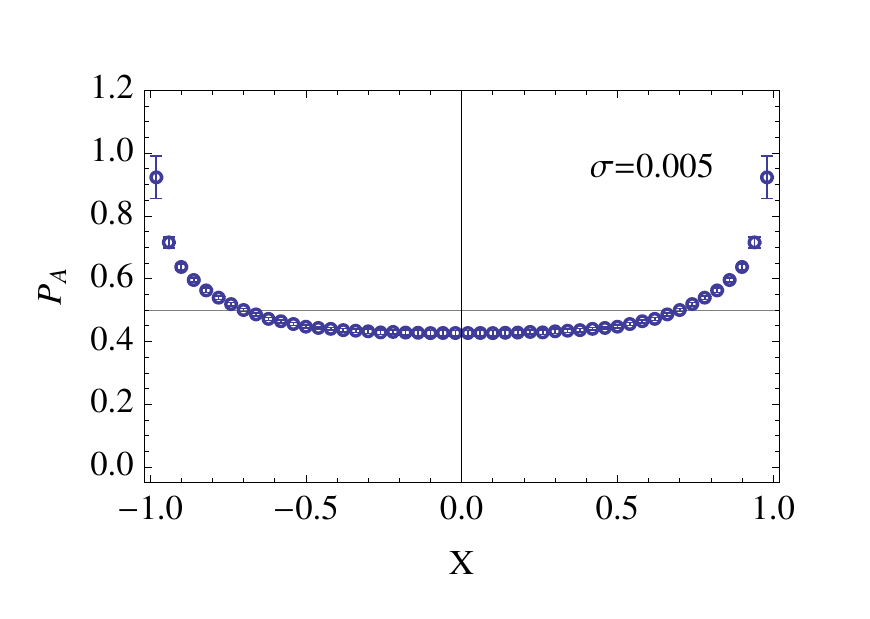}
    \hskip -0.00in
    \includegraphics[width=8.0truecm,angle=0]{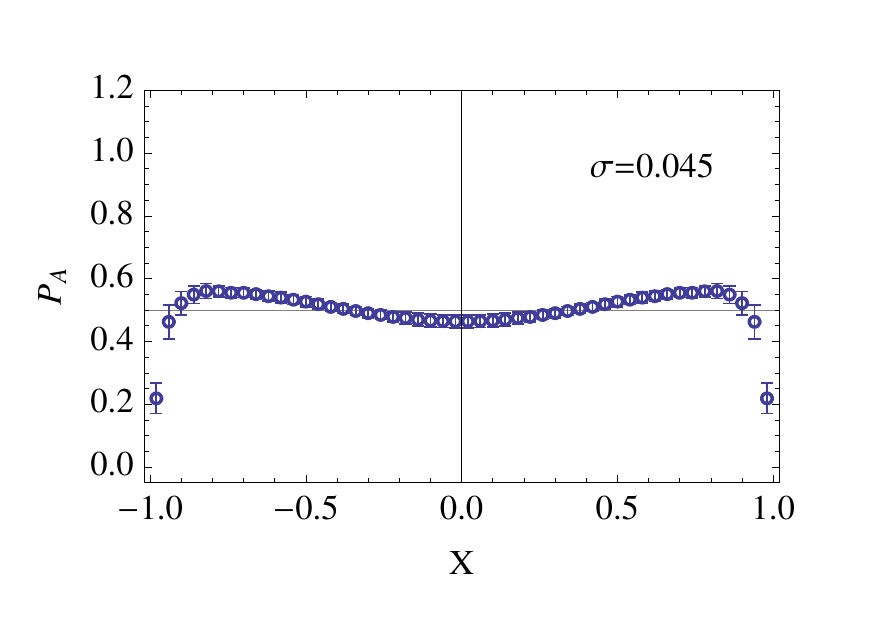}
     }
     \vskip -0.15in
    \centerline{
    \hskip 0.00in
    \includegraphics[width=8.0truecm,angle=0]{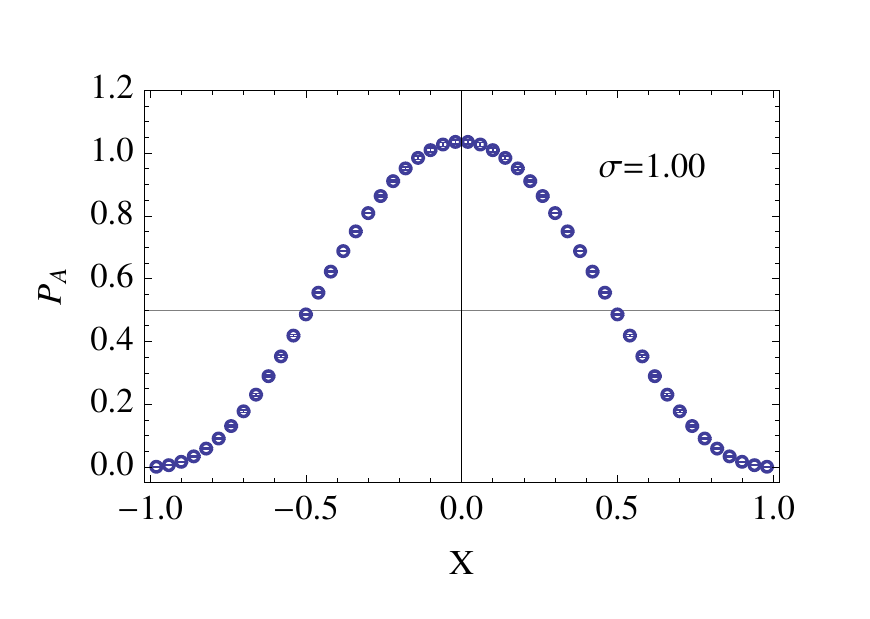}
    \hskip -0.00in
    \includegraphics[width=8.0truecm,angle=0]{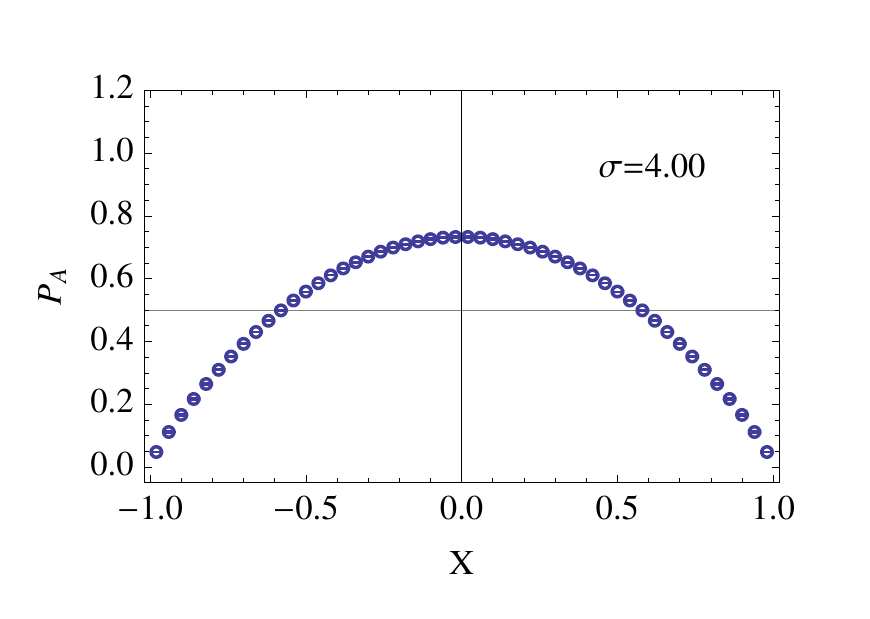}
     }
     \vskip -0.1in
     \caption{Absolute \Xg--distributions in ensemble $G_3$ for increasing $\sigma$ 
     representing four kinds of behavior found. The upper two results were obtained
     with coarse--graining parameter $\delta=0.01$ (23 modes), 
     while $\delta=0.2$ (460 modes) was used for the lower two.}
     \label{fig:pa_n32_nt7}
    \vskip -0.35in
\end{center}
\end{figure}

We finally turn to the ``mixed phase'' ($T_c < T < T_{ch}$). This dynamics 
exhibits chiral polarization ($\Omega>0$), which in case of confined vacuum happens 
to be synonymous with convexity of $\xd_A(\Xg)$ for $\sigma < \Omega$. 
However, {\em Conjecture 5b} suggests that 
deconfinement is tied to indefinite convexity of absolute \Xg--distributions. 
How then is the coexistence of vSChSB and deconfinement, and thus of chiral 
polarization and indefinite convexity, realized in the mixed phase? The specific 
arrangement we found is exemplified via ensemble $G_3$ in Fig.~\ref{fig:pa_n32_nt7}. 
The spectrum starts with a layer of convex modes (top left), like at low 
temperatures, but the distribution loses its definite convexity for 
$\Omega' < \sigma < \Omega_1$, after which it becomes concave (bottom right). 
There is a band $\Omega_0 < \sigma < \Omega_1$ within the convex--indefinite regime, 
where modes are of the type found at $T>T_{ch}$, i.e. their $\xd_A(\Xg)$ has one 
concave-to-convex inflection point (bottom left). We thus propose the following.

\medskip

\noindent {\bf Conjecture 5c}

\smallskip

\noindent {\sl Consider lattice--regularized N$_f$=0 theory at finite temperature
and overlap valence quarks. There exist lattice cutoffs $\Lambda_{lat}$ such
that $T_c(\Lambda_{lat}) < T_{ch}(\Lambda_{lat},V_3)$ for sufficiently large $V_3$. 
At temperatures $T_c < T < T_{ch}$ there are $\Omega' < \Omega_0 < \Omega_1$ 
such that $\xd_A(\Xg,\sigma)$ is (i) convex for $0 < \sigma < \Omega'$, 
(ii) convex--indefinite but not of type (iii) for $\Omega' < \sigma < \Omega_0$, 
(iii) has one inflection point of concave-to-convex type for $\Omega_0 < \sigma < \Omega_1$, 
and (iv) is concave at least for some band $\sigma > \Omega_1$. 
Moreover, $\Omega' < \Omega < \Omega_0$.}

\medskip

\noindent It should be pointed out that the convex--indefinite spectral band of 
{\sl (ii)} may just be the ``reversed'' version of {\sl (iii)}, namely that 
the associated $\xd_A(\Xg)$ has a single inflection point of convex-to-concave type.
However, our statistics is not large enough to support this aspect with sufficient 
certainty. Note also that the above formulation doesn't explicitly exclude 
the possibility that $\Omega'=0$, neither in finite volume nor in the infinite volume 
limit.\footnote{This doesn't necessarily contradict {\em Conjecture 4} which is only 
concerned with the limit $\xd_A(\Xg,\sigma\to 0)$.} However, $\Omega_0$ is predicted 
to be positive in both cases since $\Omega$ is.

\subsection{Dirac Mode Landscape at Finite Temperature}
\label{ssec:convexity}

\begin{figure}[t]
\begin{center}
    \centerline{
    \hskip 0.0in
    \includegraphics[width=16.0truecm,angle=0]{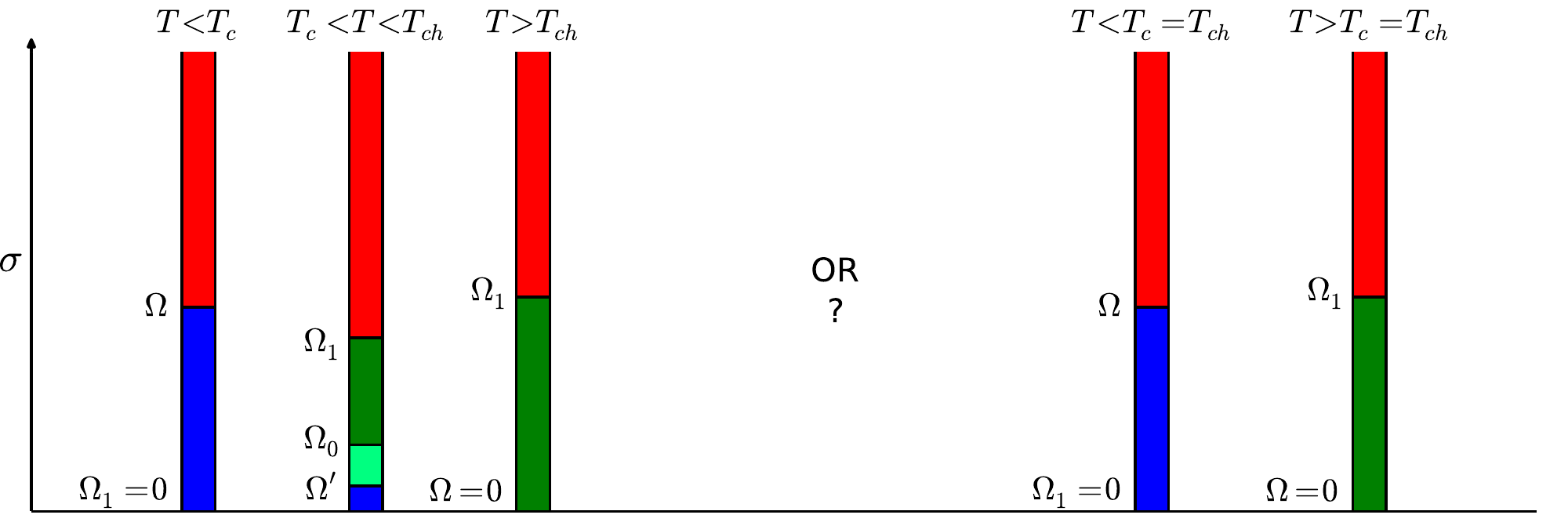}
    }
    \vskip -0.00in
    \caption{Schematic view of Dirac eigenmodes at finite temperature according to 
    their absolute \Xg--distributions with blue representing convexity, red concavity, 
    and shades of green indefinite convexity. The left side corresponds to continuum--limit 
    scenario with $T_c<T_{ch}$ when the mixed phase (middle bar) exists, while the right 
    side to scenario with $T_c=T_{ch}$.}
    \label{fig:illus2}
    \vskip -0.35in
\end{center}
\end{figure}

According to vSChSB--ChP correspondence, chiral symmetry restoration at finite
temperature is the process of chiral depolarization (positive $\cop_A$ becoming
negative) in the Dirac spectrum. However, results of the previous section suggest 
that more detailed polarization characteristic -- absolute \Xg--distribution -- encodes 
{\em both} major effects of thermal agitation: valence chiral symmetry restoration and 
deconfinement. This information is stored in convexity 
properties of $\xd_A(\Xg)$ which can in this case even be invoked on their own, without 
explicit reference to $\cop_A$. The resulting eigenmode ``convexity landscape'' 
is schematically shown in Fig.~\ref{fig:illus2} with blue and red color marking 
purely convex and purely concave behavior of $\xd_A(\Xg)$ respectively. The two shades 
of green represent convex--indefinite $\xd_A(\Xg)$, with darker version signifying 
the presence of a single concave-to-convex inflection point. 

Notice that Fig.~\ref{fig:illus2} offers two scenarios for possible behavior
in the vicinity of QCD phase transition. Indeed, whether {\em continuum} N$_f$=0 QCD 
exhibits the mixed phase or not remains an open issue, and the formulation of 
{\em Conjecture 5c} reflects this status. Thus, the option with mixed phase is shown
on the left, and the one without it on the right. Regardless of which possibility
is realized in the continuum however, the results provide for definite and 
somewhat unexpected new characterization of confinement via chirality properties.

\medskip

\noindent {\em In N$_f$=0 QCD at finite temperature, confined vacuum supports 
$\xd_A(\Xg)$ of definite convexity in overlap Dirac modes, while deconfined 
vacuum produces a band of modes with convex--indefinite $\xd_A(\Xg)$ at 
low $\sigma$.}

\medskip

With regard to the situation at high temperature ($T>T_{ch}$), one should mention 
a possible connection of these findings to recent works, proposing that QCD phase
transition can be viewed as a version of Anderson localization, with increasing 
temperature controlling the degree of randomness~\cite{Gar06A,Kov10A,Kov12A}. 
In this scenario, there
is an analogue of Anderson's mobility edge, below which Dirac modes are localized.
Given that absolute \Xg--distribution is a fully dynamical characteristic of 
the modes, it is reasonable to put forward the hypothesis that the suggested mobility 
edge is associated with ``concavity edge'' $\Omega_1$ from the above analysis. 
Denoting by $\lambda=\Lambda_{ch1}$ the spectral scale corresponding 
to $\Omega_1$, it is natural to expect that the mobility edge in fact coincides 
with $\Lambda_{ch1}$. The merits of this hypothesis need to be examined in a detailed 
dedicated study.

\vfill\eject

\section{Effects of Many Light Flavors}
\label{sec:light}

The second and qualitatively different route toward chiral symmetry restoration
within ${\eusm T}$ proceeds via including increasing number of dynamical quark flavors. 
While the general tendency is significantly more general, the usual setup deals with 
N$_f$ massless flavors at zero temperature. Owing to numerous lattice studies, as well 
as other considerations, there is very little doubt that the canonical N$_f$=2 case 
exhibits SChSB. However, massless fermions weaken the running of the gauge coupling, 
and it is expected that there is a critical number of flavors N$_{f,cr}$, beyond which 
chiral symmetry remains unbroken. The existence of N$_{f,cr}$ is connected to a larger 
issue, namely the existence of a ``conformal window'' in flavor, containing theories 
with infrared fixed point~\cite{Ban82A}. Indeed, this is expected to occur at 
N$_{f,cr}<\;$N$_f<16.5$ for SU(3), and the reliable determination of N$_{f,cr}$ is of 
an ongoing interest. The specific issue whether N$_f$=12 theory belongs to the conformal 
window gained a particular attention recently, as discussed e.g. in 
reviews~\cite{Nei12A,Gie12A,Ito13A}.
  
Rather than entering the discussion of unresolved problems such as the above, our aim 
is to check the plausibility of vSChSB--ChP correspondence in this important corner of 
quark--gluon dynamics. It should be kept in mind that the validity of the proposed 
relation is to be examined for any given lattice regularization of any given 
theory from ${\eusm T}$: it either holds or not for the regularized system at hand. 
We will thus not be much concerned with extrapolating quark mass to zero, or detailed
issues of continuum limit. Instead, our goal is to check the correspondence in the situation 
where the effect of many light fermions is apparent, and the possibility for valence 
chiral restoration exists.

\subsection{Lattice Setup}

For purposes of this pilot inquiry, we obtained some of the previously generated N$_f$=12 
staggered fermion ensembles described in Ref.~\cite{Has12B}. More specifically, 
the regularization in question uses a negative adjoint plaquette term (coupling $\beta_A$) 
in addition to fundamental plaquette (coupling $\beta_F$), and nHYP--smeared staggered 
fermions. This arrangement helps with ameliorating the problems of spurious UV fixed 
points and the unphysical lattice phases encountered in theories with many light flavors. 
Such issues have been carefully studied in this particular setting~\cite{Has12A}, and 
are not expected to arise for ensembles listed in Table~\ref{tab:stag_ensemb}.

\begin{table}[b]
   \centering
   \begin{tabular}{@{} cccccccccc @{}} 
      \toprule
      Ensemble  &  Size  &  $\beta_F$  &  $\beta_A/\beta_F$  &  am  &  N$_{conf}$  &
      $|\lambda|_{min}^{av}$  &  $|\lambda|_{min}$ &  
      $|\lambda|_{max}^{av}$  &  $|\lambda|_{max}$\\
      \midrule
     $S_{1}$  &  $16^{3}\times 32$  & 2.8  & -0.25  &  0.0200  &  100  &  
                                      0.0098  &  0.0007  & 0.5190  &  0.5259\\
     $S_{2}$  &  $32^{3}\times 64$  & 2.8  & -0.25  &  0.0025  &  30   &  
                                      0.0007  &  0.0001  &  0.2016  &  0.2035\\
     $S_{3}$  &  $24^{3}\times 48$  & 2.8  & -0.25  &  0.0025  &  50   &
                                      0.0193  &  0.0001  &  0.3046  &  0.3075\\  
      \bottomrule
   \end{tabular}
   \caption{Ensembles of N$_f$=12 lattice QCD with nHYP smeared staggered fermions and 
   fundamental--adjoint ($\beta_F\,$--$\,\beta_A$) gauge plaquette action~\cite{Has12B}.}
   \label{tab:stag_ensemb}
\end{table}

Simulations of many--light--flavor systems have to deal with the fact that the equilibrium 
gauge fields are quite rough at currently accessible lattice cutoffs. Incorporating
smearing into definition of lattice Dirac operator helps to make such calculations
feasible. To define our overlap chiral probe, we use the same smearing procedure that 
was used in Monte Carlo generation of the ensembles. Nevertheless, it is prudent
to exercise some care when using overlap operator even on fields that are moderately
rough. Indeed, the physical branch of the Wilson--Dirac spectrum can shift significantly
away from the origin on such backgrounds, and the mass parameter $\rho \in (0,2)$ 
in overlap construction needs to be chosen sufficiently large to contain it. 
To avoid potential issues of this kind, we set $\rho=1.55$, which is somewhat larger
than $\rho=26/19 \approx 1.37$ used in our ``real world QCD'' simulations.
Performing small statistics calculations with ensemble $S_1$, we verified that overlap 
low--mode abundances are reasonably stable in the vicinity of this value. 

At fixed N$_f$, the mass $m$ of degenerate quarks provides for the only parameter 
distinguishing various physical behaviors in this set of theories. At N$_f$=2, 
there is valence spontaneous chiral symmetry breaking at arbitrary $m$, while
at N$_f$=12, there could be a transition to chirally symmetric vacuum when $m$ is 
sufficiently small. The logic behind the choice of ensembles in 
Table~\ref{tab:stag_ensemb} is that system $S_1$, characterized by larger mass, was 
found in Ref.~\cite{Has12B} to be mode condensing (vSChSB) with respect to staggered 
Dirac operator, while the system represented by $S_2$ and $S_3$ appeared non--condensing 
(valence chiral symmetry) at this cutoff. Note that $S_2$ and $S_3$ only differ by volume 
to give a sense of finite--volume effects.

\begin{figure}[t]
\begin{center}
    \centerline{
    \hskip 0.00in
    \includegraphics[width=8.0truecm,angle=0]{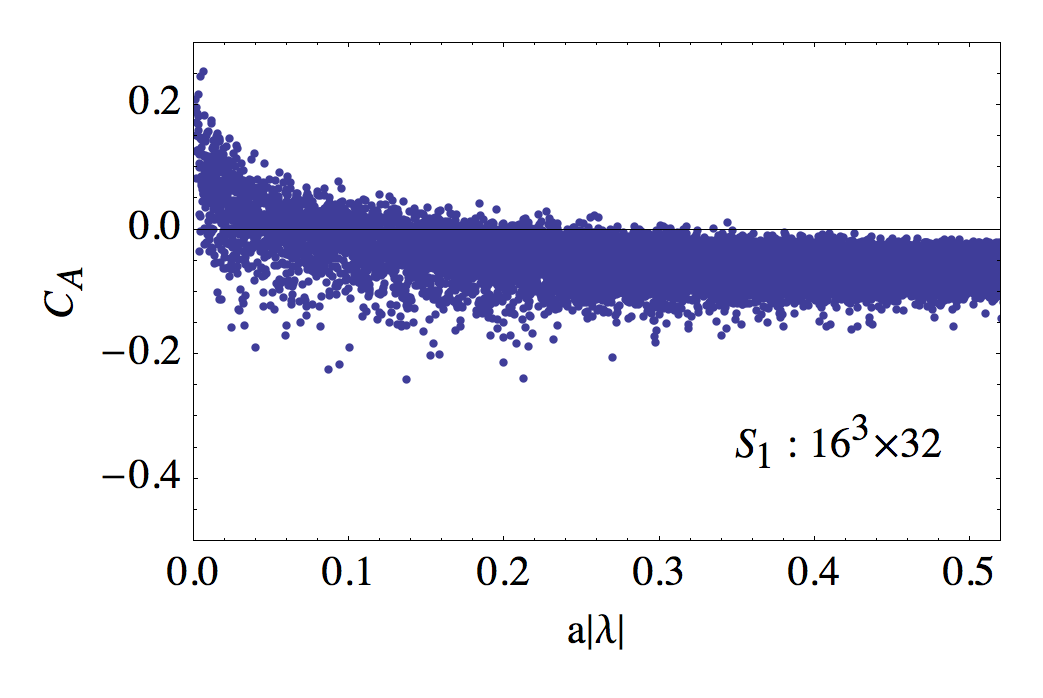}
    \hskip -0.00in
    \includegraphics[width=8.0truecm,angle=0]{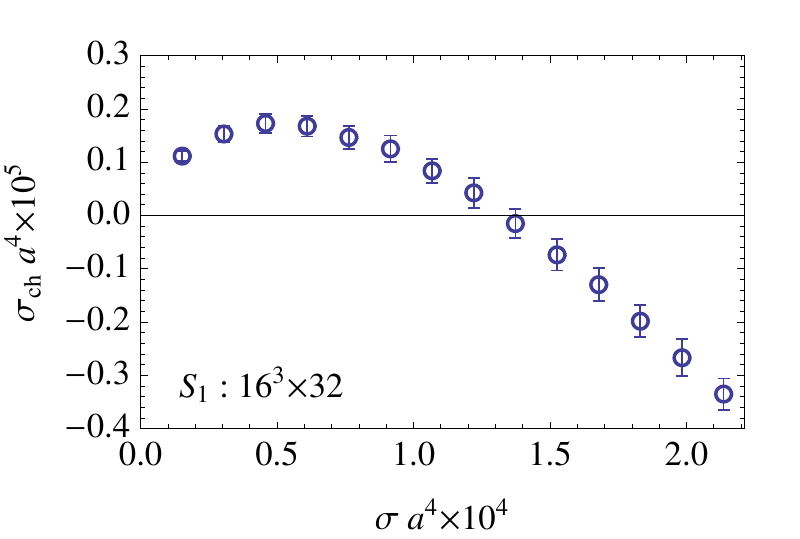}
     }
     \vskip -0.05in
    \centerline{
    \hskip 0.00in
    \includegraphics[width=8.0truecm,angle=0]{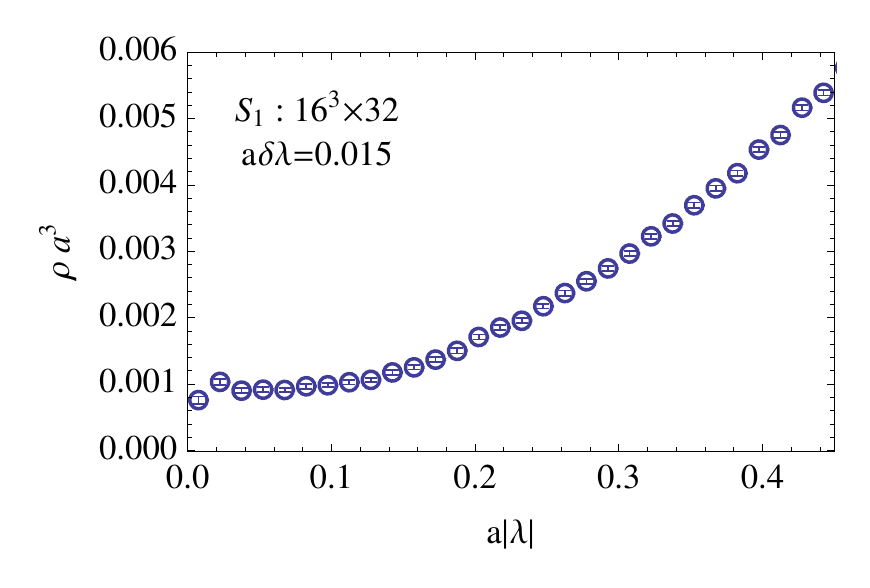}
    \hskip -0.00in
    \includegraphics[width=8.0truecm,angle=0]{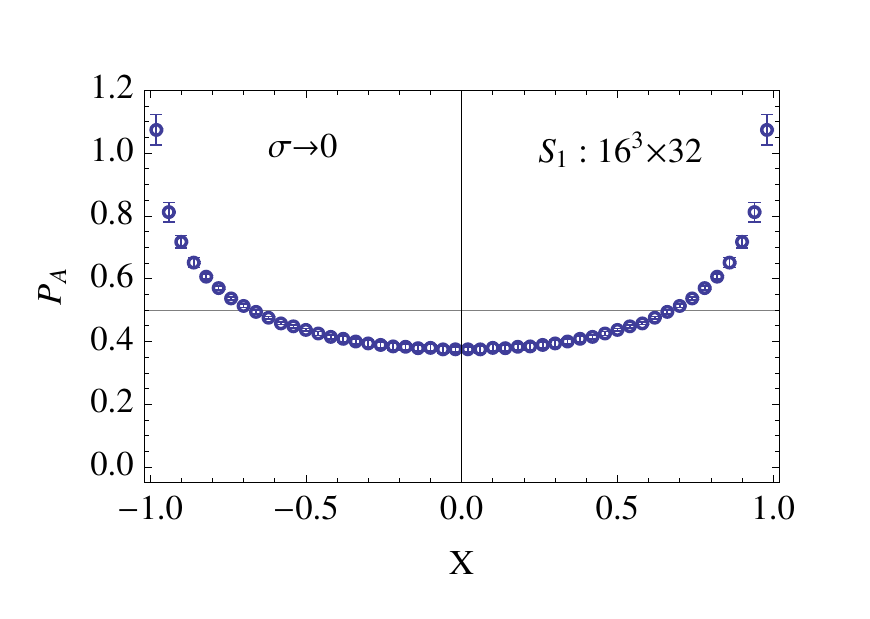}
     }
     \vskip -0.1in
     \caption{Chiral polarization characteristics for ensemble $S_1$. See the discussion
     in the text.}
     \label{fig:S1}
    \vskip -0.35in
\end{center}
\end{figure}

\subsection{Selected Results}

We now proceed to discuss the results from the perspective of 
vSChSB--ChP correspondence. Our extensive treatment of formalism in 
Sec.~\ref{sec:background} and analysis of finite--temperature data in 
Sec.~\ref{sec:fintemp} served in part to identify effective ways 
to perform a study of this type. Here we show the raw data for $\cop_A$ 
vs $\lambda$, together with the plots of $\sigma_{ch}(\sigma)$, 
$\rho(\lambda)$, and $\xd_A(\Xg,\sigma \to 0)$. The former three 
are designed to reveal whether vSChSB and ChP are tied together, 
while the latter serves as a first step to explore the newly proposed 
convexity connections in this particular corner of quark--gluon dynamics.

Fig.~\ref{fig:S1} shows the above set of characteristics for the system
at larger of the two quark masses, represented by ensemble $S_1$. A mere 
glance at the raw data (top left) reveals the presence of chiral 
polarization at low energy without noticeable depletion of eigenvalues 
near the origin. The system thus shows simultaneous signs of chiral 
polarization and mode condensation in accordance with vSChSB--ChP 
correspondence. This is confirmed by the behavior of $\sigma_{ch}(\sigma)$
(top right) and $\rho(\lambda)$ (bottom left). Indeed, the former shows
the positive bump at low $\sigma$, characteristic of chiral polarization, 
while the behavior in the latter is typical of mode--condensing theory.

\begin{figure}
\begin{center}
    \centerline{
    \hskip -0.05in
    \includegraphics[width=8.0truecm,angle=0]{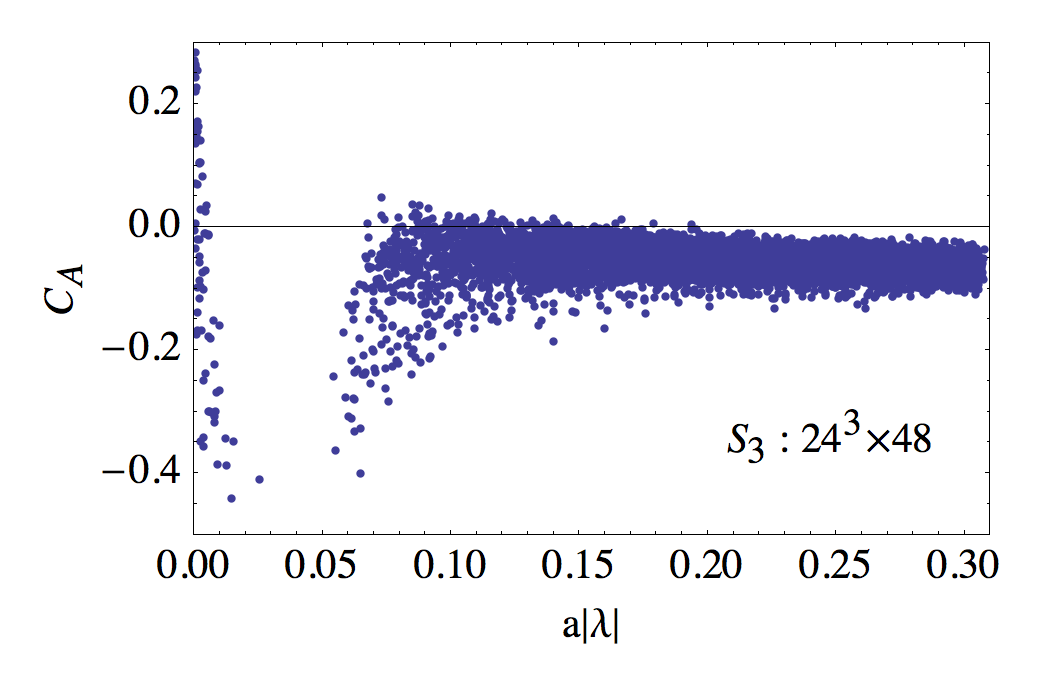}
    \hskip -0.00in
    \includegraphics[width=8.0truecm,angle=0]{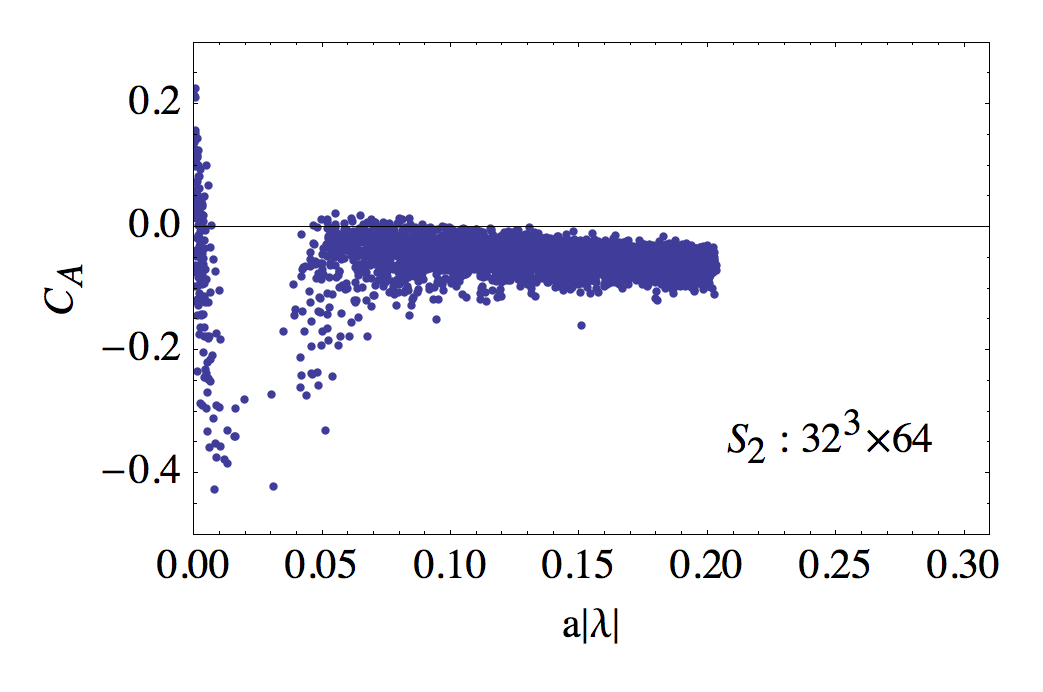}
     }
     \vskip -0.15in
    \centerline{
    \hskip 0.14in
    \includegraphics[width=8.0truecm,angle=0]{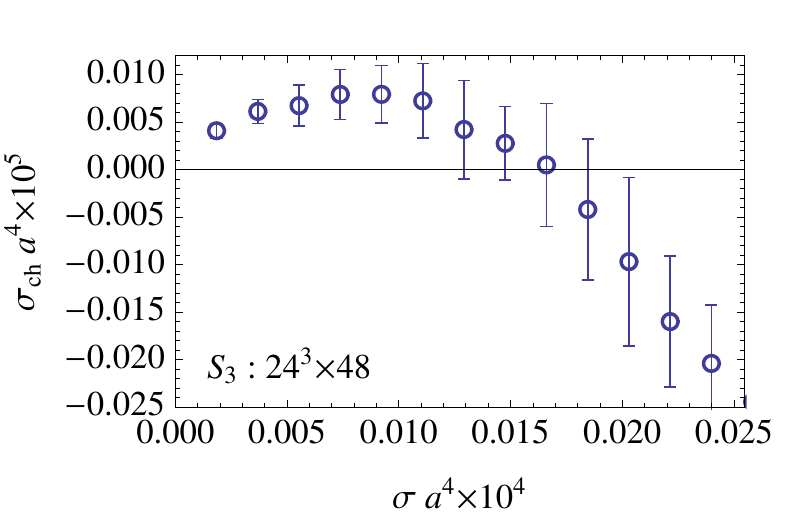}
    \hskip -0.00in
    \includegraphics[width=8.0truecm,angle=0]{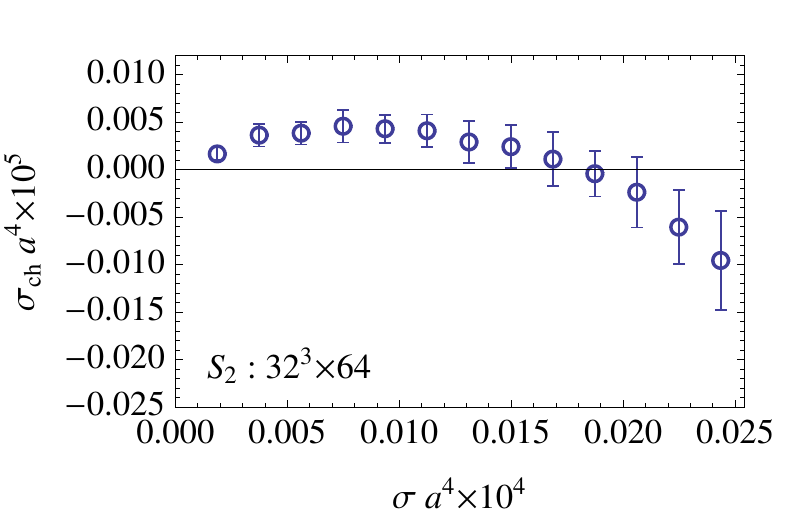}
     }
     \vskip -0.02in
    \centerline{
    \hskip 0.00in
    \includegraphics[width=8.0truecm,angle=0]{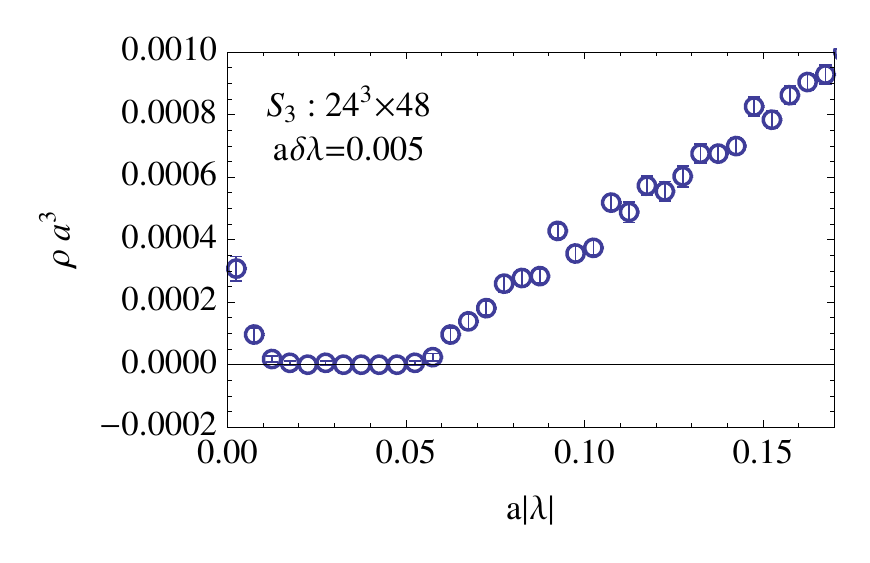}
    \hskip -0.00in
    \includegraphics[width=8.0truecm,angle=0]{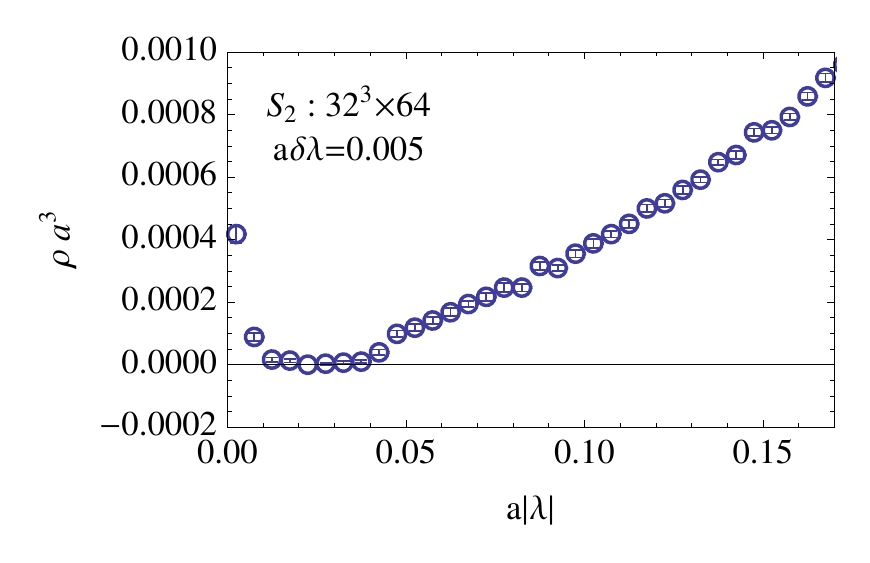}
     }
     \vskip -0.15in
    \centerline{
    \hskip 0.20in
    \includegraphics[width=8.0truecm,angle=0]{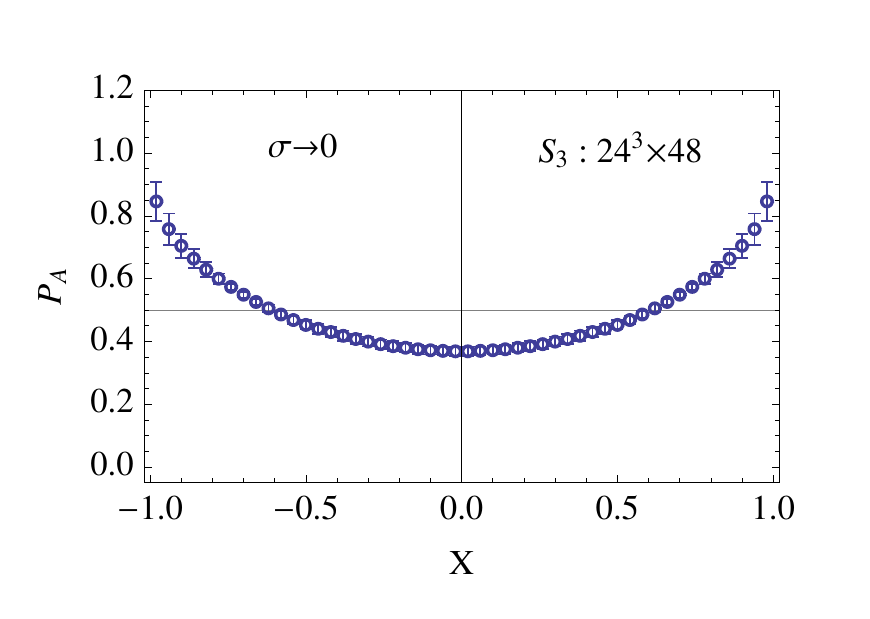}
    \hskip 0.05in
    \includegraphics[width=8.0truecm,angle=0]{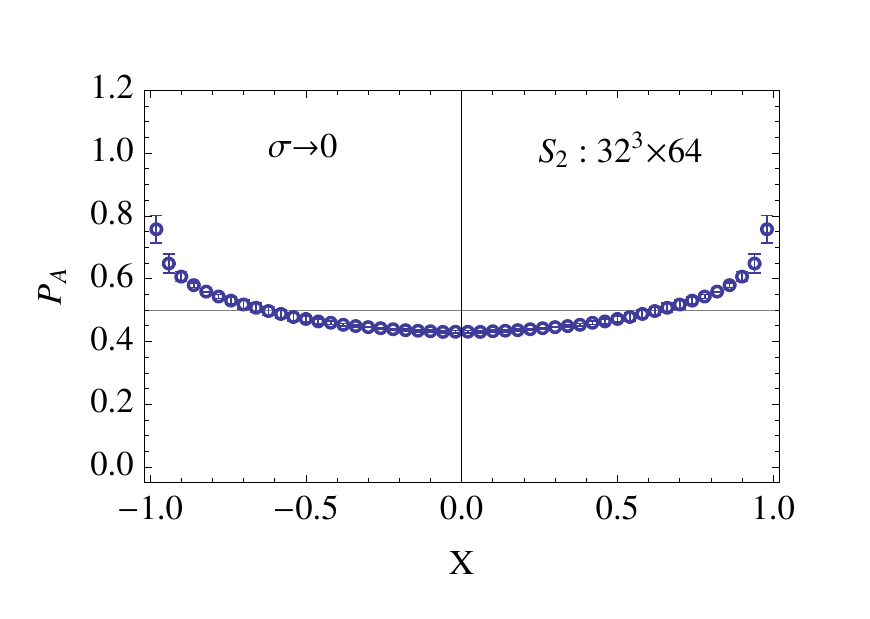}
     }
     \vskip -0.1in
     \caption{Chiral polarization characteristics for ensemble $S_2$ (right column) 
     and $S_3$ (left column). See the discussion in the text.}
     \label{fig:S2S3}
    \vskip -0.15in
\end{center}
\end{figure}

Focusing now on the situation at smaller mass, the same set of characteristics 
are shown in Fig.~\ref{fig:S2S3}, with the smaller volume ($S_3$) in the left 
column and the larger volume ($S_2$) in the right column. From the raw data
alone (top row) one can immediately see that a qualitative change in the Dirac 
spectrum indeed occurred. However, it is not a simple depletion of eigenvalues in 
the vicinity of the origin as one would naively expect. Rather, there is a break
in the spectrum, characterized by significant depletion, together with the 
accumulation of modes very close to zero. Given that, vSChSB--ChP correspondence 
predicts chiral polarization at the low end of the spectrum. This is indeed 
featured in the raw data qualitatively, and is properly quantified via the behavior 
of $\sigma_{ch}(\sigma)$ (second row).  

While the options are limited for finite--volume scaling with only two volumes
available, there is little doubt that the theory in question is overlap mode
condensing. Indeed, as one can see in the third row of Fig.~\ref{fig:S2S3},
the peak in the density near the origin is actually growing as the volume
is increased. On the chiral polarization side, the positive maximum in 
$\sigma_{ch}(\sigma)$, namely $\Omega_{ch}$, visibly shrinks at larger volume. 
Thus, a detailed finite volume study is required to decide whether 
the infinite--volume correspondence in the form of {\sl Conjecture 2} or 
{\sl Conjecture 2'} holds true.
However, the position of the maximum in $\sigma_{ch}(\sigma)$, namely $\Omega$, 
in fact mildly increases. We thus expect that the correspondence in the form 
of {\sl Conjecture 2''} certainly holds in this case. Needless to say, our results 
are also in agreement with chiral polarization characteristics being 
the finite--volume order parameters of vSChSB, and thus concur with 
{\sl Conjectures 3,3'}.

Lastly, we comment on the behavior of absolute \Xg--distributions for lowest
modes, i.e. $\xd_A(\Xg,\sigma\to 0)$. These results are shown in the lower right 
plot of Fig.~\ref{fig:S1} and the last row of Fig.~\ref{fig:S2S3}. Only the lowest
15 modes from each ensemble were used in the computation, leading to 
coarse--graining parameter $\Delta \ll \Omega$ in each case. All three 
systems exhibit convex behavior, thus following the same pattern observed in 
case of N$_f$=0 at finite temperature for chirally polarized theories ($T<T_{ch}$). 
It should be mentioned here that the same holds for the N$_f$=2+1 systems close
to ``real world QCD'' studied in Ref.~\cite{Ale12D}. It is thus reasonable
to expect that vSChSB is equivalent to convexity of $\xd_A(\Xg,\sigma\to 0)$ 
over the whole base set ${\eusm T}$.

\subsection{Intermediate Phases}

The above discussion of finite--temperature and many--flavor results evokes
certain analogy between the mixed phase in N$_f$=0 QCD and the situation 
found in N$_f$=12 at lighter mass (ensembles $S_2$ and $S_3$). It appears 
that decreasing quark mass for N$_f$=12 has a dynamical effect similar to 
increasing temperature for N$_f$=0 in that a phase with narrow band of 
near--zeromodes separated from the bulk is generated, at least for certain 
range of cutoffs. In thermal case, the onset of this behavior appears 
to coincide with deconfinement, and thus $T_c$. In many--flavor case,
where confinement is non--trivial to define, there is presumably a mass
$m_{in}$ below which this starts occurring.\footnote{If this behavior survives 
the continuum limit, then some universal parametrization of the transition
point e.g. in terms of ratios of certain hadron masses, will be more 
appropriate. The same applies to $m_{ch}$.}  
Regardless of whether the existence of $m_{in}$ is a lattice artifact,
we can define an analog of $T_{ch}$, namely $m_{ch}$, below which valence chiral 
symmetry gets restored in N$_f$=12. The aforementioned analogy would then apply
to the regime $T_c<T<T_{ch}$ of N$_f$=0 and $m_{ch}<m<m_{in}$ of N$_f$=12,
which we refer to as ``intermediate phases'' in what follows. It should be 
emphasized that whether $m_{ch}$ is zero or non--zero is currently an open issue.

In what follows we summarize few observations on the two intermediate phases
and compare them. Given that the available data on the many--flavor side is 
rather limited, this should be considered an initial assessment which can 
serve as a starting point for more detailed investigation.

\medskip

\noindent {\bf (i) Separation from the bulk.} Both intermediate phases are characterized 
by anomalous behavior of spectral density $\rho(\lambda)$, wherein the narrow 
peak forms near the origin, and is clearly distinguished from the rest of the spectrum.
This is ``anomalous'' in the sense that, by virtue of the above, $\rho(\lambda)$
becomes non--monotonic. Note that in the thermal case, data indicates that the peak 
of near--zeromodes becomes of $\delta$--function type in the infinite volume limit
while it is currently not clear what happens in N$_f$=12.

\medskip

\noindent {\bf (ii) Inhomogeneity.} The anomalous near--zeromodes are highly 
inhomogeneous in both intermediate phases. By this we mean that the bulk of their
norm is carried by very small fraction of space--time points. Here we will not 
focus on quantifying this feature, but it will certainly become a characteristic
of interest if either one of the intermediate phases turns out to be the reality of 
the continuum limit.

\medskip

\begin{figure}[t]
\begin{center}
    \centerline{
    \hskip 0.00in
    \includegraphics[width=8.0truecm,angle=0]{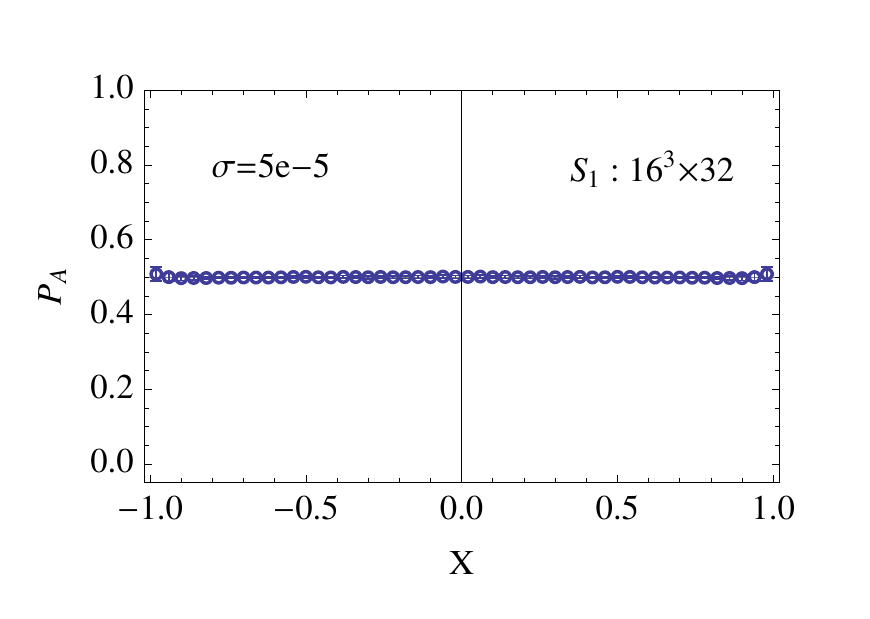}
    \hskip -0.00in
    \includegraphics[width=8.0truecm,angle=0]{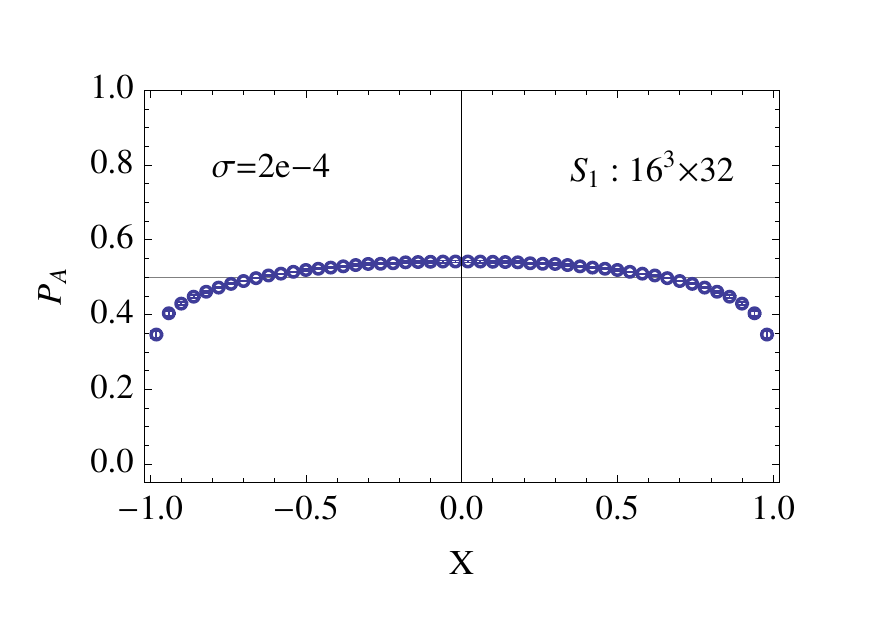}
     }
     \vskip -0.15in
    \centerline{
    \hskip 0.00in
    \includegraphics[width=8.0truecm,angle=0]{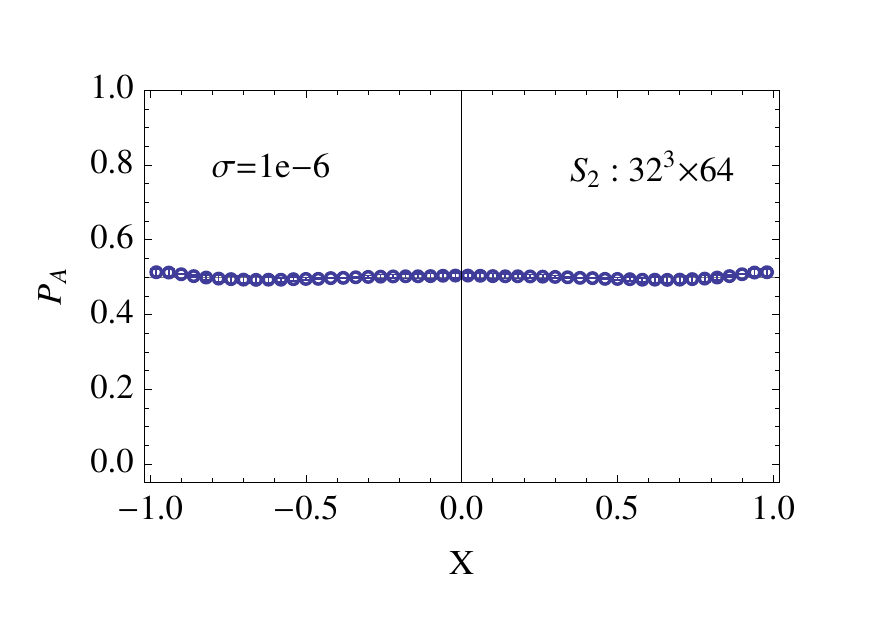}
    \hskip -0.00in
    \includegraphics[width=8.0truecm,angle=0]{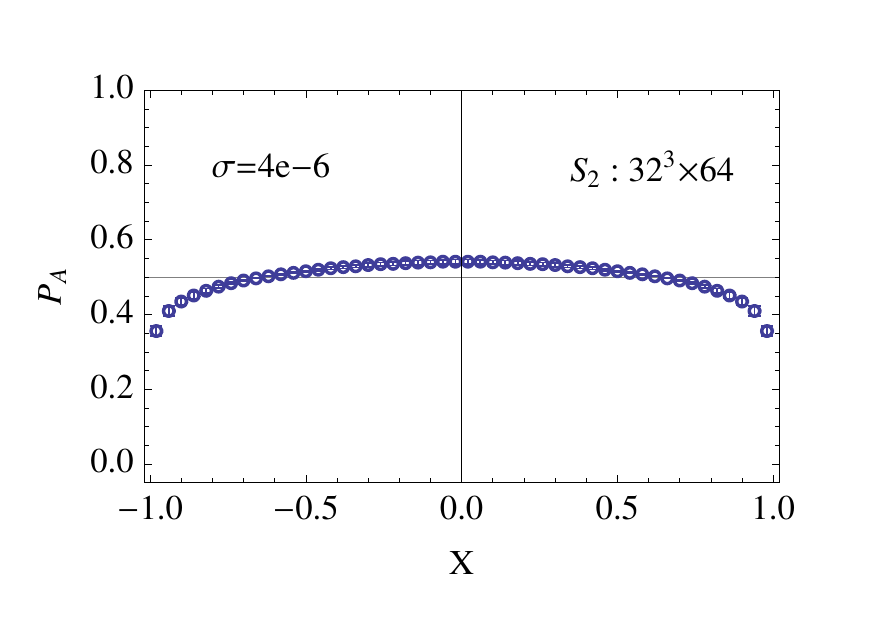}
     }
     \vskip -0.1in
     \caption{Absolute \Xg--distributions $\xd_A(\Xg,\sigma)$ for ensemble $S_1$ (top) 
     and $S_2$ (bottom). In the left column $\sigma\approx \Omega$ while in the right
     column $\sigma > \Omega$. Coarse--graining parameter $\delta=10^{-5}$ was 
     used for $S_1$ and $\delta = 2 \times 10^{-7}$ for $S_2$.}
     \label{fig:pa_nf12_trans}
    \vskip -0.35in
\end{center}
\end{figure}

\noindent {\bf (iii) Indefinite convexity.} As discussed in 
Secs.~\ref{ssec:Xd},\ref{ssec:convexity}, thermal transition to intermediate (mixed)
phase in N$_f$=0 is characterized by the appearance of modes with indefinite convexity
of absolute \Xg--distributions. In case of N$_f$=12 we do not find a clear evidence
of this happening. In fact, the spectral transition from chirally polarized to chirally
anti--polarized regime looks more like a direct transition from strictly convex to strictly 
concave distribution. In Fig.~\ref{fig:pa_nf12_trans} (left column) we show 
$\xd_A(\Xg,\sigma)$
for $\sigma \approx \Omega$. As one can see both for heavier mass (top) and the lighter
mass in intermediate phase (bottom), the absolute \Xg--distribution is flat which is 
characteristic of the direct transition. The right column in the figure illustrates the 
concave behavior at larger values of $\sigma$. Thus, to the extent that indefinite 
convexity of $\xd_A$ reflects deconfinement even in the situation with light dynamical
quarks, the intermediate phase in N$_f$=12 exhibits not only valence chiral symmetry 
breaking but also signs of confinement. Detailed inquiry at yet lower mass should clarify 
this further.

\section{Discussion}

The main purpose of this work is to test the idea that, for quark--gluon interactions 
of QCD type, generating chiral condensate is the same thing as generating the layer 
of chirally polarized Dirac eigenmodes at low end of the spectrum~\cite{Ale12D}. 
When viewed as a feature characterizing the state of vacuum correlations, this 
connection is not limited to theories containing massless quarks, but applies 
generically, even when all quarks are massive. In such generalized context,
vacuum correlations are probed by a pair of external (valence) massless quarks
and the role of chiral symmetry breaking is assumed by its valence counterpart.
Examining the temperature effects in N$_f$=0 QCD, and effects of many light flavors 
at $T\!=\!0$ (for N$_f$=12), we find a complete agreement with this vSChSB--ChP 
correspondence. 

One relevant aspect of this connection, both practically and conceptually, 
is that it is perfectly well defined even at lattice--regularized level. Indeed, if 
lattice fermions with exact chiral symmetry provide the massless probe, then vSChSB 
has a full-fledged lattice representation.\footnote{ChP is in fact well--defined 
even if probing fermion is not exactly chiral away from continuum limit.} 
The associated conjectures can then be formulated directly for lattice theories, 
as done here, but with compliance only expected sufficiently close to the continuum 
limit. Nevertheless, we found the agreement for every lattice system studied, even in 
corners where relevance for continuum physics hasn't yet been established, e.g. in 
the ``mixed phase'' of N$_f=0$ QCD. It is thus conceivable that the connection is 
even more robust than originally expected.

Important motivation to identify the correspondence of vSChSB--ChP type is to narrow 
down options in searches for specific mechanism of the breaking phenomenon. Indeed, 
the two aspects involved are not required to be locked together by general
principles and are in fact quite different in nature. Like that of any broken symmetry, 
the definition of vSChSB is intimately tied to thermodynamic limit (infinite volume). 
On the other hand, ChP is defined in any fixed finite volume which is quite natural
given its role to characterize and detect chiral symmetry breaking nature of 
the interaction at hand. Indeed, such indicator is not expected to turn itself on 
in infinite volume only, but rather when volume is sufficiently large to contain 
relevant finite scales of the theory.

Above considerations lead to the notion of {\em finite--volume ``order parameter''} 
introduced in Sec.~\ref{ssec:finvol}. This is intended to be a dynamical quantifier 
assuming non--zero value in sufficiently large {\em finite} volumes if and only if 
the symmetry in question is broken (in infinite volume). The width of chiral 
polarization layer ($\Lambda_{ch}$), the volume density of total chirality 
($\Omega_{ch}$), and the volume density of polarized modes ($\Omega$), are viewed 
as prototypes of such objects associated with vSChSB (see {\sl Conjectures 3,3'}$\,$), 
and all available data is consistent with this proposition. In fact, we have not yet 
encountered a reversal of dynamical tendency for chiral polarization due to the volume 
being too small. In this vein, it is instructive to think of chiral polarization 
framework as an attempt to construct a ``predictor'' of valence chiral symmetry 
breaking for QCD--like theories. Such predictor, say $\Omega$, will be most efficient 
if it is also a finite--volume order parameter. Indeed, a well--founded wager on 
vSChSB can then be made based on the value of $\Omega$ in a single otherwise 
acceptable volume. 

Conceptual simplicity in the notion of finite--volume order parameter is further 
underlined by the fact that all three ChP--based characteristics are (non)zero 
simultaneously in finite volume. Thus, although the state of chiral polarization 
is characterized by the triplet
\begin{equation}
  \mbox{\rm ChP}_{V<\infty}  \quad \longleftrightarrow \quad
  \bigl(\,\Lambda_{ch}(V),\, \Omega_{ch}(V),\, \Omega(V) \,\bigr)
\end{equation}
the associated vSChSB inquiry in fact involves a single object, namely
\begin{equation}
   \chi_p(V) \,\equiv\, \sgn \bigl( \Lambda_{ch} (V) \bigr) 
             \,=\,      \sgn \bigl( \Omega_{ch} (V) \bigr) 
             \,=\,      \sgn \bigl( \Omega(V) \bigr) \; \in \{0,1\} 
\end{equation}
Note that, as finite--volume order parameters, elements of ChP$_{V<\infty}$ are not 
required to have non--zero infinite--volume limits when symmetry is broken, but 
$\chi_p \!\equiv\! \lim_{V\to\infty} \chi_p(V)=1$.

If, contrary to initial evidence, the notion of finite--volume order parameter on 
$\euT$ doesn't materialize, vSChSB--ChP correspondence can still be based on order 
parameters of more traditional variety: those indicating the symmetry breakdown 
by positive {\em infinite--volume} value. The situation becomes more structured
though, since the classification of possible ChP behaviors in infinite volume is 
given by the heptaplet of non--negative parameters
\begin{equation}
  \mbox{\rm ChP}_{V=\infty} \quad \longleftrightarrow \quad
  (\Lambda_{ch}, \Lambda_{ch}^\infty, \Omega_{ch}, \Omega_{ch}^\infty,
  \Omega, \Omega^\infty,\chi_p)
  \label{eq:5.010} 
\end{equation}
with variety of mixed--positivity scenarios allowed in principle. In this language,
the hypothetical lack of finite--volume order parameter implies that $\chi_p$ is 
not a good infinite--volume indicator of vSChSB. Nevertheless, in the vSChSB--ChP 
correspondence being constructed, $\chi_p\!=\!0$ still indicates symmetric vacuum 
since it implies vanishing of all elements in ChP$_{V=\infty}$ and thus  
absence of any polarized behavior. However, to formulate the equivalence, one needs 
to specify which forms of ChP$_{V=\infty}$ with $\chi_p\!=\!1$ are associated with 
vSChSB. This heavily depends on the scope of polarized behaviors generated 
by theories in $\euT$.

To this effect, we formulated three versions of such infinite--volume correspondence 
that are not mutually exclusive, but include an increasingly large variety of chiral 
polarization. The most restrictive form, {\sl Conjecture 2}, assumes that no singular 
cases occur sufficiently close to the continuum limit~\cite{Ale12D}. Here 
``singular'' refers to any combination of the first six elements in ChP$_{V=\infty}$ 
with mixed positivity, or with paired characteristics not matching 
(e.g. $\Lambda_{ch} \ne \Lambda_{ch}^\infty$). In this case, each element of 
ChP$_{V=\infty}$ (except $\chi_p$) is individually a valid infinite--volume order 
parameter of vSChSB. Such scenario was only found to be violated in the narrow mixed 
phase of N$_f$=0 QCD, but complete agreement may hold closer to continuum limit,
which is sufficient. In {\sl Conjecture 2'} we minimally expanded the range of 
anticipated ChP behaviors in order to include the singular option encountered above, 
namely $\Lambda_{ch}\!=\!\Lambda_{ch}^\infty\!=\!0$ and $0\!<\!\Omega_{ch}\!<\!\Omega$. 
Here $\Omega_{ch}$, $\Omega$ and their partners are each an infinite--volume 
order parameter of vSChSB. Most generally, {\sl Conjecture 2''} admits all singular 
ChP behaviors, in which case $\Omega$ becomes the sole reliable indicator of 
vSChSB: the layer of chirally polarized modes is expected to be physically relevant 
only when total volume density of participating modes remains positive in 
the infinite--volume limit.

We emphasize that having differently focused versions of vSChSB--ChP correspondence
simply reflects differently aimed benefits of the underlying relationship. Indeed, 
the more restrictive the form of correspondence turns out to be valid, the more 
information on the continuum behavior of vSChSB it conveys. On the other hand, 
the more generic the formulation becomes, the wider its applicability becomes 
in terms of cutoff theories. In an extreme case, the relationship might turn out
to be as generic as the vSChSB--QMC correspondence, i.e. valid at arbitrary 
non--zero cutoff. 

The analysis in this paper (and the discussion above) focuses attention to $\Omega$ 
as a central characteristic of ChP in relation to vSChSB. Indeed, not only 
does $\Omega$ provide for the simple and most generic way to express vSChSB--ChP 
correspondence, it is also expected to be a universal quantity characterizing QCD 
vacuum (see Sec.~\ref{ssec:universal}). Moreover, our numerical experiments show that
the most practical scheme for detecting chiral polarization proceeds via computing 
the dependence of $\sigma_{ch}$ on $\sigma$, from which $\Omega$ (and $\Omega_{ch}$) 
is directly determined.

During the course of our main inquiry we encountered few results that are noteworthy 
in their own right. The first one relates to the issue of thermal mixed phase
($T_c < T < T_{ch}$) in N$_f$=0 QCD, i.e. the existence of deconfined system in real 
Polyakov line vacuum but with broken valence chiral symmetry~\cite{Edw99A}. This is 
a well--posed and interesting question even at given cutoff, but its resolution 
requires a careful volume study. Our data involves several volumes and the results 
indicate that the mixed phase with overlap valence quarks does indeed exist, at least 
for some range of cutoffs. The associated vSChSB proceeds via band of spatially 
inhomogeneous and chirally polarized near--zeromodes, well separated from the rest of 
the spectrum. The width of the band appears to vanish in the infinite--volume limit, 
possibly involving $\delta(\lambda)$ singularity in spectral density. We found that 
a similar phase also exists in N$_f$=12 theory at light quark mass, which deserves 
a dedicated study.

The last side result we wish to discuss suggests novel characterization of 
(de)confinement. While at $T<T_c$ only convex or concave absolute 
$\Xg$--distributions of Dirac modes are found in N$_f$=0 QCD, the band of 
convex--indefinite modes appears at $T>T_c$. Thus, at least in this setting, 
the existence of such band seems to play the same role for deconfinement as 
the existence of chirally polarized band plays for vSChSB. Both layers are present 
in the mixed phase as they should. The situation at high temperatures brings up 
an interesting question, namely how does the ``concavity edge'', marking the transition 
from convex--indefinite to concave behavior in Dirac spectra at $T>T_c$, relate to 
``mobility edge'' feature discussed in Refs.~\cite{Gar06A,Kov10A,Kov12A}? 
With natural expectation being the coincidence of the associated scales, computations 
needed to explore this issue are straightforward to set up.

\bigskip

\noindent{\bf Acknowledgments:} 
We are indebted to Anna Hasenfratz and David Schaich for sharing their N$_f$=12 
staggered fermion configurations for the purposes of this work. Thanks to Terry Draper 
for discussions, and to Mingyang Sun for help with graphics. A. A. is supported by U.S. 
National Science Foundation under CAREER grant PHY-1151648. I.H. acknowledges 
the support by Department of Anesthesiology at the University of Kentucky.

\bigskip\medskip

\begin{appendix}

\section{Absolute Polarization and Dynamical Chirality}
\label{app:chirality}

In this Appendix we briefly describe the absolute (dynamical) polarization method
of Ref.~\cite{Ale10A}. In the present context the elementary object of study is 
a chirally decomposed eigenmode $\psi=\psi_L+\psi_R$. Given that we are only 
interested in local (on--site) relationship between left and right, the sufficient 
information is stored in the probability distribution $\df(\psi_L,\psi_R)$ of its 
values $(\psi_L(x),\psi_R(x))$. This setup coincides with the starting point of 
a general approach that considers arbitrary stochastic quantity $Q$ with values 
in a vector space decomposed into a pair of equivalent orthogonal subspaces 
($Q = Q_1 + Q_2$, $Q_1 \cdot Q_2 = 0$), and whose ``dynamics'' is described by 
symmetric probability distribution $\df(Q_1,Q_2)=\df(Q_2,Q_1)$. 

Given the goal of characterizing the (normalized) asymmetry between the two subspaces 
in favored values of $Q$ (polarization), the method proceeds by first marginalizing 
the full distribution $\df(Q_1,Q_2)$ to the distribution of magnitudes $\db(q_1,q_2)$. 
Indeed, the weight of a given subspace in sample $Q$ can be assessed via magnitude 
$q_i \equiv |Q_i|$ of its component. One possibility for a normalized variable 
expressing the desired relationship is~\cite{Hor01A}
\begin{equation}
    \rpc \,=\, \frac{4}{\pi} \, \tan^{-1}\Bigl(\frac{q_2}{q_1}\Bigr) \,-\,1   
          \equiv \Fgr(q_1,q_2) 
    \label{eq:3.100}
\end{equation}
namely the {\em reference polarization coordinate}. Note that $\rpc \in [-1,1]$ 
with extremal values taken by samples strictly polarized into one of the subspaces, 
and zero value assigned to strictly unpolarized samples ($q_1=q_2$). Probability 
distribution of $\rpc$ in $\db(q_1,q_2)$, namely
\begin{equation}
    \dop(\rpc) \,=\,  
      \int_0^\infty d q_1 \int_0^\infty d q_2 \,
      \db(q_1,q_2) \; \delta\Bigl( \rpc - \Fg_r(q_1,q_2) \Bigr)
      \label{eq:3.110}
\end{equation}
is called the {\em reference \Xg--distribution}, and represents detailed polarization 
characteristic of dynamics $\df(Q_1,Q_2)$ with respect to polarization measure $\Fgr$.

The large freedom in choosing the polarization measure makes the above characteristic 
highly non--unique and thus kinematic. Various ``reference frames'' of polarization 
can be represented by suitably constructed {\em polarization functions} $\Fg(\rpc)$. 
In this language, the reference \Xg--distribution is associated with polarization 
function $\Fg_r(\rpc)=\rpc$. The main idea driving the absolute polarization method 
is to adjust the polarization function characterizing $\db(q_1,q_2)$ so that it
measures polarization relative to its ``own statistical independence'', namely 
relative to the stochastic dynamics described by 
\begin{equation}
    \db^u(q_1,q_2) \,\equiv\, p(q_1)\, p(q_2)   \qquad\qquad
    p(q) \equiv  \int_0^\infty d q_2 \, \db(q,q_2) =
                      \int_0^\infty d q_1 \, \db(q_1,q)
   \label{eq:3.120}
\end{equation}
One can show~\cite{Ale10A} that this is accomplished by utilizing the polarization function
\begin{equation}
   \Fg_A(\rpc) \equiv 2 \int_{-1}^\rpc d y \, \dop^u(y) - 1 
   \label{eq:3.130}   
\end{equation}
where reference \Xg--distribution $\dop^u(\rpc)$ is associated with uncorrelated 
dynamics $\db^u(q_1,q_2)$. The corresponding distribution of polarization values, namely
\begin{equation}
    \xd_A(\Xg) \,\equiv\,  
      \int_{-1}^1 d \rpc \,
      \dop(\rpc) \; \delta\Bigl( \Xg - \Fg_A(\rpc) \Bigr)
      \,=\, \frac{1}{2} \, 
                  \frac{\dop   \Bigl(  \Fg_A^{-1}\,(\,\Xg\,) \Bigr)}
                       {\dop^u \Bigl(  \Fg_A^{-1}\,(\,\Xg\,) \Bigr)}
      \label{eq:3.140}
\end{equation}
is called the {\em absolute \Xg--distribution}. By construction, and as seen from
the above explicit form, $\xd_A(\Xg)$ is a differential measure quantifying polarization 
tendencies relative to statistical independence. Moreover, it is unique: arbitrary 
choice of the reference polarization coordinate (function) leads to the same
absolute \Xg--distribution~\cite{Ale10A}.\footnote{This uniqueness of the ``correlational'' 
approach is in fact why the method is referred to as {\em absolute}.}
Consequently, absolute \Xg--distribution is viewed as a genuinely dynamical concept.
 
Differential information contained in $\xd_A(\Xg)$ can be integrated into the 
{\em correlation coefficient of polarization} $\cop_A \in [-1,1]$, namely
\begin{equation}
   \cop_A \,\equiv \,
   2\,\int_{-1}^{1} d \Xg \, |\Xg| \, \xd_A(\Xg) \,-\,1
   \label{eq:3.150}
\end{equation}
Statistical meaning of $\cop_A$ is clarified by noting that the integral in the above 
expression is the probability for sample drawn from $\db(q_1,q_2)$ to be more polarized 
than sample drawn from $\db^u(q_1,q_2)$. Consequently, positive correlation means that 
stochastic dynamics enhances polarization relative to statistical independence, while 
negative correlation (anti--correlation) implies its suppression. $\df(Q_1,Q_2)$ is 
said to support {\em dynamical polarization} in the former case and 
{\em dynamical anti--polarization} in the latter.
In the context of Dirac eigenmodes and their chiral decomposition, 
expressions like ``mode is chirally polarized'' vs ``anti--polarized'', 
``mode is chirally correlated'' vs ``anti--correlated'', or 
``mode supports dynamical chirality'' vs ``anti--chirality'',
are all verbal descriptions of $\cop_A>0$ vs $\cop_A<0$.

\section{Generalities on Spectral Definitions}
\label{app:spectral}

The primary entities involved in spectral definitions are the cumulative 
densities $\sigma(\lambda)$ and $\sigma_{ch}(\lambda)$. At the regularized level, 
$\sigma$ is proportional to certain cumulative probability function of $\lambda$ 
(non--decreasing, bounded) and can thus have at most countably many finite 
discontinuities. All of its possible behaviors are then contained in the form 
\begin{equation} 
   \sigma(\lambda,M,V) \,=\, 
   \sum_j A_j\, H(\lambda - \alpha_j) \;+\; \hat{\sigma}(\lambda,M,V)
   \label{eq:a.010}
\end{equation}
where $\alpha_j=\alpha_j(M,V) \ge 0$ are the points of discontinuity,
$A_j=A_j(M,V) \ge 0$, and $H(x)$ is the left--continuous version of the Heaviside 
step function ($H(x)=0$ for $x \le 0$ and $H(x)=1$ for $x>0$). $\hat{\sigma}$ is 
a continuous non--decreasing function of $\lambda$ and, as such, it can only be 
non--differentiable on the set of Lebesgue measure zero. However, the exotic 
``Cantor function''--like cases, where the subset of non--differentiability is 
uncountable, are very unlikely to appear in this physical context. Differentiable 
functions producing derivatives that are discontinuous on uncountable subsets can 
be quite safely omitted for the same reason. Thus, the ``standard model'' of 
cumulative mode density, expected to cover all theories considered, 
is \eqref{eq:a.010} with $\hat{\sigma}(\lambda)$ being a continuous non--decreasing 
function that is continuously differentiable except for countably many (thus isolated) 
points. The differential representation (generalized function) then exists
and is given by
\begin{equation}
   \bar{\rho}(\lambda,M,V) \;=\; 
   \sum_j A_j\, \delta(\lambda - \alpha_j) \;+\; \hat{\rho}(\lambda,M,V)
   \label{eq:a.020}
\end{equation}
where the ordinary function $\hat{\rho}(\lambda) = d \hat{\sigma}(\lambda)/d_+\lambda$ 
exists and is continuous except for points $a_k=a_k(M,V) \ge 0$ where it can have 
an integrable divergence or simple discontinuity. Note that there is no a priori
relation between sets $\{\alpha_j\}$ and $\{a_k\}$.

The definition of cumulative chirality density is a priori less constraining on 
its behavior. Indeed, while $\sigma_{ch}(\lambda)$ is bounded in absolute value 
by $\sigma(\lambda)$, it is not necessarily monotonic,
and the most general form analogous to \eqref{eq:a.010} is thus not strictly 
guaranteed. On the other hand, more singular behavior of $\sigma_{ch}(\lambda)$ would 
require $\cop_A(\lambda)$ -- a dynamical property -- to be discontinuous on uncountable 
subsets which is highly unlikely in this physical setting. Thus, the most general 
form of $\sigma_{ch}(\lambda)$ expected to occur is
\begin{equation} 
   \sigma_{ch}(\lambda,M,V) \,=\, 
   \sum_j B_j\, H(\lambda - \beta_j) \;+\; \hat{\sigma}_{ch}(\lambda,M,V)
   \label{eq:a.030}
\end{equation}
where $B_j=B_j(M,V)$ has indefinite sign, $\beta_j=\beta_j(M,V) \ge 0$, 
and $\hat{\sigma}_{ch}$ is a continuous function of $\lambda$. In slightly
more restrictive ``standard model'', guaranteeing differential representation, 
$\hat{\sigma}_{ch}(\lambda)$ is also continuously differentiable except for 
countably many points, i.e.
\begin{equation}
   \bar{\rho}_{ch}(\lambda,M,V) \;=\; 
   \sum_j B_j\, \delta(\lambda - \beta_j) \;+\; \hat{\rho}_{ch}(\lambda,M,V)
   \label{eq:a.040}
\end{equation}
The function $\hat{\rho}_{ch}(\lambda) = d \hat{\sigma}_{ch}(\lambda)/d_+\lambda$ 
exists and is continuous everywhere except for points $b_k=b_k(M,V) \ge 0$ 
where it can have an integrable divergence or a simple discontinuity. Note that 
the sets $\{\beta_j\} \subseteq \{\alpha_j\}$ need not be identical, and neither
do the sets $\{a_k\}$, $\{b_k\}$. 

There is very little doubt that the behavior of $\sigma(\lambda)$ and 
$\sigma_{ch}(\lambda)$ in all theories considered falls under the ``standard model'' 
description specified above. In fact, it appears very likely that only more 
restricted forms will actually appear. Nevertheless, it is interesting to note
that the key concepts utilized in our discussion, namely that of chiral polarization
scale $\chps$ and low--energy chirality $\Omega_{ch}$, are well defined without 
any assumptions placed on $\sigma_{ch}(\lambda)$ beyond its existence. More
precisely, below we put forward definitions that assign definite characteristics 
to any real--valued function $\sigma_{ch}(\lambda)$ such that 
$\sigma_{ch}(\lambda) \equiv 0$ for $\lambda \le 0$, and bounded on any 
$(-\infty,\Lambda]$.  

We first define $\chps$ as the ``largest'' $\Lambda$, such that
$\sigma_{ch}(\lambda)$ is strictly increasing on $[0,\Lambda]$, i.e.
\begin{equation}
   \chps[\sigma_{ch}] \,\equiv\, \sup \, \{\, \Lambda \, | \, 
   \sigma_{ch}(\lambda_1) < \sigma_{ch}(\lambda_2) 
   \;\, \mbox{\rm for all} \;\,
   0 \le \lambda_1 < \lambda_2 \le \Lambda \,\}
   \label{eq:a.050}    
\end{equation}
when $\sigma_{ch}$ is not strictly increasing on $[0,\infty)$, and $\chps=\infty$
otherwise. Note that when positive $\chps$ doesn't exist, the defining condition 
\eqref{eq:a.050} is vacuously satisfied by $\chps=0$ which is then its assigned
value. Thus, $\chps[\sigma_{ch}]$ always exists and is non--negative.

Next, associate with $\sigma_{ch}(\lambda)$ its ``running maximum function''
\begin{equation}
   \sigma_{ch}^m(\lambda) \,\equiv\, 
   \sup \, \{ \sigma_{ch}(\lambda') \,|\, \lambda' \le \lambda \,\}
   \label{eq:a.060}
\end{equation}
which is a non--negative non--decreasing function bounded on any $(-\infty,\Lambda]$.
Thus, it can only be discontinuous via countably many finite jumps, and one--sided
limits exist everywhere. The low energy chiral polarization is then defined as
\begin{equation}
   \Omega_{ch}[\sigma_{ch}] \,\equiv\, 
   \lim_{\lambda \to \chps^+} \sigma_{ch}^m(\lambda) 
   \label{eq:a.070}    
\end{equation}
We emphasize that the above definitions of $\chps$ and $\Omega_{ch}$ coincide with those
given in the main text when $\sigma_{ch}(\lambda)$ is of ``standard form''.

\end{appendix}

\end{document}
\bye